\documentclass[rmp,aps,twocolumn,nofootinbib,floatfix]{revtex4}
\pdfoutput=1
\usepackage[usenames]{color}
\usepackage{graphics}
\usepackage{bbm}
\usepackage{amsmath}
\usepackage{verbatim}
\usepackage{graphicx}
\usepackage{dsfont}

\newcommand{\ra}{\rangle}
\newcommand{\la}{\langle}
\newcommand{\rperp}{\vec{r}_\perp}
\newcommand{\rvec}{\vec{r}}
\newcommand{\eplus}{\vec{E}^{(+)}}
\newcommand{\eminus}{\vec{E}^{(-)}}%
\newcommand{\Dplus}{\vec{D}^{(+)}}
\newcommand{\Dminus}{{\vec{D}^{(-)}}}
\newcommand{\ket}[1]{|#1\rangle}
\newcommand{\p}[2]{P_{L_{#1},\mathrm{\,#2}}}
\newcommand{\x}[2]{X_{L_{#1},\mathrm{\,#2}}}
\newcommand{\xa}[1]{X_{A,\,\mathrm{#1}}}
\newcommand{\pa}[1]{P_{A,\,\mathrm{#1}}}
\newcommand{\y}[1]{Y_\mathrm{\,#1}}
\newcommand{\q}[1]{Q_\mathrm{\,#1}}
\newcommand{\mean}[1]{\langle#1\rangle}

\begin{document}

\title{Quantum interface between light and atomic ensembles}

\author{Klemens Hammerer}
\affiliation{Institute for Theoretical Physics, University of Innsbruck, and Institute for Quantum Optics
and Quantum Information, Austrian Academy of Science, Technikerstrasse 25, 6020
Innsbruck, Austria}
\author{Anders S. S{\o}rensen and Eugene S. Polzik}
\affiliation{Niels Bohr Institute, Copenhagen University,
Blegdamsvej 17, Copenhagen 2100, Denmark}

\begin{abstract}
During the past decade the interaction of light with multi-atom ensembles has attracted a lot of attention as a basic building block for quantum information processing and quantum state engineering. The field started with the realization that \emph{optically thick} free space ensembles can be efficiently interfaced with quantum optical fields. By now the atomic ensemble - light interfaces have become a powerful alternative to the cavity-enhanced interaction of light with single atoms.  We discuss various mechanisms used for the quantum interface, including quantum nondemolition or Faraday interaction, quantum measurement and feedback, Raman interaction, photon echoe and electromagnetically induced transparency. The paper provides a common theoretical frame for these processes, describes basic experimental techniques and media used for quantum interfaces, and reviews several key experiments on quantum memory for light, quantum entanglement between atomic ensembles and light, and quantum teleportation with atomic ensembles. We discuss the two types of quantum measurements which are most important for the interface: homodyne detection and photon counting. The paper concludes with an outlook on the future of atomic ensembles as an enabling technology in quantum information processing.
Published in Reviews of Modern Physics \textbf{82}, 1041 (2010)\\ \url{http://link.aps.org/doi/10.1103/RevModPhys.82.1041}
\end{abstract}

\maketitle \tableofcontents


\section{Introduction}

\subsection{History and motivation}
Quantum features of atom-light interaction have been among the
central issues in physics since the early days of quantum mechanics.
Starting in the 1960s, with the development of quantum optics - the field where second
quantization of light is central -
quantum
electrodynamics became part of optical and atomic physics. For
decades after that the inherently quantum features of atom-light
interaction have been studied primarily within the framework of
cavity Quantum Electrodynamics (QED) where light can be efficiently
coupled to a few atoms or even to a single atom. Despite the
spectacular progress achieved in this direction, the complexity and
technical challenges associated with an atom strongly coupled to a
high-finesse cavity were calling for alternative approaches.

A new approach to the matter-light quantum interface came with the
realization of the fact that a large collection of atoms - an atomic
ensemble - can be efficiently coupled to quantum light if a
collective superposition state of many atoms can be utilized for the
coupling. The simplest example of such coupling is presented in Fig. \ref{elemlevels}a. A collection of atoms in the ground state is illuminated with
two modes of light $a_+$ and $a_-$. As shown by
\textcite{Kuzmich:1997}, if the modes possess quantum correlations
(entanglement) and light is absorbed by the atomic
ensemble, the quantum correlations can be mapped on the collective
superposition of the two final states of the atoms. The strong coupling
condition in this case amounts to the requirement of large resonant
optical depth $d$ of the atomic ensemble. It  later turned out, that
the requirement $d\gg1$ is the most significant
requirement for all types of the quantum interface between atomic
ensembles and light known up to now. The experiment demonstrating
that a quantum feature of radiation (squeezing) can be transferred
onto atoms via the process shown in Fig. \ref{elemlevels}a was performed by \textcite{Hald:1999}. This approach has been further developed using photon echo ideas \cite{Moiseev:2003}.

A natural next step was to utilize long lived atomic ground states for the interface via a Raman interaction in a $\Lambda$-scheme, as proposed by \textcite{Kozhekin:2000} for storage of squeezed states. The Raman process together with  Electromagnetically Induced Transparency (EIT) \cite{Fleischhauer:2005,Boller:1991,Lukin:2003} have soon become important routines for quantum interfaces. After \textcite{Hau:1999} demonstrated that EIT allows for very slow propagation of light  through an atomic ensemble it was quickly realized that  reducing the group velocity to zero  would enable an atomic memory for light \cite{Lukin:2000,Fleischhauer:2000} and  the first experimental demonstrations of this for classical pulses have been presented by  \textcite{Phillips:2001,Liu:2001}.

Quantum nondemolition (QND) measurement \cite{Braginsky:1996} based on light-matter interaction has emerged as a powerful tool for quantum state engineering, first in the cavity QED setting and then in the atomic ensemble context as an efficient method for generation of spin squeezing \cite{Kuzmich:1998}. Shortly thereafter QND interaction with atomic ensembles has become one of the main instruments for the quantum interface.

The process depicted in Fig. \ref{elemlevels}a is a
rudimentary example of one of the main routines for atoms-light
quantum interface: the quantum state transfer from light to atoms,
or the quantum memory for light. The ability for mapping, storing,
and retrieving quantum states of light -- the natural long-distance carrier of information
-- onto the material storage medium is one of the major enabling
procedures in quantum information processing. In this review we
cover various approaches to the quantum memory, including the
Raman process, EIT, photon echo, and the QND measurement and feedback. We review methods which provide a long term quantum memory with the fidelity
better than any classical procedure can achieve, as well as approaches which allow to preserve entanglement in the process of storage and retrieval. The interface can be implemented
either via interaction only or by the
teleportation-like procedure involving generation of entanglement,
Bell measurement on light and quantum feedback onto atoms. In this
article we will discuss both approaches.

The second most important routine for the quantum interface is
generation of entanglement between light and atoms. The light/atoms
entanglement in turn enables generation of entanglement between
remote atomic ensembles, as well as atomic teleportation and entanglement
swapping protocols. Furthermore the light--atom entanglement also allows for  quantum memory through light--atom
teleportation.

The quantum interface can be formulated either in the
Schr\"{o}dinger or in the Heisenberg picture. For example, the transformation of a quantum state of light into
a quantum state of atoms in the Schr\"{o}dinger
picture
$\hat{U}|\Psi_{L}0_{A}\rangle\rightarrow|0_{L}\Psi_{A}\ra$ corresponds to the operator transformation $\hat{U}^{\dag}\hat{a}_{A}\hat{U}=a_L$ in the
Heisenberg picture. The two pictures are equivalent,
and we will mostly use the Heisenberg picture throughout this
review.

In this review we discuss protocols based on both homodyning of light and photon counting. The most dramatic difference between the two approaches is that a single homodyne measurement does not necessarily distinguish between a vacuum and a non-vacuum state, whereas an ideal photon counter does. This makes ideal photon counting insensitive to losses if the protocol is conditioned on a click of the detector. This feature is important as an elementary purification mechanism, but it also makes protocols which use it probabilistic. On the other hand, homodyning always yields a measurement result and is thus deterministic. Another difference between the two measurements is that photon counting yields a discrete variable result, whereas homodyning yields a continuous variable outcome. From a practical perspective detectors used for homodyning are almost perfect in their quantum efficiency and dark current, whereas photon counters usually are less than perfect (although the progress in their development driven by quantum information applications is remarkable). The distinction between the two approaches is, however, not strict, as discussed later in this review. For example, photon counting can be used as a deterministic characterization of a protocol if the absence of a photon count (the vacuum contribution) is included in the analysis. Continuous variable outcome of a homodyne measurement can be "digitized" if suitable superposition states are employed. Various figures of merit are used for characterization of quantum interfaces, as discussed in Sec. \ref{sec:figure_of_merit}.

Probabilistic
protocols based on generation of a single collective atomic
excitation of an atomic ensemble following detection of a photon
emitted by the ensemble have been actively developed in the recent
years \cite{vanderWal:2003,Kuzmich:2003,Chou:2004,Matsukevich:2004,Chen:2006,Matsukevich:2006,Chaneliere:2007,Chou:2007}. This approach has been motivated by a proposal for a quantum repeater with atomic ensembles \cite{Duan:2001}. The research on quantum repeaters deserves a separate review paper and will only be rather briefly discussed in this article.

The requirements for the atomic memory may differ depending on
the particular application. A distributed quantum computer network
requires the complete set of memory capabilities: mapping of the
light state onto memory, storage and operations on the
memory state, and retrieval of the memory state back onto light for
further processing. Applications in quantum communications, which
involve local operations on stored states and classical
communications between partners, often only require a measurement of the memory state
in a specific basis, i.e., no full retrieval of the quantum state of the memory back onto light is
necessary. Yet other proposals, such as linear optics quantum computing which uses off-line entanglement resources require only the retrieval of the atomic state onto light but it has to be rather efficient and with high fidelity \cite{Menicucci:2006}.

\subsection{Elementary level schemes}

Naturally, the atomic levels used for storage of a quantum state
should be long lived, with particular requirements for the lifetime
depending on applications. For example, for a
memory used in a long distance communication protocol, the
memory lifetime usually should be longer than the time required for
classical communication over this distance. For a few hundred
kilometers this time is of the order of $10^{-3}$ sec. Short distance
applications may require shorter memory time but one should keep in
mind that a low loss fiber loop can be a strong competitor for a
short term atomic quantum memory for light  \cite{Pittman:2002}. A $5$ km fiber loop can,
in principle, store a photon for $25~\mu{\rm sec}$ with only $20\%$
losses at the telecom wavelength. However even for short storage times
atoms have the important advantage of being able to provide on-demand retrieval and may in addition be
advantageous if a nontrivial operation has to be performed on the stored quantum states.

\begin{figure}[ht]
  \includegraphics[width=8.3cm]{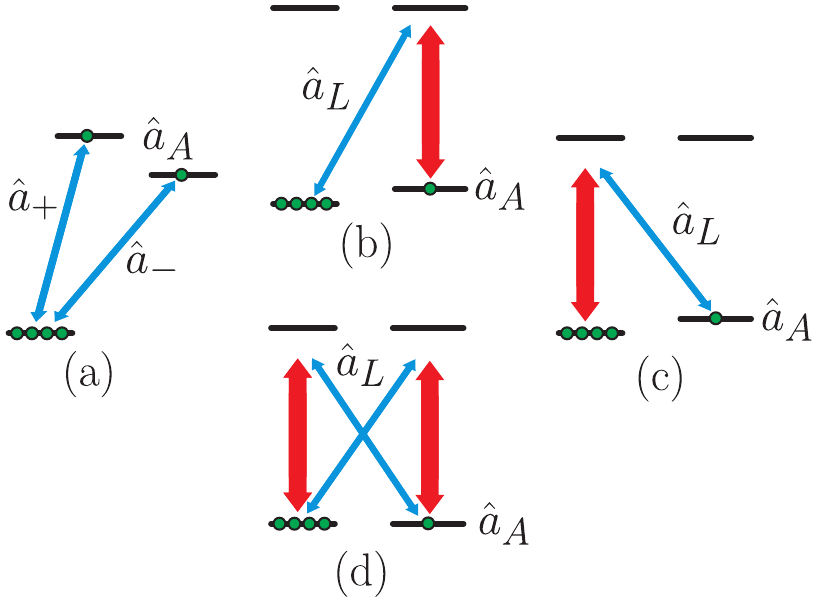}
  \caption{Elementary level schemes: (a) A simple absorption scheme with the quantum fields $a_{\pm}$ mapped onto the atomic states $a_{A}$, (b) Beam splitter type interaction - basis for Raman and EIT memory schemes, (c) Parametric gain type interaction - basis for entanglement schemes, (d) Quantum non-demolition (or Faraday) interaction - basis for entanglement, memory and teleportation schemes}\label{elemlevels}
\end{figure}

Preferably the optical atomic transitions used for coupling light to the storage
ground states of the atoms should be strong in order to have a large bandwidth of the memory. 
Therefore strong dipole allowed transitions are typically used for
the interaction, but some experiments, in particular in solid state systems, compensate for weak optical transition by having a large number of atoms.
 Figs. \ref{elemlevels}(b), (c), and (d)
present the atomic level schemes typically used for the interface.
The (b) and (c) parts of the figure present the $\Lambda$ - scheme
used in the Raman  and  EIT memory schemes as well for entanglement generation. The (d) part shows the so-called Faraday interaction which is sometimes also referred to as
QND interaction for
reasons discussed below.

In (b) the atoms are prepared in the ground state
coupled to the quantum field $\hat{a}_{L}$ (thin line), with the bold line showing a strong coupling field. With a large detuning from the optically excited states, the interaction
Hamiltonian for such a system, after adiabatic elimination of the
excited state, can be cast in the form
$\hat{H}=\chi_{BS}\hat{a}_{L}\hat{a}_{A}^{\dag}+ {\rm H.C.}$ In quantum optics this
Hamiltonian is often referred to as the beam-splitter Hamiltonian \textcite{Leonhardt:2003}. The interface mixes the input atom and light states as a "beamsplitter" and the "reflection coefficient" of unity corresponds to a perfect state swapping between light and atoms.
The detailed derivation of this Hamiltonian for the light-atoms
interaction will be given later in the article, but the intuitive
picture is obvious - if a single photonic excitation is removed
(annihilated) from the field $\hat{a}_{L}$, a single collective atomic
excitation $\hat{a}_{A}^\dag$ is created. If this process is
efficient, it works as the Raman-type quantum memory for light introduced for atomic ensembles by \textcite{Kozhekin:2000} and described in Sec. \ref{sec:raman_eit}. The
same level scheme can be used for the EIT-based memory where the fields are resonant with the optical transition \cite{Fleischhauer:2000,Lukin:2000}, although in this limit it is essential to account for spontaneous emission and the effective beam splitter Hamiltonian is less applicable (see the detailed theory in Secs.
II
). EIT based quantum memory experiments are also described in Sec. \ref{sec:raman_eit}.

Part (c) of the figure shows the same atomic structure but now the
fields are arranged in a way which can be used for the atoms-light
entangling interaction with the Hamiltonian
$\hat{H}=\chi_{P}\hat{a}_{L}\hat{a}_{A}+{\rm H.C.}$ The Hamiltonian is formally
identical to the parametric gain interaction Hamiltonian \textcite{Leonhardt:2003}
which has been a workhorse for studies of entangled and
non-classical states of light since the 1960s. The important
new feature in the present case is that two entangled operators
belong to a light mode and an atomic mode respectively. This kind of
entanglement has been used for unconditional light-to-atoms
teleportation  experiment described in Sec. \ref{sec:teleportation} and its
probabilistic version discussed in Sec. \ref{sec:probabilistic} is the basis
for the repeater protocol of \textcite{Duan:2001}.

To complete the discussion of elementary level schemes used for
basic interface routines we consider the four-level scheme shown in
figure Fig. \ref{elemlevels}(d). The Hamiltonian for this interaction can be obtained by combining the beam-splitter Hamiltonian (Fig. \ref{elemlevels} (b))
 and the parametric
entangling Hamiltonian with equal coupling constants, $\chi_{BS}=\chi_{P}$ (Fig. \ref{elemlevels} (c)) provided that the two quantum fields
(thin lines) belong to the same mode:
$\hat{H}\sim \chi\hat{P}_{L}\hat{P}_{A}$ where the canonical operators
for light and atoms which obey the
canonical commutation relation $[\hat{X},\hat{P}]=i$ have been introduced. This interaction allows
for a QND measurement of the atomic operator $P_{A}$ by means of
detection of the light operator $X_{L}$. As discussed
in detail in Sections \ref{sec:Entanglement} A,B the QND measurement projects atoms into
an entangled state. The same interaction is often called the quantum Faraday interaction because in case of magnetic levels it leads to polarization rotation of light. The QND-Faraday interaction of atoms and light followed by the measurement on the light and the feedback conditioned on the measurement applied onto
atoms was used to demonstrate quantum memory for light as described in Section \ref{sec:memory}.

As shown theoretically and experimentally in \cite{Wasilewski:2009}, a more general combination of the beam-splitter Hamiltonian and the parametric entangling Hamiltonian with \emph{unequal} weights performs both those operations simultaneously.

To summarize, the basic features of most quantum interface protocols
to date can be understood by analyzing simple 3-- or 4--level atoms. Besides the condition $d\gg1$ mentioned above, another
unifying feature for all approaches which use multi-atom ensembles is
the possibility to initialize the ensemble, e.g., by optical pumping,
in one of the ground substates. Choosing suitable atomic transitions, and polarizations and frequencies of the quantum and classical coupling fields one can
choose between various routines, such as memory, entanglement and
Faraday interaction, as shown in Fig. \ref{elemlevels}. In the following theoretical description we will provide a unified approach to all these types of interfaces.

In the literature the interface protocols are often divided into those for states of continuous variables and those for discrete variables, or qubits. The former are usually based on the Faraday interaction and are described most conveniently in terms of $X,P$ operators measured by homodyne detection, while the latter are commonly based on the $\Lambda$-type interactions in combination with counting of single photons and are most easily described in terms of $a,a^\dagger$ representations. One of the goals of this review is to show that continuous and discrete variable protocols can in many aspects be treated on equal footing, and that the choice of variables is defined by the convenience of description and the type of measurements involved. E.g., in the ideal limit a memory protocol which is most conveniently described by $X,P$ operators could be used to store a single photon with perfect fidelity. On the other hand, the state of the atomic memory in protocols which use $a,a^\dagger$ representation can be conveniently analyzed by atomic tomography in the $X,P$ basis \cite{Fernholz:2008,Sherson:2006c} (for a recent review on quantum tomography see \cite{Lvovsky:2009}. The type of errors which appear under non-ideal conditions is of course different for different protocols, but irrespectively of the specific protocol, the condition $d\gg1$, leads to fewer errors. Which protocol to use, for a given optical density, therefore depends on the specific application and a detailed analysis of the imperfections should be made in each situation.

This review strives to provide a coherent picture of the work on the quantum interface between light and atomic ensembles using various approaches, atomic media, and protocols. It includes the discussion of major experimental achievements to date and concludes with the analysis of the current limitations and future goals.


\section{Theoretical Background}
%
\label{sec:theory}
\subsection{Description of light and atoms}\label{sec:description}

\paragraph*{Harmonic oscillators}
Throughout this review we shall be dealing with single modes of the electromagnetic field and collective spin excitations of atomic ensembles which can be well approximated by harmonic oscillators with canonical position and momentum operators $X_n$ and $P_n$, where $n$ refers to the mode number. In most cases we shall omit the hats on the operators in what follows. These canonical operators are dimensionless with the standard commutator
\begin{equation}
[ X_n,P_m]=i\delta_{mn}.
\label{eq:xp-commutator}
\end{equation}
The harmonic oscillators can also be described in terms of the annihilation  operators
\begin{equation}
\begin{split}
a_n&=\frac{1}{\sqrt{2}}{\left( X_n+iP_n\right)},
\end{split}
\label{eq:xpa-trans}
\end{equation}
which have commutation relation $[ a_n,a_m^\dagger]=\delta_{nm}.$

 Instead of labeling by a discrete number $n$ the modes can be denoted by a continuous parameter, e.g., the position vector $\rvec$ with the commutation relations
\begin{eqnarray}
[ X(\rvec),P(\rvec\, ')]&=&i\delta(\rvec-\rvec\,'),\\
{\left[ a(\rvec),a^\dagger(\rvec\,')\right]}&=&\delta(\rvec-\rvec\,').
\end{eqnarray}
 In some cases we deal with the storage or transfer of a set of $n$ modes. Such discrete modes can be constructed from the continuous modes by introducing a complete orthogonal set of mode functions $\{u_m(\rvec)\}$ satisfying
\begin{eqnarray}
\int d\rvec\ u_m^*(\rvec) u_n(\rvec)&=&\delta_{mn}\\
\sum_m u_m^*(\rvec) u_m(\rvec\,')&=&\delta(\rvec-\rvec ').
\label{eq:completeness}
\end{eqnarray}
If we now define the discrete annihilation operators
\begin{equation}
a_m=\int d\rvec\ u_m^*(\rvec) a(\rvec)
\label{eq:discrete_from_cont}
\end{equation}
they have the appropriate commutator $[a_n,a_m^\dagger]=\delta_{mn}$.

 \paragraph*{Light}
 Light beams travelling in the $z$-direction can be described (in c.g.s. units) in the paraxial approximation by a quantized  electric field
 \begin{equation}
\vec{E}(\rvec)=\sum_{m,\sigma,k} \sqrt{\frac{2\pi \omega_0}{l}}
\vec{e}_{\sigma} u_m(\rperp;z) {\rm e}^{ikz} a_{L,m\sigma k} + {\rm
H.C. }
\end{equation}
Throughout this review we set $\hbar=1$. $l$ is the length of  the quantization volume, and the sum is over the polarization
$\sigma$, the transverse mode number $m$, as well as the
longitudinal wave vector $k$. The mode functions $u_m(\rperp;z)$, where $\rperp=(x,y)$
describes the transverse profile of the beam, form a complete
orthogonal set in the plane transverse to the propagation direction
\begin{equation}
\int d^2\rperp u_m^*(\rperp;z)
u_{m'}(\rperp;z)=\delta_{m,m'}.
\label{eq:orthog_um}
\end{equation}
In the above we have assumed that the fields belong to a narrow frequency band such that for all modes the frequency under the square root is $\omega_0$. Secondly, the field should in general be expanded into a complete set of modes  $\vec{u}_k(\rvec)$, but in the paraxial approximation, we have assumed that we can factor out a polarization vector $\vec{e}_\sigma$ as well as ignore the $k$ dependence of the transverse mode function $u_m(\rperp;z)$.

Instead of using longitudinal wave vectors we use a slowly varying position space annihilation
operator defined by
\begin{equation}
a_{L,m\sigma}(z)=\sqrt{\frac{c}{l}}\sum_k {\rm e}^{i(k-k_0)z+i\omega_0 t} a_{L,m\sigma k},
\label{eq:a_light_z}
\end{equation}
where $c$ is the speed of light and $k_0=\omega_0/c$.
In the continuum limit $l\rightarrow \infty$ this operator has the  commutation relation
  \begin{equation}
[a_{L,m\sigma}(z),a_{L,m'\sigma'} ^\dagger
(z')]=c \delta_{m,m'}\delta_{\sigma,\sigma'} \delta(z-z').
\end{equation}
Note that we have chosen here a normalization with $c$ appearing in the commutator. With this normalization (i) the travelling fields $a_{L,m\sigma}(z,t)=a_{L,m\sigma}(z-ct)$ considered below have the commutation relation appropriate for operators which are a function of $t$ ($[a_{L,m\sigma}(z,t),a_{L,m'\sigma'}^\dagger(z,t')]=\delta_{m,m'}\delta_{\sigma,\sigma'}\delta(t-t')$), and (ii) with this normalization $a_{L,m\sigma}(z)^\dagger a_{L,m\sigma}(z)$ describes the flux of photons in mode $m$ with polarization $\sigma$ at position $z$.

In terms of this operator the electric field is given by
\begin{equation}
\vec{E}(\rvec)=\sqrt{\frac{2\pi\omega_0}{c}} \sum_{m,\sigma}  \vec{e}_{\sigma}
u_m(\rperp;z) {\rm e}^{i(k_0z-\omega_0 t)} a_{L,m\sigma}(z) + {\rm H.C. }
\label{eq:E_multi}
\end{equation}
Below we mainly deal with a single transverse field mode and a single polarization, so that the sum in the expression above can be omitted. For an introductory textbook to continuous mode quantum optics we refer to \cite{Loudon:2004}.

 \paragraph*{Atoms}
We first discuss the theory for atoms with two stable ground states $|0\ra$ and $|1\ra$.  In subsec. \ref{sec:real} we show how one can in many cases reduce the description for multilevel atoms to two state atoms. The two ground states are conveniently described in terms of angular momentum operators. We shall here use the $x$ as the quantization axis for consistency with \textcite{Julsgaard:2001,Julsgaard:2004} and \textcite{Sherson:2006a}.  The angular momentum operators describing the $m$th atom are
\begin{eqnarray}
j_{x,m}&=&\frac{1}{2}{\left( |0\ra_m\la 0|-|1\ra_m\la 1|\right)},\\
j_{+,m}&=&j_{y,m}+i j_{z,m} =|0\ra_m\la 1|,
\end{eqnarray}
where $j_{+,m}$ is the operator which raises $j_{x,m}$ by unity.

We shall be interested here in collective variables for an ensemble containing many atoms. In the simplest case such collective operators are given by the total angular momentum operators
$J_l=\sum_m j_{l,m}$
(with $l=x,y,z$), which fulfil the standard angular momentum commutation relation
\begin{equation}
[J_y,J_z]=i J_x.
\label{eq:J-commute}
\end{equation}
The collective state with all atoms in state $|0\ra$, then  corresponds to the state $|J=N_A/2,M_x=N_A/2\ra$ with total angular momentum quantum number $J=N_A/2$ and an eigenvalue of   $J_x$ equal to $N_A/2$, where $N_A$ is the number of atoms.  If we  consider a large number of atoms and only weakly perturb the system (only change the state of a few atoms), we can approximate the $J_x$ operator by its expectation value $J_x\approx \la J_x\ra$. For readers who feel uneasy about replacing an operator by its mean value, a more rigorous formulation can be made by using the so called Holstein-Primakoff transformation \cite{holstein:1940,Kittel:1987} or in terms of a Wigner group contraction \cite{Arecchi:1972}. Without loss of generality we can assume the expectation value $\la J_x\ra$ to be positive and we can then introduce new canonical position and momentum operators by
\begin{align}\label{eq:xp_sym}
X_A&=\frac{J_y}{\sqrt{\la J_x\ra}}, &
P_A&=\frac{J_z}{\sqrt{\la J_x\ra}}.
\end{align}
From (\ref{eq:J-commute}) we immediately see that these operators satisfy the standard commutation relation for position and momentum (\ref{eq:xp-commutator}). The collective annihilation operator is then
\begin{equation}
a_A=\frac{X_A+iP_A}{\sqrt{2}}=\frac{\sum_m j_{+,m}}{\sqrt{2\la J_x\ra}}=\frac{\sum_m |0\ra_m\la 1|}{\sqrt{2\la J_x\ra}}.
\label{eq:collective_aatom}
\end{equation}
To get a feeling for this operator, let us consider the action of the creation operator $a_A^\dagger$. If we apply this operator to the initial state, where all atoms are in state $|0\ra$, we create a symmetric superposition of one atom being flipped
\begin{equation}
a_A^\dagger|0,0,0,...,0\ra=\frac{1}{\sqrt{N_A}}\sum_m |0,0,....,0,1_m,0,...,0\ra.
\label{eq:w_state}
\end{equation}
Here $|0,0,....,0,1_m,0,...,0\ra$ is the state, where all atoms except the $m$th atom are in state $|0\ra$.

The collective operators introduced above are convenient for describing the entire ensemble. We shall, however, also be dealing with situations where we need to consider collective operators which do not involve all atoms with an equal weight \cite{kuzmich:2004}.  In the literature such situations are often described by dividing the ensemble into small boxes and constructing collective operators for each box \cite{Raymer:1981,Fleischhauer:1995}.

Here we shall use a slightly different formalism \cite{Sorensen:2008}. For a collection of atoms at positions $\vec{r}_1,  \vec{r}_2, ...., \vec{r}_{N_A}$  we define the  density distribution function
\begin{equation}
n(\vec{r})=\sum_m \delta(\vec{r}-\vec{r}_m).
\label{eq:densitydistribution}
\end{equation}
In condensed matter physics this density distribution function may be used to describe scattering from structures \cite{Chaikin:1995}: averaging this density distribution over the random positions of the atoms gives the average number density of the atoms $\bar n(\vec{r})=\langle n(\vec{r})\rangle$, whereas higher order correlations like  $\langle n(\vec{r})n(\vec{r}\,')\rangle$ describe the correlation responsible for Bragg scattering (in this context the classical function (\ref{eq:densitydistribution}) is sometimes refereed to as the density operator \cite{Chaikin:1995}, but to avoid confusion with quantum mechanical operators we shall avoid this terminology). Similarly we may introduce continuous atomic spin operators by
\begin{equation}
j_k(\vec{r})=\sum_m \delta(\vec{r}-\vec{r}_m) j_{k,m},
\end{equation}
where $k=x,y,z,+,-$. A position dependent atomic annihilation operator can then be introduced by
\begin{equation}
a_A(\vec{r})=\frac{1}{\sqrt{2\langle j_x(\vec{r})\rangle}} j_+(\vec{r}),
\label{eq:cont_aatom}
\end{equation}
 where the average spin density $\langle j_x(\vec{r})\rangle$, which we assume to be positive, is the quantum mechanical expectation value with respect to the internal state averaged over the random (classical) position of the atoms.
The annihilation operator has the commutation relation
 \begin{equation}
[a_A(\rvec),a_A^\dagger(\rvec\,')]=\delta(\rvec-\rvec\,') \frac{ j_x(\rvec)}{\langle j_x(\vec{r})\rangle}\approx \delta(\rvec-\rvec\,'),
\label{eq:aatom_commutator}
 \end{equation}
where we in the last step have assumed that the fluctuations in the mean spin are much smaller than its average. We shall use this approximation throughout this article. 

The operators (\ref{eq:cont_aatom}) will be convenient for describing the spatial dependence of various operators. To relate them to single mode operators such as Eq. (\ref{eq:collective_aatom}), we can introduce a normalized set of mode functions $u_n(\rvec)$ fulfilling the orthogonality and completeness relations in Eq. (\ref{eq:completeness}).

We can then construct single mode operators as in Eq. (\ref{eq:discrete_from_cont}).
In particular let us consider the normalized mode $u_{\rm sym}(\rvec)=\sqrt{\la j_x(\rvec)\ra/\langle J_x\rangle}$. If we use this mode to construct a collective operator, we see from Eq. (\ref{eq:cont_aatom}) that this produces the symmetric operator defined in Eq. (\ref{eq:collective_aatom}).

 The $j_x(\rvec)$ operator will appear in the Hamiltonians below, although sometimes it is more convenient to have expressions which only involve the annihilation operator $a_A(\rvec)$.
%
To find the equations of motion we need to take the commutator of $a_A(\rvec)$ with the Hamiltonian, but from the commutation relation
\begin{equation}
\begin{split}
[a_A(\rvec),j_x(\rvec\,')]=&-\frac{\delta(\rvec-\rvec\,')}{ \sqrt{2 \la j_x(\rvec)\ra}}\sum_m\delta(\rvec-\rvec_m) j_{+,m}\\ =&-\delta(\rvec-\rvec\,') a_A(\rvec),
\end{split}
\end{equation}
we see that we get the same result if we make the replacement
\begin{equation}
j_x(\rvec)\rightarrow \frac{n}{2}-a_A^\dagger(\rvec)a_A(\rvec)
\label{eq:densityone}
\end{equation}
and use the commutation relation in Eq. (\ref{eq:aatom_commutator}) (the first term is included to ensure that $j_x$ has the right value in the vacuum state of $a_A$, where all atoms are in state $|0\ra$). This replacement holds even as an exact operator identity in the framework of the Holstein-Primakoff transformation \cite{holstein:1940}.

\subsection{Interaction of light with model atoms}
\label{sec:model_atoms}

We consider atoms with stable ground states denoted by $|g_m\ra$ and excited states denoted by $|e_m\ra$.
The Hamitonian $H=H_{\rm L}+H_{\rm A}+H_{\rm int}$
  describing this system is the sum of the field energy $H_{\rm L}$, the atomic energy $H_{\rm A}$, and the interaction Hamiltonian $H_{\rm int}$.
Below we first consider the atomic and interaction parts of the Hamiltonian and derive an effective interaction Hamiltonian involving only the ground states of the atoms. We then include the Hamiltonian $H_{\rm L}$ responsible for the propagation of the light field, and derive coupled equations of motion for the light and  atomic operators.

\paragraph*{Interaction with a single atom}
To describe the atomic part and the interaction let us first consider only a single atom at location $\vec{r}$. In a rotating frame with respect to the laser frequency the atomic Hamiltonian is given by
$
H_{\rm A}=\sum_m \Delta_m |e_m\ra\la e_m|,
$  
 where $\Delta_m$  is the detuning of the $m$th excited state with respect to the laser frequency.

In the dipole approximation the interaction between light and atoms is described by the Hamiltonian
$H_{\rm int}=-\vec{E}\cdot \vec{D},
$  
 where $\vec{D}$ is the electric dipole operator for the atom.
In Appendix \ref{sec:adiabat} we describe the adiabatic elimination of the excited state and derive an effective ground state Hamiltonian
\begin{equation}
\begin{split}
H_{{\rm int}}'=& \sum_{m,m'} V_{m',m}(\rvec) |g_{m'}\ra\la g_{m}|,
\end{split}
\label{eq:Hadiabat}
\end{equation}
where the coupling matrix $V_{m',m}$ is given by
\begin{equation}
V_{m',m}(\rvec)= - \sum_{m''} {\frac{{\left(\eminus(\rvec) \cdot \Dplus_{m',m''}\right)}{\left( \Dminus_{m'',m}\cdot\eplus(\rvec)\right)}}
{\Delta_{m''} 
}}.
\end{equation}
Here superscripts $(+)$ and $(-)$ refer to the positive and negative frequency components of the electric field and dipole operators (the positive frequency part, is the part of the operators which removes an excitation, e.g. $\Dplus_{m',m}=\la g_{m'}|\vec{D}|e_{m}\ra$) \cite{Loudon:2004}. (Note that $H_{{\rm int}}'$ is not the same as $H_{\rm int}=-\vec{E}\cdot \vec{D},
$ since $H_{{\rm int}}'$ contains a contribution from $H_{{\rm A}}$ as discussed in Appendix \ref{sec:adiabat}  )

\paragraph*{Interaction with many atoms}
\label{subsec:many}
Let us now consider the situation, where we have many atoms. We are mainly interested in describing the interaction of a weak quantum field with an ensemble driven by a strong classical field. In this subsection we only describe the interaction between the atoms and the forward propagating quantum fields and ignore the spontaneous emission due the coupling to all other e.-m. modes. We shall include the decoherence caused by spontaneous emission in subsec. \ref{subsec:decoherence}. To simplify the theory we shall only derive equations of motion for the initially fully polarized ensemble $\langle J_x(\rvec,t)\ra\approx n(\rvec)/2$ (we ignore the difference between the density distribution $n(\rvec)$ and its average value $\bar{n}(\rvec)$). The atomic operators defined in Eq. (\ref{eq:cont_aatom}) are, however, well behaved annihilation operators even for an ensemble which is not fully polarized provided that the fluctuations of the mean spin are small. This will for instance be the case if an ensemble containing many atoms is prepared with imperfect optical pumping.

The interaction Hamiltonian can be obtained by summing the single atom Hamiltonian (\ref{eq:Hadiabat}) over all atoms. This sum may be replaced by an integral by introducing the continuous atomic annihilation operator defined in Eq. (\ref{eq:cont_aatom})
\begin{equation}
\begin{split}
H=\int d^3\rvec\ & {\Bigg [}{\left(\sqrt{n(\rvec)} a_A(\rvec) V_{01}(\rvec)+{\rm H.C.}\right)}\\
&+n(\rvec) V_{00}+a_A^\dagger(\rvec)a_A (\rvec)(V_{11}-V_{00}){\Bigg ]},
\label{eq:general_3d_ham}
\end{split}
\end{equation}
where we have used the replacement (\ref{eq:densityone}) for the spin operator $j_x(\rvec)$.
In Appendix \ref{sec:3d_ham} we use this general Hamiltonian to derive Hamiltonians for the three different model systems in Fig. \ref{elemlevels} (b)-(d).

\begin{figure}[ht]
  \includegraphics[width=8.5cm]{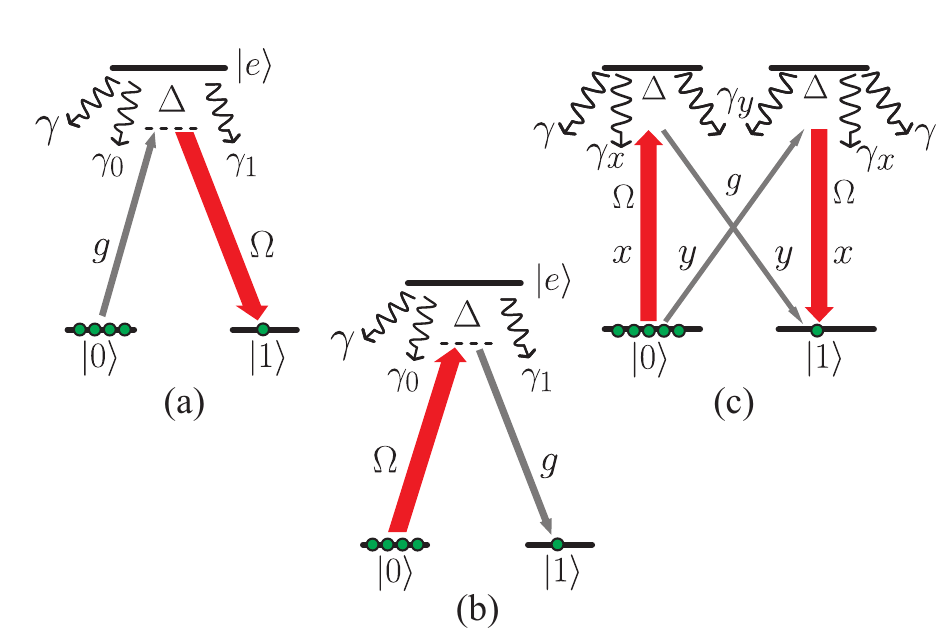}
  \caption{$\Lambda$ configuration with dominant population of level $\ket0$: (a) Beam splitter interaction with the quantum field of the single photon Rabi frequency $g$ on the $\ket{0}\rightarrow\ket{e}$ and the strong light with the Rabi frequency $\Omega$ on the $\ket{e}\rightarrow\ket{1}$ transition. The two fields are in two-photon resonance with a detuning $\Delta$ from the excited state $\ket{e}$. Decay due to spontaneous emission goes back to one of the ground states at the rates $\gamma_{0(1)}$ or to other levels (not shown) summing up to a total rate $\gamma$. (b) Parametric gain interaction. (c) Faraday interaction: Atoms are polarized to $\ket{0}$, the strong field is linearly polarized along $x$ and drives the up-transitions with the Rabi frequency $\Omega$ and the quantum field in $y$ polarization couples to the cross-transitions with the single photon Rabi frequency $g$. The fields are in two photon resonance with a detuning $\Delta$ from the excited states. Decay due to spontaneous emission goes back to one of the ground states at the rates $\gamma_{x(y)}$ or to other levels (not shown) at the rate $\gamma$. }\label{fig:lambda}
\end{figure}

%
For some situations the 3D Hamiltonians derived in Appendix \ref{sec:3d_ham} can be reduced to one dimension. Let us perform this reduction for the beam splitter interaction Hamiltonian in Eq.~(\ref{eq:H_lambda_3}), corresponding to the level configuration as shown in Fig.~\ref{fig:lambda}(a), where both classical and quantum field is traveling in the $z$ direction. Lets us now assume that the density is independent of the transverse coordinate  $n(\rvec)=n(z)$, and that the classical laser field is also    constant transverse to the propagation direction.  Since the mode functions $u_m(\rperp;z)$ form a complete set in the plane, we can expand the atomic operator $a_A(\rvec)$ in the same set
\begin{equation}
a_A(\rvec)=\sum_m u_m(\rperp;z)a_{A,m}(z),
\end{equation}
where
\begin{equation}
a_{A,m}(z)=\int d^2\rperp\ u_m^*(\rperp;z)a_A(\rvec).
\label{eq:aatom_transverse_lambda}
\end{equation}
These new operators will then have the appropriate commutation relation
$
[a_{A,m}(z),a^\dagger_{A,m'}(z')]=\delta(z-z')\delta_{m,m'}.
$ 
We insert  this expansion into the Hamiltonian (\ref{eq:H_lambda_3}), and  then use the orthogonality relation (\ref{eq:orthog_um}) integrate over the transverse coordinate to obtain the one dimensional Hamiltonian
\begin{equation}
\begin{split}
H_{BS}=\int  &dz\ {\Bigg[} \frac{-|\Omega(z,t)|^2}{4
\Delta
}
\sum_m a_{A,m}^\dagger(z)a_{A,m} (z)
\\ -& \frac{|g(z)|^2}{\Delta
}\sum_m a^\dagger_{L,m}(z)a_{L,m}(z)\\
-& 
{\left(\frac{g^*(z)\Omega(z,t)}{2\Delta} \sum_m  a^\dagger_{L,m}(z)a_{A,m}(z)+{\rm H.C.}\right)}{\Bigg ]},
\end{split}
\label{eq:H_lambda_1}
\end{equation}
 where the  the coupling constant $g(z)$ and slowly varying resonant Rabi frequency $\Omega(z,t)$ are given by
 \begin{equation}
\begin{split}
g(z)&=\sqrt{\frac{2\pi\omega n(z)}{c}}D_0,\\
\Omega(z,t)&=2 \Dminus_{e,1}\cdot \la \eplus\ra\exp(-i(k_0z-\omega_0 t))
\end{split}
\label{eq:coupling}
\end{equation}
and the dipole element for the $|0\ra$--$|e\ra$ transition with the polarization of the quantum field $\vec e_{{\rm q}}$ is given by $D_0= \Dminus_{e,0}\cdot{\vec e_{{\rm q}}}$. The Hamiltonian above consists of three terms: the first line is the AC-Stark shift of the atomic ground state, the second is the index of refraction of the gas, and the last
line, which is the most important for our discussion here, describes the exchange of excitations between atoms and light.

 Note  that each of the transverse modes of the light field in (\ref{eq:H_lambda_1}) talks to a single transverse mode of the atoms, which then couples back to the same transverse light mode. Since the dynamics is actually the same for all the involved transverse modes, the atomic ensembles may in fact be used as a memory for  for multiple transverse modes \cite{Camacho:2007,Vudyasetu:2008,Shuker:2008,Vasilyev:2007}. A similar description of the reduction from three to one dimension is also presented in \textcite{Andre:2005}.

We have used here the same central frequency $\omega_0$  for both the quantum and classical fields, which is true for degenerate $|0\ra$ and $|1\ra$ states. If they are not degenerate the energy difference between the two states can be accounted for by changing the frequency $\omega_0'$ of the classical field. In this case, however, an additional phase factor $\exp(i(k_0'-k_0)z)$ associated with the difference of the $k$ vectors appears.  We will discuss the effect of this phase factor in Sec. \ref{sec:other_strat} and Sec. \ref{sec:errors}.

For the parametric gain Hamiltonian (\ref{eq:H_gain_3}), cf. Fig.~\ref{fig:lambda}(b), the reduction from three to one dimension can be achieved using the same procedure as above. The only difference is that instead of Eq. (\ref{eq:aatom_transverse_lambda}) we should now  define the discrete atomic operator by
\begin{equation}
a_{A,m}(z)=\int d^2\rperp\ u_m(\rperp;z)a_A(\rvec).
\label{eq:aatom_transverse_gain}
\end{equation}
Although the omission of the complex conjugate in this expression compared to Eq. (\ref{eq:aatom_transverse_lambda}) may seem of minor importance, it actually does play a role for several experiments as we will discuss in Sec. \ref{sec:probabilistic}. Omitting the sum over multiple modes, we arrive at the Hamiltonian
\begin{equation}
\begin{split}
H_G=\int  &dz\  {\Bigg [} \frac{|\Omega(z,t)|^2}{4
\Delta
}
a_{A}^\dagger(z)a_{A} (z)
\\ 
-& 
{\left(\frac{g^*(z)\Omega(z,t)}{2\Delta}  a^\dagger_{L}(z)a^\dagger_{A}(z)+{\rm H.C.}\right)}{\Bigg ]}.
\end{split}
\label{eq:H_gain_1}
\end{equation}

Finally we derive the one dimensional Hamiltonian for the QND (Faraday) interaction as shown in Fig. \ref{fig:lambda}(c), where the $x$-polarized classical field couples the vertical transitions and a quantum field in $y$ polarization couples diagonal transitions.  From the figure we see that the Faraday interaction is essentially a combination of the beam splitter and the parametric gain interaction taken with the same coupling strength. In fact, as discussed in Appendix \ref{sec:3d_ham}  the three dimensional Hamiltonian for the Faraday interaction is simply $H_F=(H_{BS}-H_G)/\sqrt{2}$. To reduce the problem to one dimension we would like to introduce atomic operators similar to Eqs. (\ref{eq:aatom_transverse_lambda}) and (\ref{eq:aatom_transverse_gain}), but now there is an ambiguity as to which of the two forms one should use. To avoid this ambiguity we assume that the mode functions $u_m(\rperp;z)$ are real [for Hermite-Gaussian modes \cite{Milonni:1988}, this condition can only be satisfied if the Fresnel number is much greater than unity $F=w_0^2/\lambda L\gg 1$, where $w_0$ is the beam waist, $L$ is the length of the medium, and $\lambda$ is the wavelength of light \cite{Sorensen:2008,Muller:2005}], and that $g^*(\rvec)\Omega(\rvec,t)$ has a constant phase, which we take to be zero. With these assumptions the Hamiltonian can be reduced to the simple form
\begin{equation}
\begin{split}
H_F=- \int  &dz\   \frac{g^*(z)\Omega(z,t)}{\sqrt{2}
\Delta
}
p_L(z)p_A(z),
\end{split}
\label{eq:H_Faraday_1}
\end{equation}
 We have omitted here the index of refraction of the gas for the reasons discussed in Appendix \ref{sec:3d_ham}.

We emphasize that the above 3D to 1D reduction provides a highly simplified treatment of  the propagation of light through an atomic gas. In particular a more general treatment should include the spontaneous emission, the density-density correlation of the atoms and the optically induced dipole-dipole interaction of the atoms.
For the Faraday interaction a detailed study of the reduction from three to one dimension is presented by \textcite{Sorensen:2008}. In essence this study confirms the treatment presented here provided that the gas is ideal and that the modefunctions $u_m(\rperp;z)$ are the solutions to the propagation equation including the index of refraction. Little work has been  done so far on the optically induced dipole interactions.

\paragraph*{Equations of motion}
The addition of the Hamiltonian for light $H_L$ to the atomic part of the Hamiltonian $H_A$ and the interaction $H_{\rm int}$ discussed above allows to describe the propagation of light through the ensemble. As shown in Appendix \ref{sec:propagation} the equation of motion is derived by introducing a rescaled time $\tau=t-z/c$ and becomes a differential equation in space ($z$) instead of time $t$.

  For the beam splitter interaction, following this recipe, we find the equations of motion by calculating the commutator with $H_{BS}$
\begin{equation}
\begin{split}
\frac{\partial}{\partial z}a_L(z,t)
&=i \frac{|g(z)|^2}{\Delta
}a_L(z,t)
+i \frac{g^*(z)\Omega(t)}{2
\Delta
}a_A(z,t),\\
\frac{\partial}{\partial t}a_A(z,t)&=i \frac{|\Omega(t)|^2}{4
\Delta
}a_A(z,t) +i \frac{g(z)\Omega^*(t)}{2
\Delta
}a_L(z,t).
\end{split}
\label{eq:lambda_nodecay}
\end{equation}
The above equations can be solved analytically \cite{Kozhekin:2000,Gorshkov:2007c,Nunn:2007,Mishina:2007,Kupriyanov:2005}. We will defer a discussion of the solutions till Sec. \ref{subsec:decoherence} where the spontaneous emission is included.

Similarly we can find the equations of motion for the parametric gain:
\begin{equation}
\begin{split}
\frac{\partial}{\partial z}a_L(z,t)
&=
i \frac{g^*(z)\Omega(z,t)}{2
\Delta
}a_A^\dagger(z,t),\\
\frac{\partial}{\partial t}a_A(z,t)&=-i \frac{|\Omega(z,t)|^2}{4
\Delta
}a_A(z,t)\\
&\qquad
+i \frac{g^*(z)\Omega(z,t)}{2
\Delta
}a_L^\dagger(z,t).
\end{split}
\label{eq:gain_nodecay}
\end{equation}
Again these equations have an analytical solution \cite{Raymer:1981,Carman:1970}.

For  the Faraday interaction the equations of motion are much simpler when expressed in terms of $x$ and $p$ and read
\begin{equation}
\begin{split}
\frac{\partial}{\partial z}x_L(z,t)&=
- \frac{g^*(z)\Omega(t)}{\sqrt{2}
\Delta
}
p_A(z,t),\\
\frac{\partial}{\partial z}p_L(z,t)&=0,\\
\frac{\partial}{\partial t}x_A(z,t)&=
-\frac{g^*(z)\Omega(t)}{\sqrt{2}
\Delta
}p_L(z,t), \\
\frac{\partial}{\partial t}p_A(z,t)&=0.
\label{eq:faraday_nodecay}
\end{split}
\end{equation}
Because the two momentum operators are conserved quantities these equations describe a QND interaction, where one can, e.g., make a measurement of the position operator $x_L$ after the interaction and thereby obtain a QND measurement of the atomic momentum operator $p_A$, as will be further explained in Sec.~\ref{sec:SpinSqu}.

The presence of the conserved quantities $p_A$ and $p_L$ makes it straightforward to solve the equations of motion. The only $z$ dependence in the above expressions comes from the $z$ dependence of the density $n(z)$. We can then define the symmetric operators by
\begin{equation}
\begin{split}
X_A&=\frac{\int dz\ \sqrt{n(z)} x_A(z)}{\sqrt{\int dz\ n(z)}},\\
P_A&=\frac{\int dz\ \sqrt{n(z)} p_A(z)}{\sqrt{\int dz\ n(z)}}.
\end{split}
\end{equation}
If the transverse mode function includes all the atoms in the ensemble, these operators are equivalent to the symmetric operators defined in Eq. (\ref{eq:xp_sym}).
Similar integrated operators for the light field are
\begin{equation}
\begin{split}
X_L&=\frac{\int dt\  \Omega(t) x_L(t)}{\sqrt{\int dt \Omega^2(t)}},\\
P_L&=\frac{\int dt\  \Omega(t) p_L(t)}{\sqrt{\int dt \Omega^2(t)}},
\end{split}
\end{equation}
where we assume that $\Omega$ is real. Expressed in terms of these variable the equations of motions have the simple solutions
\begin{equation}
\begin{split}
X_{L,{\rm out}}&=X_{L,{\rm in}}+\kappa P_{A,{\rm in}},\\
P_{L,{\rm out}}&=P_{L,{\rm in}},\\
X_{A,{\rm out}}&=X_{A,{\rm in}}+\kappa P_{L,{\rm in}},\\
P_{A,{\rm out}}&=P_{A,{\rm in}}.
\label{eq:faraday_nodecay_result}
\end{split}
\end{equation}
The subscript ''in'' and ''out'' means input and output variables, e.g.,  $X_{L, {\rm in}}= X_L(z=0)$ and $X_{L, {\rm out}}=X_L(z=L)$ for light and  $X_{A, {\rm in}}=X_A(t=0)$ and $X_{A, {\rm out}}=X_A(t=T)$ for atoms.
The coupling constant $\kappa$ is given by
\begin{equation}
\begin{split}
\kappa^2&=  \int dt\ \frac{|\Omega(t)|^2}{2\Delta^2} \int dz\ |g(z)|^2\\
&=  \frac{\pi\omega D_-^2}{c}\int\ dt \frac{|\Omega(t)|^2}{\Delta^2} \int dz\ n(z).
\end{split}
\label{eq:kappa}
\end{equation}
In the last line we have inserted the expression for the coupling constant $g(z)$ (\ref{eq:coupling}) and have denoted the dipole matrix element by $D_-$ to indicate that it is the coupling constant for $\sigma-$ polarized light.

As we shall see the coupling constant $\kappa$ (and the generalization of it) will play an important role for characterizing the strength of the interaction regardless of the level scheme being used. Note, that $\kappa$  only depends on the total integrated density. This property can be shown to apply also to the $\Lambda$ schemes by a simple rescaling of the $z$ coordinate \cite{Gorshkov:2007c}. We shall therefore for simplicity only consider a constant density below.

\subsection{Theory Including Spontaneous emission}
\label{subsec:decoherence}
\label{sec:decoherence}

In subsec. \ref{sec:model_atoms} the spontaneous emission was omitted. In this subsection we will give the theory including the spontaneous emission and discuss the solutions to the equations.

 Instead of repeating the calculations in subsec. \ref{sec:model_atoms} now with a non-zero decay, we note that if the spontaneous emission from each atom is independent, it can be accounted for by making the substitution
\begin{equation}
\Delta_m\rightarrow \Delta_m\pm i\frac{\gamma_m}{2}
\label{eq:substitute_delta}
\end{equation}
in all the calculations performed so far. The choice of the sign is discussed in the Appendix \ref{sec:decay_derivation} and the results are stated below. It is also important to note that the quantity $\la j_x (\vec{r})\ra$ used to define the operator $a_A(\vec{r},t)$ in Eq. (\ref{eq:cont_aatom}), is not necessarily a constant. To ensure that the operator $a_A(\vec{r},t)$ has the right normalization we should always normalize by the time dependent expectation value $\la j_x (\vec{r},t)\ra$. The time derivative of $\la j_x (\vec{r},t)\ra$ will introduce extra terms as discussed in Appendix \ref{sec:decay_derivation}.


\paragraph*{Beam splitter interaction}
As discussed in Appendix \ref{sec:decay_derivation} for the beam splitter interaction we should use the minus sign in  the substitution (\ref{eq:substitute_delta}), and the time derivative of $\la j_x(\rvec)\ra$ can be neglected because the strong classical field talks to an almost empty level.  For a constant atomic density the equations of motion then become
\begin{equation}
\begin{split}
\frac{\partial}{\partial z}a_L(z,t)
&=\frac{i |g|^2}{\Delta-i\frac{\gamma}{2}}a_L(z,t)
+ \frac{i g^*\Omega(t)}{2
\Delta-i
\gamma
}a_A(z,t),\\
\frac{\partial}{\partial t}a_A(z,t)&= \frac{i|\Omega(t)|^2}{4(\Delta-i\frac{\gamma}{2})}a_A(z,t) + \frac{ig\Omega^*(t)}{2\Delta-i
\gamma
}
a_L(z,t),
\label{eq:lambda}
\end{split}
\end{equation}
where $\gamma$ is the total decay rate of the excited state and we have omitted the the noise operators given in Eq. (\ref{eq:lambda_noise}). The admixed noise is vacuum in both equations of motion, and since the  equations only couple annihilation operators to annihilation operators we can ignore the noise for the calculation of any normally ordered products (see below). In the above equations the imaginary part of the first terms on the right hand side of each line refers to the phase shift caused by the index of refraction of the medium and the AC-Stark shift of the atoms, and the real part of these terms represent damping by spontaneous emission. The last term on each line represents the coupling of light and atoms which is our main interest here.


Solving these equation for a resonant field $\Delta=0$ with no classical field $\Omega=0$ we find that the intensity is reduced by a factor of $\exp(-d)$, with the optical depth $d$ given by
\begin{equation}
d=L \frac{4|g|^2}{\gamma}= \frac{3\lambda^2\gamma_0}{2\pi\gamma}  n L=n\sigma_0 L,
\label{eq:d_lambda}
\end{equation}
where we have used the expression for the coupling constant $g$ (\ref{eq:coupling}) and have introduced the spontaneous decay rate $\gamma_0=4\omega^3|D_0|^2/3c^3$ from the excited state $|e\ra$ into the ground state $|0\ra$ \cite{Milonni:1988} and the
 absorption cross section for an atom $\sigma_0=3\lambda^2\gamma_0/2\pi\gamma$  \cite{Jackson:1975}.

It is sufficient in our case to solve the operator equations of motion (\ref{eq:lambda}) as "classical equations" with the operators replaced by complex functions \cite{Gorshkov:2007b,Raymer:1981}.  The reason is that the equations are linear in the operators. The solutions to the operator equations will therefore be of the form
\begin{equation}
\hat a_{L, {\rm out}}(t)=\int_0^L dz\ m(\Omega(t');t,L-z) \hat a_{A,{\rm in}}(z)+.....,
\label{eq:light_out_bs_opr}
\end{equation}
where the operators are identified with hats for clarity, and the argument $\Omega(t')$ indicates that the solution depends on the driving field at all times.  The remaining terms denoted by dots in Eq. (\ref{eq:light_out_bs_opr})  are similar linear combinations of the input light and noise operators.  If we for instance solve the equations of motion with a complex function $a_{A,{\rm in}}(z)$ as the initial condition, we will due to the linearity of the equations get the same solution only without the hats
\begin{equation}
 a_{L, {\rm out}}(t)=\int_0^L dz\ m(\Omega(t),t,L-z)  a_{A,{\rm in}}(z) .
\label{eq:light_out_bs}
\end{equation}
We can thus obtain most of the solution (\ref{eq:light_out_bs_opr}) by simply inserting hats in the solution. Because we have ignored some input operators, however, the resulting operators will not necessarily have the right commutation relation. If there is no incident light all of these other modes will be in vacuum, and one can obtain the right commutation relation by adding a suitable amount of vacuum noise \cite{Gorshkov:2007b}. Another way to see why the complex number equations are sufficient to obtain full information about the dynamics is to note that if all input modes are in classical coherent states we can take expectation values of the equations of motion (\ref{eq:lambda}) and obtain the same equations of motion for the mean values. These classical equations of motion are thus identical to the quantum equations of motion \cite{Raymer:1981}. The equations of motion correspond to a general beam splitter relation, so that with coherent states as input states the output quantum states will also be a set of coherent states with amplitudes given by the mean values. Since any initial state can be expanded on the set of coherent states, the knowledge about the evolution of the coherent states obtained by solving the equations of motion for the mean values gives us complete information about the evolution for any quantum state.

The equations of motion can in fact be solved analytically \cite{Gorshkov:2007c}. If we consider the situation where the only non vanishing initial value is $a_A$ we find the solution for the output light in Eq.  (\ref{eq:light_out_bs}) with
\begin{equation}
\begin{split}m(\Omega(t');t,z)=& \sqrt{\frac{\gamma d}{L}}\frac{(-i)\Omega(z,t)}{4{\left(\Delta-i\frac{\gamma}{2}\right)}}{\rm e}^{i{\left[\frac{2\Delta+i\gamma}{2\gamma}\frac{h(0,t)}{d}+\frac{d\gamma z}{4L(\Delta-i\gamma/2)} \right]}}\\
& \times I_0{\left(-i {\rm e^{i\phi}} \sqrt{h(0,t)\frac{z}{L}}\right)},
\label{eq:lambda_kernel}
\end{split}
\end{equation}
where $I_0(x)$ denotes the 0-th order Bessel function of the first kind.
Here we have assumed that the coupling constant is real, which can always be done by absorbing any phase into the definition of $\Omega$, and we
have replaced the coupling constant by the optical depth through the relation (\ref{eq:d_lambda}). The function $h(t,t')$ is defined by
\begin{equation}
h(t,t')=\int_t^{t'} dt''\ \frac{d\gamma |\Omega(t'')|^2}{4\Delta^2+\gamma^2},
\label{eq:ht}
\end{equation}
and as we shall see below, this is also the function characterizing the strength of the Faraday interaction. Finally the phase $\phi$ is defined by
\begin{equation}
\tan(\phi)=\frac{\gamma}{2\Delta}.
\label{eq:phase}
\end{equation}

Similarly we find for the output atomic variables
\begin{equation}
a_{A,{\rm out}}(z)=\int_0^T dt\ m(\Omega^*(T-t);T-t,z) a_{L,{\rm in}}(t).
\label{eq:atom_bs_out}
\end{equation}
The fact that the kernel $m(\Omega^*(T-t);T-t,z)$ is similar to the kernel in Eq. (\ref{eq:light_out_bs}) is a direct consequence of time reversal symmetry \cite{Gorshkov:2007c}.

While the above solutions for the equations of motion are exact , they are sufficiently complicated and it is hard to gain any physical intuition from them. More insight can be gained by applying to the equations the Laplace transform in space using
\begin{equation}
a_A(u,t)=\frac{1}{\sqrt{L}}\int_0^\infty dz\ {\rm e}^{- u z/L}a_A(z,t),
\end{equation}
where we have chosen the normalization so that $u$ is a dimensionless number of order unity. We can then derive the equations of motion for the Laplace transformed variables, which  only couple operators with the same parameter $u$. (Note, however that because  the atomic operators only have support on $z\in [0,L]$ and not $[0, \infty[$ different Laplace components are not orthogonal and care should be taken when applying these formulas \cite{Gorshkov:2007c}).
Since the Laplace transformed equation for the light field does not involve derivatives, we can then eliminate the light field and obtain a single equation for the atomic operator
\begin{equation}
\begin{split}
\frac{\partial}{\partial t}a_A(u,t)=& \frac{i|\Omega(t)|^2}{4{\left(\Delta-i\frac{\gamma}{2}{\left(1+\frac{d}{2u}\right)}\right)}}a_A(u,t) \\
&
+ \frac{ig\sqrt{L}\Omega^*(t)}{2u{\left(\Delta-i\frac{\gamma}{2}{\left(1+\frac{d}{2u}\right)}\right)}}a_{L,{\rm in}}(t).
\label{eq:lambda_laplace}
\end{split}
\end{equation}
This equation now has a simple interpretation. Let us suppose that the incident light field is in vacuum so that we can ignore the last line in Eq. (\ref{eq:lambda_laplace}). The initial atomic state can then decay through two different mechanisms: either through spontaneous emission, or through coherent interaction with the forward light mode. Consider now the fraction in the first line of Eq. (\ref{eq:lambda_laplace}).
  The imaginary part of the first term, proportional to the detuning $\Delta$ will give rise to an unimportant phase, while the real part describes decay of the atomic excitation at an effective rate (proportional to) $\gamma(1+d/2u)$. Here the decay rate $\gamma$ is due to spontaneous emission whereas the second term $\gamma d/2u$ is due to the coherent interaction with the forward light mode. The optical depth $d$ has been introduced in this expression through its relation with the coupling constant $g$ in Eq. (\ref{eq:d_lambda}), and thus characterizes the strength of the coherent interaction with the forward light mode.    For a sufficiently high optical depth $d\gg 1$ and sufficiently smooth atomic excitations $u\sim 1$ the coherent interaction will dominate the spontaneous emission $\gamma d/u\gg \gamma$ and we can obtain an efficient interface between atoms and light.

\paragraph*{Parametric gain interaction}
In case of the parametric gain interaction we need to use the plus sign in one of the substitutions as well as include some terms associated with the time derivative of $\la j_x(\rvec,t)\ra$ as discussed in Appendix \ref{sec:decay_derivation}.  We then find
\begin{equation}
\begin{split}
\frac{\partial}{\partial z}a_L(z,t)
&=
i \frac{g^*(z)\Omega(z,t)}{2\Delta-i\gamma}a_A^\dagger(z,t),\\
\frac{\partial}{\partial t}a_A(z,t)&=-i \frac{ |\Omega(z,t)|^2\Delta }{4 \Delta^2 +\gamma^2}a_A(z,t)\\
&\qquad
+i \frac{g^*(z)\Omega(z,t)}{2\Delta-i\gamma}a_L^\dagger(z,t),
\end{split}
\label{eq:gain_full}
\end{equation}
where the $z$ dependence of the Rabi-frequency is given by the $\exp[i\int dz' \ |g(z')|^2/(\Delta-i\gamma/2)]$ dependence associated with the change in the propagation of the classical field caused by the index of refraction and scattering of the medium. To arrive at these equations, we have assumed that the decay from the excited state goes into some states $|a_m\ra$ different from $|0\ra$ and $|1\ra$, cf. Fig.~\ref{fig:lambda}(b).
This model for the decay is not always best for real systems used for parametric-type interaction, for example, in DLCZ-type applications \cite{Duan:2001}  a much stronger decay to auxiliary states would decrease the effective optical depth, and a large optical depth is crucial for the interface. The model used here is, however, the simplest possible model, and we therefore restrict ourself to this situation. The interpretation of the above equation is similar to the interpretation of the equation of motion for the beam splitter interaction (\ref{eq:lambda}), only in these equations there is no coupling of the light to itself because the light talks to an almost empty transition. The reason why there is no decoherence term in the atomic equation  is that spontaneous emission drives  atoms from state $|0\ra$ into some other state $|a\ra$, where they are lost from the system. Collective states like the one in Eq. (\ref{eq:w_state}) are, however, immune to removing one of the atoms in state $|0\ra$ and this source of decoherence has therefore no effect within the approximations we are using here. See the work by \textcite{Mewes:2005} for a more detailed discussion of the robustness of collective atomic states.


These equations of motion can also be solved analytically \cite{Raymer:1981,Carman:1970}. The solution is similar to the solution for the beam splitter interaction and we shall not go further into it here. Again it can be shown that it is possible to obtain a large coherent coupling with a large optical depth,  by Laplace transforming the equations. By doing so one finds a strong  gain term for the Laplace component with argument $u$, which is $\sim d/u$ times the single atom scattering rate. For a large optical depth $d$ we are thus  dominated by the coherent interaction, regardless of the assumption about the final state after the decay, which we  made above. Unlike the beamsplitter interaction, where one can work at any detuning, this strong coherent interaction only works in the far off-resonant regime, because the classical beam would be completely depleted if one works on resonance in a medium with large optical depth.

\paragraph*{Faraday interaction}
The Faraday interaction is a combination of the beam splitter interaction and the parametric gain with equal weights, and  the easiest way to obtain the equations of motion is therefore to just combine the results obtained in the previous two subsections. As discussed in Appendix \ref{sec:decay_derivation} the resulting equations of motion are then (assuming $\Omega$ and $g$ to be real)
\begin{equation}
\begin{split}
\frac{\partial}{\partial z}x_L(z,t)=&
 -\frac{2\sqrt{2}\Delta g(z)\Omega(t)}{4
\Delta^2+\gamma^2
}
p_A(z)-\frac{\gamma g^2}{4
\Delta^2+\gamma^2 }x_L(z),\\
\frac{\partial}{\partial z}p_L(z,t)=& \frac{\sqrt{2}\gamma g(z)\Omega(t)}{4
\Delta^2+\gamma^2
}
p_A(z)-\frac{\gamma g^2}{4
\Delta^2+\gamma^2}p_L(z),\\
\frac{\partial}{\partial t}x_A(z,t)=& -
\frac{2\sqrt{2}\Delta g(z)\Omega(t)}{
4\Delta^2+\gamma^2}p_L(z) -\frac{\gamma\Omega^2}{2{\left(4
\Delta^2+\gamma^2\right)}}x_A(z),\\
\frac{\partial}{\partial t}p_A(z,t)=& \frac{\sqrt{2}\gamma g(z)\Omega(t)}{4
\Delta^2+\gamma^2
}
p_L(z)-\frac{\gamma\Omega^2}{2{\left(4
\Delta^2+\gamma^2\right)}}p_A(z),
\label{eq:faraday_full}
\end{split}
\end{equation}
where we have omitted the noise operators, which are given in Eq. (\ref{eq:faraday_noise}). Again these equation are derived under the assumption that the decay goes to some auxiliary state $|a_m\ra$. For a treatment with decay back to the interface levels, see for instance \textcite{Duan:2000,Hammerer:2006,Madsen:2004}.

%

Let us now consider the solution of Eq. (\ref{eq:faraday_full}) in the limit of a small damping.
The Faraday interaction is only used with far off resonant light $\Delta\gg \sqrt{d}
\gamma$ since the classical field would be completely absorbed if we were working close to resonance. We will therefore ignore the first term in the evolution of the momentum operators $p_L$ and $p_A$, which is much smaller than the similar term in the evolution of $x_L$ and $x_A$. Furthermore, a time dependent driving $\Omega(t)$ can be accounted for by a simple rescaling (see Appendix \ref{sec:dimensionless}), so we only consider a constant driving field $\Omega$.
 Because the quantities $p_L$ and $p_A$ appearing in the coupling to the operators $x_A$ and $x_L$ are conserved quantities apart from the small decays, the dynamics effectively only involve the integrated operators
\begin{equation}
\begin{split}
X_L&=\frac{1}{\sqrt{T}}\int_0^{T} d t\ x_L(t),\\
P_L&=\frac{1}{\sqrt{T}}\int_0^{T} d t\ p_L(t),\\
X_A&=\frac{1}{\sqrt{L}}\int_0^L d z\ x_A(z),\\
P_A&=\frac{1}{\sqrt{L}}\int_0^L d z\ p_A(z),
\end{split}
\label{eq:sym_modes}
\end{equation}
where the normalization is chosen as for single mode operators $[X,P]=i$. In the limit of small scattering the resulting dynamics are then given by
\begin{equation}
\begin{split}
X_{L,{\rm out}}&\approx {\rm e}^{-\eta_L/2} X_{L,{\rm in}}+\kappa
P_{A,{\rm in}},\\
P_{L,{\rm out}}&\approx {\rm e}^{-\eta_L/2} P_{L,{\rm in}},\\
X_{A,{\rm out}}&\approx {\rm e}^{-\eta_A/2} X_{A,{\rm in}}+\kappa P_{L,{\rm in}},\\
P_{A,{\rm out}}&\approx{\rm e}^{-\eta_A/2} P_{A,{\rm in}}.
\label{eq:faraday_weakdecay_result}
\end{split}
\end{equation}

Apart from the decay, this solution is the same as the results derived in Eq. (\ref{eq:faraday_nodecay_result}) with a minor modification of the coupling constant (\ref{eq:kappa}) which is now given by
\begin{equation}
 {\kappa}^2=h(0,t)\frac{4\Delta^2}{4\Delta^2+\gamma^2},
\label{eq:kappa_full}
\end{equation}
where the function $h(0,t)$ is the same as the function introduced for the beam splitter interaction in Eq. (\ref{eq:ht}). In the limit of large detuning  $\Delta$  the coupling constant here is the same as the one derived in Eq. (\ref{eq:kappa}).

In the solutions above, the light and atomic operators are damped by factors of  $\exp(-\eta_L/2)$  and  $\exp(-\eta_A/2)$ respectively. The damping factor for the light is given by
\begin{equation}
\eta_L=\frac{\gamma|g|^2L}{4{\left(\Delta^2+\frac{\gamma^2}{4}\right)}}=\frac{d}{2}\frac{\gamma^2}{4\Delta^2+\gamma^2}.
\end{equation}
Note that the optical depth used here is for linear polarization. The definition of $d$ therefore differs by a factor of 2 from (\ref{eq:d_lambda})  because we have taken $g$ to be the coupling constant for circular polarized light and not for linear polarization.
From the expression above it is clear that we can ignore the decay of the light field if we use a sufficiently large detuning $\Delta\gg \sqrt{d}\gamma$. A large detuning also reduces the coupling constant but this can be compensated by using stronger laser fields.

The atomic decoherence can be related to the coupling constant through
\begin{equation}\label{eq:eta_kappa}
d\, \eta_A=\frac{\gamma d \int dt |\Omega|^2}{4\Delta^2+\gamma^2}= \kappa^2.
\end{equation}
This relation between $d$, $\eta_A$, and $\kappa$ is valid regardless of which assumption one makes about the final state after a decay.  But different decay channels will have different effects in the equations  of motion, and Eq. (\ref{eq:faraday_full}) is only directly applicable for the particular model considered here.
Nevertheless, with a fixed interaction strength $\kappa^2$ we can always obtain negligible decoherence with a sufficiently large optical depth $d$. Conversely, for a given optical depth $d$ there is always an optimal damping factor $\eta_L$ balancing the losses against the gains in some figure of merit (such as e.g. state transfer efficiency, light-matter entanglement etc.) which depends on the coupling strength $\kappa$, see for example \cite{Hammerer:2004}.

\paragraph*{Scaling with atom and photon number}
Further insight into the connection between the coupling constant, atomic decoherence and dissipation of light can be obtained by expressing these parameters through the number of atoms $N_A$ and photons $N_P$ of the classical field.
Let us assume that the classical beam has a square profile with a cross section of area  $A$. Using the definitions of the electric field (\ref{eq:E_multi}) and Rabi frequency  (\ref{eq:coupling}) we find
\begin{equation}
\int dt |\Omega|^2=\gamma_c \frac{3\lambda^2}{2\pi A}N_P,
\end{equation}
where $\gamma_c$ ($\gamma_q$) is the decay rate between the two states coupled by the classical (quantum) field with an emitted photon of the same polarization as the classical (quantum) field.
The function $h(0,T)$ characterizing the strength of the interaction in the far off-resonant limit can be then expressed as
\begin{equation}\label{eq:scaling}
\begin{split}
h(0,T)&={\left(\frac{3\lambda^2}{2\pi A}\right)}^2\frac{\gamma_c \gamma_q}{4\Delta^2+\gamma^2}N_A N_P\\
&=\frac{\sigma_c\sigma_q}{A^2} \frac{\gamma^2}{4\Delta^2+\gamma^2}N_A N_P,
\end{split}
\end{equation}
where in the last line we have introduced the resonant cross sections  for the scattering of the quantum and  classical fields ($\sigma_m=3\lambda^2\gamma_m/2\pi\gamma$).
The atomic and light decoherence can then be related by
\begin{equation}
N_A\eta_A=N_P \eta_L,
\end{equation}
which simply reflects the fact that the number of scattered photons is the same as the number of atoms which have scattered a photon. If the number of input classical photons is much larger than the number of atoms $N_P\gg N_A$ the decay of light is much smaller than the atomic decoherence and can be neglected.

The function $h(0,t)=\kappa^2$ characterizes the solution of all three  model systems considered here. Further insight into the reason for this can be gained by rewriting the equations in terms of rescaled dimensionless variables (Appendix \ref{sec:dimensionless}).

\newcommand{\SixJ}[6]{\left\{\begin{smallmatrix}#1&#2&#3\\#4\!\!&#5&#6\end{smallmatrix}\right\}}
\newcommand{\tens}[1]{\protect\@ontopof{#1}{\leftrightarrow}{1.15}\mathord{\box2}}

\subsection{Realistic Multilevel Atoms}\label{sec:real}\label{sec:realatoms}


In the previous section we derived the main equations describing three types of light-matter interactions, the beam splitter, the parametric gain and the Faraday-QND type, Eqs. \eqref{eq:H_lambda_1}, \eqref{eq:H_gain_1} and \eqref{eq:H_Faraday_1} respectively, for simple few-level model atoms. In this section we will give examples of how these interactions are commonly realized with real atoms.

\begin{figure}[ht]
  \includegraphics[width=8.5cm]{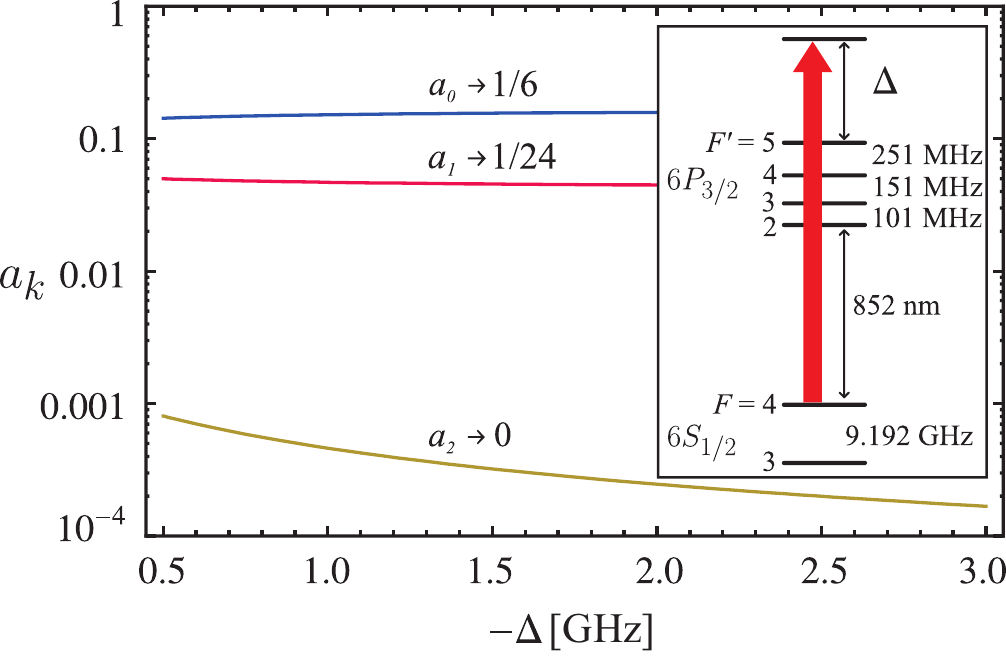}
  \caption{Coefficients $a_0,~a_1,~a_2$ for, respectively, scalar, vector and tensor polarizability versus (blue) detuning $\Delta$ of probe light driving the $6S_{1/2}\rightarrow 6P_{3/2}$ transition 
   in $^{133}$Cs. The inset shows the relevant levels and energy scales.
  }
  \label{fig:tensor}
\end{figure}

\paragraph*{Faraday interaction} Applying the Faraday interaction \eqref{eq:H_Faraday_1} to a full hyperfine level with many nearly degenerate Zeeman states would at first sight seem to violate the simple two level approximation that we have used in the theoretical derivation. For alkali atoms, however, the full theory actually reduces to what we have derived in the previous sections 
if the detuning is much larger than the hyperfine splitting in the excited state. Consider the $S_{1/2}\rightarrow P_{3/2}$ transition in alkali atoms as indicated in Fig. \ref{fig:tensor}. If atoms are optically pumped into one of their hyperfine ground state levels $F$, an off resonant probe will couple in general to three dipole allowed transitions $F\rightarrow F'=F-1,F,F+1$. Each of these transitions will contribute to the effective Hamiltonian in Eq.~\eqref{eq:Hadiabat} describing the interaction of a single atom with off-resonant light. It will be convenient to rewrite this Hamiltonian as
\[
H_{\rm int}=\eminus(\rvec)\tensor{\alpha}\eplus(\rvec),
\]
where we introduced the atomic polarizability tensor operator \cite{Happer:1972,Happer:1967,Deutsch:1998}
\[
\tensor{\alpha}=\sum_{m,m'}\sum_{F'}\sum_{m''=-F'}^{F'}
\frac{\Dplus_{m',m''}\wedge\Dminus_{m'',m}}{\Delta_{F'}}|g_{m'}\rangle\langle g_m|
\]
and $\Dplus_{m',m''}\wedge\Dminus_{m'',m}$ denotes the dyadic vector product. The polarizability operator is a rank-2 spherical tensor \cite{Edmonds:1964, Zare:1988} and can therefore be decomposed into irreducible tensor components,
\[
\tensor{\alpha}=\frac{4D_{0}^2}{\Delta}\left(a_0(\Delta)\tensor{T}^{(0)}+a_1(\Delta)\tensor{T}^{(1)}+a_2(\Delta)\tensor{T}^{(2)}\right),
\]
where the tensor operators $\tensor{T}^{(k)}$ transform under rotations as a scalar, vector and matrix for $k=0,1,2$ respectively. In this expression $\Delta=\Delta_{F'=F+1}$ is the laser detuning from the uppermost level and $2D_{0}=\langle J'||\vec{D}||J\rangle$ is the reduced dipole matrix element for the $S_{1/2}\rightarrow P_{3/2}$ transition. It relates to the spontaneous decay rate introduced in Sec.~\ref{subsec:decoherence} as $\gamma_0=4\omega^3|D_0|^2/3c^3$.

The real coefficients $a_k(\Delta)$ follow from elementary calculations \cite{Geremia:2006,Hammerer:2006,Kupriyanov:2005,Julsgaard:2003} and are given in appendix \ref{Appendix:TensorDecomposition}. The essential feature of these coefficients, which is proven in the appendix \ref{Appendix:TensorDecomposition}, is that for a laser detuning, which is large compared to the hyperfine splitting of excited states, $|\Delta|\gg|\Delta_{F+1}-\Delta_{F'}|$, the rank-2 tensor component vanishes, $a_2(\Delta)\rightarrow 0$. For the case of $^{133}$Cs the coefficients $a_k(\Delta)$ are shown in Fig. \ref{fig:tensor}.

In the asymptotic limit the interaction Hamiltonian is thus given by
\begin{multline*}
  H=\frac{D_{0}^2}{\Delta}\left[a_0\eminus(\rvec)\cdot\eplus(\rvec)\right.\\
  \left.+\frac{i}{2}a_1\eminus(\rvec)\cdot\left(\vec{j}\times\eplus(\rvec)\right)\right].
\end{multline*}
For an atomic ensemble, the first term will give rise to an index of refraction, while the second term accounts for the Faraday effect. For light propagating along $z$ and the classical light polarized along $x$ the Faraday Hamiltonian \eqref{eq:H_Faraday_1} can be easily derived.

When the detuning is larger than the hyperfine splitting, the interaction is thus essentially the same as for a spin 1/2 ground state, where $a_2$ vanishes exactly for all detunings. This can be understood by looking at how the energy levels of an alkali atom appear. The full Hamiltonian can be written $H=H_0+H_{FS}+H_{HFS} +H_{\rm int}$, where the four Hamiltonians represents the Coulomb, fine structure, hyperfine structure, and interaction with light. Normally one just considers the first three terms as an atomic Hamiltonian and does perturbation theory in the interaction Hamiltonian. When the detuning is larger than the hyperfine structure it is, however, more appropriate to do the adiabatic elimination before treating the hyperfine interaction. Without the hyperfine interaction the optical fields only talk to the electron spin, where there cannot be a rank two tensor, since the spin cannot be changed by two, and $a_2$ therefore only appears as a perturbation in $H_{HFS}/\Delta$.

The Faraday interaction of light with a true spin 1/2 ground state atom, which therefore has only scalar and vector polarizability, can be achieved with the Ytterbium isotope $^{171}$Yb, as was explored in \cite{Takeuchi:2006,Takeuchi:2007}.

\paragraph*{$\Lambda$-Systems} Although for the beam splitter and the parametric gain interaction, which require a $\Lambda$ configuration, magnetic sublevels are sometimes used \cite{Novikova:2007a}, the most common implementation makes use of $S_{1/2}(F=I\pm1/2)$ hyperfine levels as ground states $|0\rangle$ and $|1\rangle$ and one of the $P$ states as an excited state $|e\rangle$, see \cite{Chaneliere:2005,Chaneliere:2007,Chen:2008,Chen:2007,Choi:2008,Chou:2007,Chou:2004,Chou:2005, Matsukevich:2004}. This approach gives excellent results in zero magnetic field even if atoms are not optically pumped initially to one Zeeman substate. It has also been suggested by \citet{Kupriyanov:2005,Mishina:2007,Echaniz:2008} to make use of the second rank tensor polarizability to engineer effective $\Lambda$-schemes for the beam splitter and the parametric gain interactions using two degenerate Zeeman states and polarized light.

\subsection{Ensemble in Magnetic Field}
\label{sec:magnetic field}

\begin{figure}[ht]
  \includegraphics[width=5.5cm]{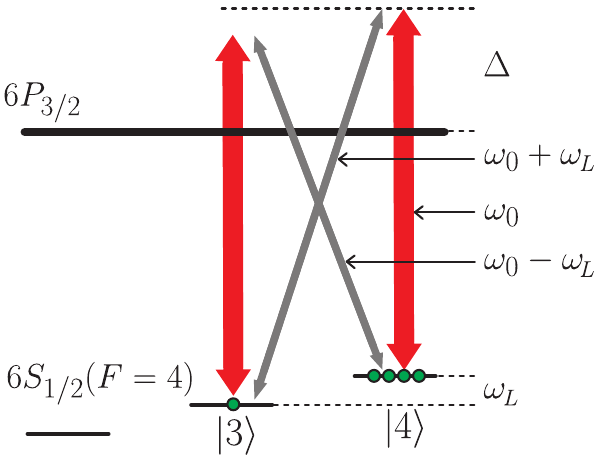}
  \caption{A magnetic field along the axis of polarization causes Zeeman splitting $\omega_L$ of ground state levels. Photons will be scattered from the classical light, driving the $\pi$ transitions, to the quantum field, coupling to the cross $\sigma$ transitions, at sideband frequencies $\omega_0\pm\omega_L$ from the carrier frequency $\omega_0$.}\label{fig:bfield}
\end{figure}

So far, our theory for light matter interaction assumed degenerate ground state levels. For pure three level $\Lambda$ schemes non-degenerate ground states do not make any difference, as the level splitting can be compensated for by choosing appropriate frequencies for light. 
In this section we will mainly  deal with the Faraday interaction for atoms in an external magnetic field along the axis of atomic polarization (Fig. \ref{fig:bfield}). The Zeeman splitting caused by this field can be advantageous in several respects. On the one hand, combined with the homodyne detection typically used in connection with the Faraday interaction, this results in low-noise AC signals, as will be detailed in sec. \ref{sec:measurement}. On the other hand, it can also simplify and enhance protocols aiming for an efficient creation of entanglement of two ensembles (Sec. \ref{sec:Entanglement}) or between an ensemble and light (Sec. \ref{sec:teleportation}).

The free Hamiltonian for an ensemble of atoms in a uniform magnetic field (Eq.~\eqref{eq:densityone}) oriented along the $x$-axis is
\begin{equation}\label{eq:magnetic_field}
H_0=\frac{\omega_L}{2}\int{\rm d}z\left(x_{A}^2(z)+p_{A}^2(z)\right),
\end{equation}
where $\omega_L$ is the Larmor frequency. This Hamiltonian generates Larmor precession of the transverse spin density components, $x_{A}(z)$ and $p_{A}(z)$, about the $x$-axis. The full Hamiltonian describing Faraday interaction in a magnetic field is $H=H_0+H_F$, where $H_F$ is given in Eq. \eqref{eq:H_Faraday_1}. In an interaction picture with respect to $H_0$ this Hamiltonian is,
\begin{multline}
H^I_F=- \int dz\frac{g^*(z)\Omega(z,t)}{\sqrt{2}\Delta}p_L(z)\\
\times\big(\cos(\omega_L t)p_A(z)+\sin(\omega_L t)x_A(z)\big).
\label{eq:H_Faraday_IP}
\end{multline}
Operators $x_{A}$ and $p_{A}$ refer now to spin components in a frame rotating at $\omega_L$ about the $x$ axis. For the sake of simplicity we use in the following the same symbols for canonical operators in both frames, as it will be clear from the context, to which one we are referring. In the rotating frame the canonical operators for transverse spin components are related to the spin components in the lab frame via
\begin{align*}
  j_y(z)/\sqrt{n(z)}&=\cos(\omega_L t)x_A-\sin(\omega_L t)p_A,\\
  j_z(z)/\sqrt{n(z)}&=\cos(\omega_L t)p_A+\sin(\omega_L t)x_A.
\end{align*}
with the number density of atoms $n(z)$.

The Maxwell-Bloch equations in the rotating frame are accordingly,
\begin{equation}
\begin{split}
\frac{\partial}{\partial z}x_L(z,t)&=
- \frac{g^*(z)\Omega(t)}{\sqrt{2}\Delta}\big(\cos(\omega_L t)p_A(z)+\sin(\omega_L t)x_A(z)\big),\\
\frac{\partial}{\partial z}p_L(z,t)&=0,\\
\frac{\partial}{\partial t}x_A(z,t)&=
-\frac{g^*(z)\Omega(t)}{\sqrt{2}\Delta}\cos(\omega_L t)p_L(z,t), \\
\frac{\partial}{\partial t}p_A(z,t)&=\frac{g^*(z)\Omega(t)}{\sqrt{2}\Delta}\sin(\omega_L t)p_L(z,t).
\label{eq:faraday_rotframe}
\end{split}
\end{equation}
Now the atomic momentum operator, i.e. the spin projection along the axis of light propagation, is not conserved anymore and the overall interaction is not of QND character. Integration of these equations becomes somewhat more involved than before. We will resort to this problem in sections \ref{sec:Entanglement} and \ref{sec:teleportation}.

\subsection{Quantum measurement and feedback}
\label{sec:measurement}

In this section we will deal with measurements which can be done on light and then fed back onto atoms. We will focus on homodyne detection of light, which is of importance for experiments using Faraday interaction and EIT but also briefly describe photon counting, which is used in combination with parametric gain and beam splitter interactions.

\paragraph*{Homodyne detection of light}

\begin{figure}[ht]
  \includegraphics[width=7.5cm]{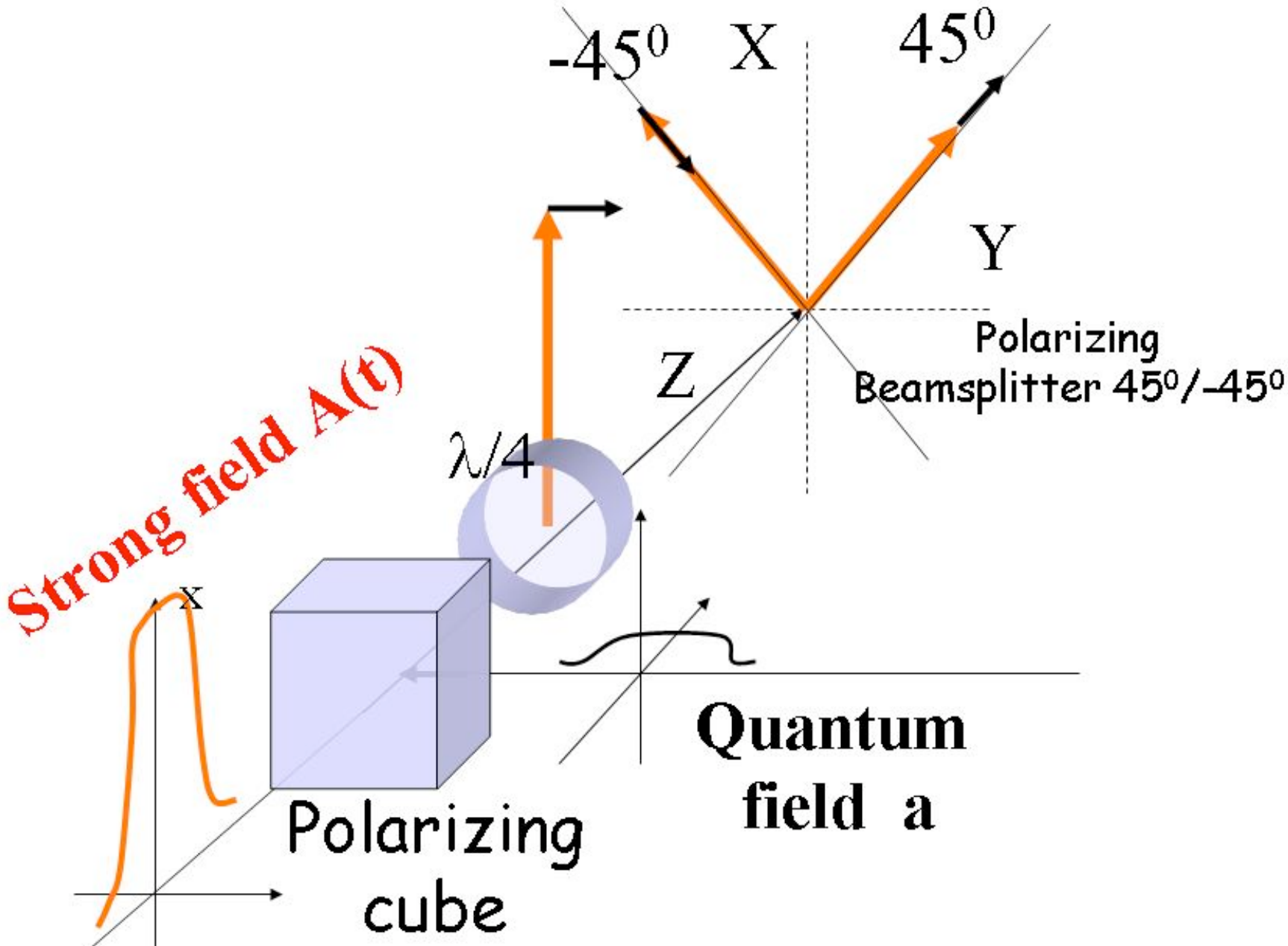}
  \caption{Polarimetric measurement of light}\label{Fig:pol_homodyning}
\end{figure}

The discussion of the quantum interface in the language of canonical variables for light is most fruitful because these variables can be measured with almost perfect efficiency by the balanced homodyne technique. We will concentrate here on the polarization homodyne version which is relevant for several protocols described in the article. In particular in the context of the Faraday interaction the polarimetric measurement of light is an important tool. Balanced homodyning employs overlapping the quantum field of interest with a strong coherent field, a local oscillator, on a $50/50$ beamsplitter and measurement of the difference of the power in the two outputs. In its polarization version as shown in Fig. \ref{Fig:pol_homodyning}, the local oscillator field and the quantum field are overlapped on a polarizing cube so that they have orthogonal polarizations and the role of the beamsplitter is played by a polarizing beamsplitter which splits the light into $45^{0}$ and $-45^{0}$ modes. The measurement of the differential power of these two modes corresponds to the measurement of the $S_{y}$ Stokes operator, whereas with an extra $\lambda/4$ plate in front of the beamslitter the $S_{z}$ Stokes operator is measured:
\begin{align*}
S_x(t)&=(a^\dagger_{L,x}a^{\phantom\dagger}_{L,x}-a^\dagger_{L,y}a^{\phantom\dagger}_{L,y})/2,\\
S_y(t)&=(a^\dagger_{L,+45^\circ}a^{\phantom\dagger}_{L,+45^\circ}-a^\dagger_{L,-45^\circ}a^{\phantom\dagger}_{L,-45^\circ})/2\\
&=(a^\dagger_{L,x}a^{\phantom\dagger}_{L,y}+a^\dagger_{L,y}a^{\phantom\dagger}_{L,x})/2,\\
S_z(t)&=(a^\dagger_{L,\sigma^+}a^{\phantom\dagger}_{L,\sigma^+}-a^\dagger_{L,\sigma^-}a^{\phantom\dagger}_{L,\sigma^-})/2\\
&=-i(a^\dagger_{L,x}a^{\phantom\dagger}_{L,y}-a^\dagger_{L,y}a^{\phantom\dagger}_{L,x})/2.
\end{align*}
The third Stokes operator $S_{x}$ is equal to the total photon number of the strong field for the case of the strong coherent field $|\alpha\rangle$ in linear $x$ polarization, such that $\langle S_x(t)\rangle=|\alpha|^2/2$. In this case the measurement of the two other Stokes operators amounts to a homodyne detection of $y$ polarized light with the coherent field in $x$ serving as the local oscillator,
\begin{align*}
\frac{S_y(t)}{\sqrt{\langle S_x\rangle}}&\simeq\frac{1}{\sqrt{2}}\left(a^{\phantom\dagger}_{L,y}(t)+a^\dagger_{L,y}(t)\right)=x_L(t),\\
\frac{S_z(t)}{\sqrt{\langle S_x\rangle}}&\simeq-\frac{i}{\sqrt{2}}\left(a^{\phantom\dagger}_{L,y}(t)-a^\dagger_{L,y}(t)\right)=p_L(t).\\
\end{align*}
The homodyne detection can also provide an excellent suppression of the technical (classical) noise if the frequency of the local oscillator and the quantum field differ by $\omega_{L}$ lying in the radio frequency domain, see Fig. \ref{fig:bfield}. In this case the relevant canonical variables are encoded in sideband modulation modes of $y$-polarized light which are read in the $\cos(\omega_L t)$ and $\sin(\omega_L t)$ components of the photodetector output:
\begin{equation}\label{eq:modulation modes}
\begin{split}
  X_{L_c}&=\sqrt{\frac{2}{T}}\int dt\cos(\omega_L t)x_L(t),\\
  P_{L_c}&=\sqrt{\frac{2}{T}}\int dt\cos(\omega_L t)p_L(t),\\
  X_{L_s}&=\sqrt{\frac{2}{T}}\int dt\sin(\omega_L t)x_L(t),\\
  P_{L_s}&=\sqrt{\frac{2}{T}}\int dt\sin(\omega_L t)p_L(t).\\
\end{split}
\end{equation}
These components of the photocurrent can be measured by lock-in amplifiers. The bandwidth of this measurement can be adjusted to $BW\approx \tau^{-1}$ where $\tau$ is the optical pulse duration. In this way fluctuations at all frequencies outside this bandwidth are effectively irrelevant. In the case where atomic ground state levels are non-degenerate, e.g., are split by the Larmor frequency $\omega_{L}$ as discussed in Sec. \ref{sec:magnetic field} and shown in Fig. \ref{fig:bfield}, the atoms couple to the sidebands of light and the entire measurement and interaction can be encoded at sideband frequencies $\pm\omega_{L}$, as in several experiments described later in the article.

\paragraph*{Feedback}

Another important tool in many quantum information protocols is the feedback of results of measurement of light onto atoms. The theory of quantum feedback is a wide field on its own, especially in the case of continuous measurement and feedback \cite{Thomsen:2002}. Here we deal with a relatively simple measurement and feedback scheme, where light observables of the type \eqref{eq:modulation modes} are measured by integrating a photocurrent over the whole pulse duration and the measurement result, a single number, is fed onto the atoms. The operations which need to be done on the collective spins are small rotations about the $y$ or $z$ axis, i.e. small tilts of the collective spin. In the language of canonical operators $X_A,~P_A$ this amounts to displacements in the phase plane \cite{Arecchi:1972}.

In this case -- feedback of integrated measurement results via displacement operations -- a simple rule can be applied for describing the overall effect on a state of the atoms. Assume that the state of the system is described by certain input-output relations of the type \eqref{eq:faraday_nodecay_result}. If a quadrature of light, say $X_L$, is measured and the corresponding measurement result $\xi$ is used to displace the atomic state, i.e. to tilt the collective atomic spin, in such a way that the mean of, say $P_A$, is transformed as $\langle P_A\rangle\rightarrow \langle P_A\rangle+g\xi$, then 
the statistics of $P_A$ after the feedback operation can be calculated from
\begin{align}\label{eq:feedback}
P_{A,{\rm final}}=P_A+gX_L,
\end{align}
that is, one simply needs to add the measured observable multiplied by the gain to the operator which is subject to the feedback. This rule for describing the feedback holds strictly as an operator identity, irrespectively of the state of the system being Gaussian or Non-Gaussian.

The proof is most easy in the Schr\"odinger picture. Assume the state of some bipartite system is $\hat\rho_{AL}$, where the indices refer, e.g., to atoms and light respectively. A measurement of $X_L$ gives a result $\xi$ with probability $p_\xi=\langle\xi|{\rm tr}_A\{\rho_{AL}\}|\xi\rangle$, where $X_L|\xi\rangle=\xi|\xi\rangle$, and the state of the system $A$ collapses to
\[\rho_A^{(1)}=p_\xi^{-1}\langle\xi|\rho_{AL}|\xi\rangle.\]
The feedback affecting the desired displacement is described by a unitary transformation of the state of the system $A$,
\[\rho_A^{(2)}=e^{ig\xi X_A}\rho_A^{(1)}e^{-ig\xi X_A},\]
which gives in the ensemble average over all possible measurement results the final state
\begin{align*}
  \rho_A^{(3)}&=\int d\xi~p_\xi\rho_A^{(2)}=\int d\xi e^{ig\xi X_A}\langle\xi|\rho_{AL}|\xi\rangle e^{-ig\xi X_A}\\
  &=\int d\xi \langle\xi|e^{igX_L X_A}\rho_{AL}e^{-igX_L X_A}|\xi\rangle \\
  &={\rm tr}_L\left\{e^{igX_L X_A}\rho_{AL}e^{-igX_L X_A}\right\}.
\end{align*}
From this equation all the moments of $P_A$ can be calculated as
\begin{align*}
\langle P_A^n\rangle&={\rm tr}_A\{P_A^n\rho_A^{(3)}\}\\
&={\rm tr}_{AL}\{(P_A+gX_L)^n\rho_{AL}\},
\end{align*}
where the cyclic property of the trace was used in the second equality. This justifies the rule given above.

Note that for Gaussian states, for which it is enough to keep track only of the first and the second moments in order to have the full knowledge of the state, the simple linear transformation of operators as above is exactly equivalent to a full description in the Schr\"odinger picture e.g. on the basis of the Wigner function. The measurement of and the feedback on more than one mode can be described by an immediate generalization of \eqref{eq:feedback} as shown in \cite{Hammerer:2005,Sherson:2006a}, provided the measurement involves commuting observables only.

\paragraph*{Photon Counting}

In some cases measurements in the Fock state basis, i.e., photon counting are convenient for characterization of the interface performance. As discussed in Sec. \ref{sec:memory},
storage and retrieval of some nonclassical states can be characterized by the measurement of the second order correlation function $g^{2}(1,2)$ which is a normalized probability of photon counts at two points in space-time \cite{Loudon:2004}. Atoms typically used for the interface are Rubidium and Caesium and the corresponding spectral lines are around $780$nm and $850$nm respectively. In this spectral domain commercial avalanche photodiodes (APDs) typically have quantum efficiency around $40-50$ \% and the dark count rate of a few hundred per second. Such parameters are sufficient to determine non-classical correlations via $g^{2}(1,2)$ which are insensitive to losses if dark counts are neglected.

\subsection{Other Strategies}\label{sec:other_strat}

In the discussion so far we have focussed on what we consider to be the main protocols in this field. There are, however, numerous variations of all of these protocols, some of which we will briefly discuss here.

\paragraph*{Non-copropagating beams.} In the derivation above we have only considered the situation, where the quantum and classical light fields are co-propagating. For many applications of the beam splitter and parametric gain interaction this assumption is, however, not necessary \cite{Balic:2005,Braje:2004,Chaneliere:2005}. For instance, an incoming photon which is absorbed in an atomic ensemble by using a co-propagating classical field generates an excitation of the form of Eq. (\ref{eq:w_state}). The reason why one can later retrieve this quantum state is constructive interference. 
During readout all atoms will radiate in phase in the direction of the classical laser and this is the effect, which allows for efficient interfaces between atoms and light in the limit of a large number of atoms (high $d$). If  the photon which is absorbed has a different direction than the classical drive field, the generated atomic excitation will still have the form of Eq. (\ref{eq:w_state}).  The only difference is that the state will have a phase factor $\exp(i\Delta\vec k \cdot \vec r_i)$ on the component, where the $i$th atom is in state $|1\ra$, with $\Delta \vec k$ being the difference between the $k$-vectors for the two fields. In order to have constructive interference in the readout process  the difference in the $k$-vectors of the outgoing photon and the classical field in the read out process should cancel the phase factor imprinted on the atoms in the first step of the protocol (see also Sec. \ref{sec:raman_eit} for a discussion of the effect of difference in the $k$-vectors).  

Expressed in different terms, the initial atomic state $|000....0\ra$ has a homogeneous phase corresponding to a zero momentum state. The initial process involving the absorption of a photon from one beam and the emission of a photon into a different beam imprints the difference momentum  $\Delta \vec k$ onto the atomic spin wave. Constructive interference is achieved for processes returning the atoms to the initial zero-momentum spin wave $|000....0\ra$  and the differential momentum in the read out process must therefore carry away the momentum in the spin wave. To achieve momentum conservation (or equivalently phase matching) the total momentum of  all the absorbed photons should thus match the total momentum of all the emitted photons.

A disadvantage of using non co-propagating beams is that the storage into non-symmetric modes limits  the storage to the time it takes atoms to move a distance $\sim 1/|\Delta \vec k|$. In particular for room temperature gasses (see Sec. \ref{sec:room_temp}) this, as well as the differential Doppler-shift, makes it undesirable to use geometries, where the beams are not nearly co-propagating. Even for cold atomic ensembles this effect limits the storage time, as was observed experimentally by \textcite{Zhao:2009}. On the other hand it is often a major experimental advantage not to have the classical and quantum beams co-propagating since this makes it much easier to count photons in the quantum beam.

\begin{figure}[ht]
  \includegraphics[width=6.5cm]{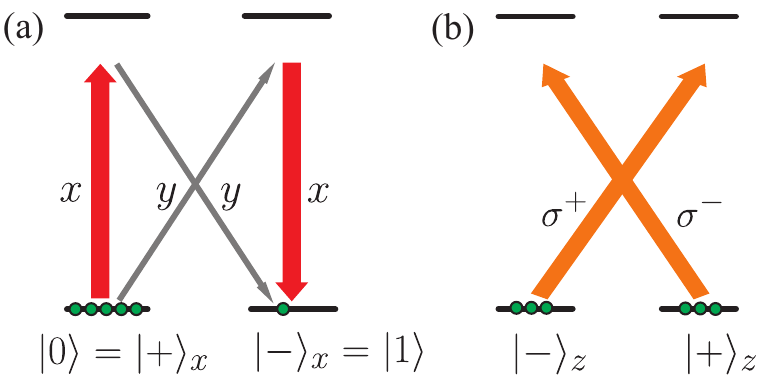}
  \caption{Level scheme for Faraday interaction: (a) For the quantization axis along $x$: Atoms are polarized to $\ket+_x$, laser light is linear polarized along $x$ and drives the up-transitions, while the quantum field in $y$ polarization couples to the cross-transitions. (b) The same interaction with the axis of quantization taken along $z$: Circular light components of equal intensity cause AC Stark shifts of equally populated states $\ket\pm_z$.}\label{fig:quantization_pictures}
\end{figure}

\paragraph*{Interaction based on phase shift}
It is instructive to have a look at the Faraday interaction from another perspective. Consider first Fig.~\ref{fig:quantization_pictures}(a) where, as before, the quantization axis is taken along $x$, the direction of polarization of both atoms and light. Selection rules for dipole transitions dictate that strong classical, $x$-polarized light with amplitude $a_x\simeq\alpha$ drives the $\ket{\pm}_x\rightarrow\ket\pm_x$ transitions, while the $\ket{\pm}_x\rightarrow\ket\mp_x$ cross transitions are coupled to the weak quantum field $a_y$ in $y$-polarization. In this picture it is evident that the Faraday interaction is the sum of the beam splitter and the parametric gain interaction, cf. Fig.~\ref{fig:lambda}. The same level configuration can also be looked at by taking the axis of quantization along the $z$ direction, as shown in Fig.~\ref{fig:quantization_pictures}(b), where $\ket\pm_z=(\ket+_x\pm\ket-_x)/\sqrt{2}$. Light propagating along this direction naturally couples with its circular components $a_{\sigma^\pm}=(a_x\pm ia_y)/\sqrt{2}\simeq (\alpha\pm ia_y)/\sqrt{2}$ to the cross transitions $\ket{\pm}_z\rightarrow\ket\mp_z$ only. The off-resonant coupling will thus give rise to AC Stark shifts on atomic levels $\ket\pm_z$ depending on light intensities of $\sigma^\pm$-polarization components. Vice versa, light polarization is thus rotated according to the difference in level populations $\ket\pm_z$.

\begin{figure}[ht]
   \includegraphics[width=6.5cm]{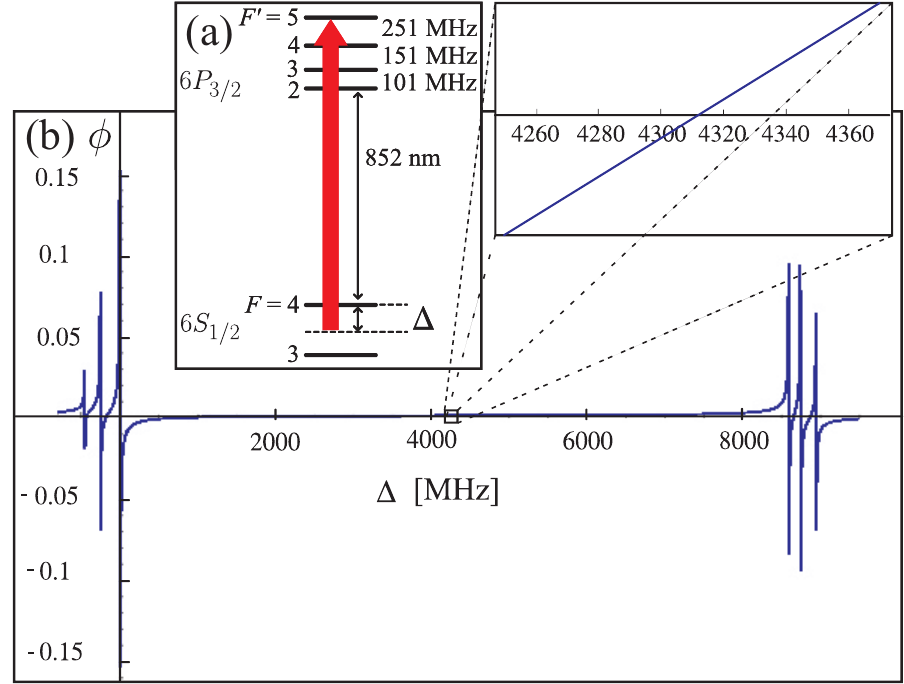}
  \caption{QND interaction for measurement of pseudospin composed of hyperfine ground states of Cs: (a) Level scheme for $^{133}$Cs with probe light tuned in between hyperfine levels $F=3,4$ of the $6^2S_{1/2}$ ground state. (b) Differential phase shift $\Phi$ due to the $F=3\rightarrow F'=2,3,4$ and the $F=4\rightarrow F'=3,4,5$. At magic frequencies the phase shift vanishes.}\label{fig:cslevels}
\end{figure}

From this observation it is clear, that all that is required for a Faraday interaction is a mechanism of non-destructive measurement of level populations via phase shifts of light. Making use of the vector polarizability for probing the collective hyperfine spin-angular momentum as described above, is therefore just one way to achieve a Faraday or QND interaction. In addition, several proposals and experiments pursue the idea to probe coherences of the pseudo-spin consisting of the two $S_{1/2}(F=I\pm1/2)$ hyperfine levels, see \cite{Chaudhury:2006,Oblak:2005,Petrov:2007, Windpassinger:2008}. As shown in Fig.~\ref{fig:cslevels} at a certain  detuning the phase shift of the probe light due to $F=I\pm1/2\rightarrow F'$ transitions exactly cancels for equal populations of $F=I\pm1/2$ levels. Any imbalance of populations will yield an interferometrically detectable phase shift.

\paragraph*{Other Hamiltonians and level structures.}
There are several possibilities involving more complicated atomic level structures than the ones shown in Fig. \ref{fig:lambda}. A particular example is the so called double $\Lambda$-systems, with two excited states. An interesting feature of this system is its potential application for four wave mixing. For a review of this see \textcite{Fleischhauer:2005}.

In the derivation of the theory we adiabatically eliminated the excited state to arrive at the effective ground state Hamiltonian. As discussed in Subsec. \ref{sec:realatoms} this adiabatic elimination in general leads to a Hamiltonian involving spherical tensors of rank zero, one, and two with strength characterized by the three coefficients $a_0$, $a_1$ and $a_2$. The three protocols that we have mainly considered thus correspond to suitable initial states and particular combinations of these spherical tensors. By adjusting the detuning as well as laser polarizations and atomic initial state there is, however, a lot of freedom in varying the relative strength and effect of the different tensors, which allow for a richer dynamics. \citet{Kupriyanov:2005} and \citet{Mishina:2007} considered how the higher order tensor operators modify the equations of motion and in particular how the Faraday interaction is influenced by the rank two tensor. For instance, \textcite{Mishina:2007} it was shown that a particular choice of detuning removes the AC-Stark for a detuned beam splitter interaction and thus removes the need to adjust the frequency of the classical driving field in order to keep the field in two photon resonance with the AC-Stark shifted transition.

To arrive at the Faraday interaction we just combined the beam splitter and parametric gain interaction with the same strength, but in \textcite{Mishina:2006} it was shown that an arbitrary combination of the parametric gain and beamsplitter interaction can be obtained by choosing suitable initial conditions, detunings and combinations of the elements of the spherical tensor.

An example of a protocol where the light-atom interface involves  excited states without adiabatic elimination is shown in Fig. \ref{elemlevels}(a) \cite{Kuzmich:1997,Hald:1999}. Another protocol of this type is considered in Sec. \ref{sec:echo}, where we discuss spin echo techniques. A disadvantage of such protocols is, however, that the storage time is limited by the coherence time of the optically excited state, which is often shorter than the coherence time of ground states.

\paragraph*{Optical cavities.}
The key parameter in characterizing the applicability of an atomic ensemble for a light matter quantum interface is its optical depth which for a free space ensemble is limited by the size and atomic interaction.  An alternative strategy is to use multiple passes of the light through the atomic ensemble by enclosing the ensemble in an optical cavity \cite{Simon:2007,Thompson:2006,Simon:2007a,Black:2005,Josse:2004,Dantan:2005}. In this case the parameter characterizing the usefulness of the system is the cooperativity parameter $C=N_A g_c^2/\kappa_c\gamma$, where $g_c$ is the coupling constant for a single atom to the cavity mode, and $\kappa_c$ is the cavity decay rate. The cooperativity can also be expressed as $C\sim\mathcal F d$, where $\mathcal F$ is the finesse of the cavity which roughly equals the number of passes that the photon makes through the cavity \cite{Gorshkov:2007b}. The gain achieved by using a cavity thus equals the number of round trips.

\paragraph*{Non Gaussian operations.}
In this review we consider the quantized light fields which are much weaker than classical control and driving light, and the quantum fluctuations of the atomic ensemble which are much smaller than the mean spin. In this limit we only include the lowest order terms in the atomic and light field operators $a_A$ and $a_L$.   Because there are no first order terms the effective Hamiltonian will be quadratic in the harmonic oscillator operators. The operations which may be performed thus fall into the class of  Gaussian operations and the solution of the equations will in general be a Bogoliubov transformation of the incident mode operators \cite{Braunstein:2005}. For any input state with a Gaussian Wigner function the output Wigner function will also be Gaussian. The main advantage of using atomic ensembles is that the dynamics resulting from these Gaussian operation are collectively enhanced so that a perfect operation is achieved in the limit of large optical depth.  While the resulting dynamics allow for a variety of quantum information protocols to be performed, such as quantum teleportation and quantum memory (sec \ref{sec:memory} and \ref{sec:teleportation}), the fact that higher order terms are not collectively enhanced limits the applications for quantum information processing. In particular it is known that Gaussian operations alone do not allow for distillation of entanglement from Gaussian states \cite{Giedke:2002,Eisert:2002,Fiurasek:2002} and that algorithms for efficient classical simulation of any evolution involving only Gaussian operation and Gaussian initial states exist \cite{Bartlett:2002,Lloyd:1999}. These limitations may, however, be avoided by combining the Gaussian operation with photon counting \cite{Neergaard-Nielsen:2006,Ourjoumtsev:2006,Genes:2006}. In the pioneering paper by \textcite{Duan:2001} such photon counting techniques were proposed as a means for quantum communication over long distances using the probabilistic entanglement protocols discussed in Sec \ref{sec:probabilistic}. Such techniques could in principle also allow for even more advanced quantum information protocols to be implemented.

The fundamental obstacle for {\it directly} achieving non Gaussian operations is that they rely on an interaction between individual excitations. Such interactions between the excitations can for instance be achieved if two photons interact with the same atom. But since this is essentially a single atom effect it is not enhanced by a large optical depth.  An approach  which allows for an enhancement of this nonlinear effect by optically imprinting a Bragg mirror which localizes excitation similar to an optical cavity is explored in \textcite{Bajcsy:2003} and \citet{Andre:2002a}.
 An  alternative approach to non Gaussian operations is to engineer  a strong  interaction between excitations stored in different atoms. Interesting proposals in this direction 
 are to use the collisional interactions of atoms in optical lattices \cite{Muschik:2008} or the so called Rydberg blockade, where the excitation of a single atom to a Rydberg level blocks the excitation of other atoms, and therefore creates a uniform long distance interaction \cite{Lukin:2001}. Alternatively one can exploit the fact that atomic ensembles are particularly well suited for "catching" traveling photons and then afterwards transfer the excitation to some other system for processing the information. A proposal along these lines is presented by \textcite{Rabl:2006} based on a transfer of excitations from an ensemble of dipolar molecules to a solid state system.
A review of techniques for achieving other types of operations, e.g., Kerr interactions, is  given by \textcite{Fleischhauer:2005}.

\subsection{Summary of the theory}
\label{sec:Theory_summary}

We have presented a detailed theory for the quantum interfaces between light and atomic ensembles. In particular we have  presented a unified theory for the three model systems presented in Fig. \ref{fig:lambda}(a)-(c), the beam splitter, the parametric gain, and the QND (Faraday) interaction. The three systems have distinct features, but are also interconnected, e.g., the Faraday interaction is just a combination of the beam splitter and parametric gain interaction with equal weights. Most importantly all three systems achieve ideal operation in the limit of high optical depth $d$. This feature can be understood as  constructive interference or collective enhancement of the coupling: the coupling between the state where all atoms are in the ground state and the collective state (\ref{eq:w_state}) scales as $\sqrt{N_A}$, whereas spontaneous emission is a single atom effect, which is independent of the atom number. We emphasize, however, that the theory is derived without the optical broadening present in many experimental realization. One therefore cannot directly replace the optical depth $d$ appearing in the equations of this section with the actual measured optical depth in the presence of inhomogeneous broadening. Nevertheless the optical depth in one version or the other remains the key parameter for characterizing the usefulness of an atomic ensemble (see Sec. \ref{sec:errors} for a discussion of inhomogeneous broadening).

For all three types of interaction, in the far off-resonant limit $\Delta\gg d \gamma$, the strength of the coupling is parametrized by exactly the same function $h(0,T)\approx\kappa^2$, cf. Eqs. (\ref {eq:light_out_bs},\ref{eq:lambda_kernel},\ref{eq:ht}) and (\ref{eq:faraday_weakdecay_result},\ref{eq:kappa_full}), which is most easily seen by rewriting the equations in the dimensionless form as in Appendix \ref{sec:dimensionless}. In this limit the decay and phase shift of the light can be ignored and the coupling constant can be related to $\eta_A$, the spontaneous emission probability per atom, through $\kappa^2=\eta_A d$, which we explicitly derived for the Faraday interaction, but which also applies to the other systems. From this expression one thus directly sees that the spontaneous emission can be eliminated for high optical depth. In a special case of the beam splitter interaction, the resonant EIT setting, these arguments are not directly applicable, but the ratio between the desired evolution and spontaneous emission, i.e. the constructive interference discussed above,  is completely independent of detuning \cite{Gorshkov:2007a,Gorshkov:2007c}.

It is instructive to discuss the bandwidth of the quantum interface. i.e., how fast an operation such as a storage or a read process can be performed. In the theory we have adiabatically eliminated the excited state, which means that the shortest time $\tau$ on which an operation can be performed  
is limited
by the low saturation parameter condition $s\ll1$ and the condition on the value of the coupling constant $\kappa$  necessary for a particular process. Bearing in mind the relations $\eta_A=\gamma \tau s$, $\tau=(BW)^{-1}$, we can draw some general conclusions on the Fourier limited bandwidth $BW$ of the process. For protocols where $\kappa^{2} \sim 1$ the low saturation condition yields the limitation on the bandwidth of the light pulse, $BW\ll\gamma d$ (again the arguments given here only directly apply in the far off resonant limit, but the conclusion is also valid for EIT). Going beyond the low saturation  regime $s\sim 1$ allows to increase the bandwidth somewhat \cite{Gorshkov:2008}, but it remains   limited by $BW\lesssim \gamma d$. For a typical alkali atomic ensemble the bandwidth of the order of $10$ MHz can be achieved. This scaling of the BW provides an upper limit,$BW\ll\gamma\sqrt{d}$,
 e.g., for an atomic EPR entanglement protocol discussed later in the article where the conditions $\kappa^{2}\gg1, \eta\approx 1/\sqrt{d}$ has to be met.
 Note, however, that the term bandwidth is also often used in connection with the number of different modes which can, e.g., be stored in an atomic ensemble. This is a different question, which we will return to in Sec. \ref{sec:echo} .

%
%
\section{Atomic media for quantum interface}\label{sec:atomic_media}

 A few common requirements can be formulated for all ensemble based interfaces described in the previous sections. A long lived (ground) state of atoms is commonly used. This could, e.g.,  be Zeeman levels or hyperfine levels.  The ensemble should be initialized to a polarized state (coherent spin state), that is, one of the ground substates should be populated by optical pumping or other means.   Most importantly, the sample should have a large resonant optical depth $d$.
  Up to date, experimental realizations of the ensemble based interfaces utilize alkali atom gases at room temperature, alkali atoms cooled and trapped at temperatures of a few tens or hundreds of microkelvin, or impurity centers in solid state. Below we describe these and other media used for quantum interfaces.

\subsection{Room temperature gases}
\label{sec:room_temp}
A gas sample of alkali atoms is one of the simplest atomic ensembles to have in the lab. Surprisingly enough such an object can also work very well as a quantum memory, if proper care of decoherence is taken. The thermal motion and associated with it Doppler broadening are not necessarily a problem.  For the Faraday interaction,
the Doppler broadening plays little role
if the detuning is much greater than the Doppler width (200-300 MHz for Cesium or Rubidium). For other protocols such as the beam splitter interaction the Doppler broadening has a detrimental but still tolerable effect as discussed in Sec. \ref{sec:inhom}.

In addition to the Doppler broadening the atomic motion also leads to changes in the atomic positions. 
 A quick glance at the 
 solution to the beam splitter interaction (\ref{eq:lambda_kernel}) and  (\ref{eq:atom_bs_out})  
reveals that the atomic operators  with different longitudinal coordinates experience different dynamics. 
The atomic motion in the process of interaction leads to washing out of these spatial modes which is a much more pronounced problem for the beam splitter interaction, as compared to the Faraday interaction.

In order to reduce the deleterious effect of atomic motion a buffer gas is usually used in the experiments which utilize the beam splitter  interaction in gas cells at room temperature \cite{Eisaman:2005,Novikova:2007}. Adding a few torr of a noble gas allows to sufficiently localize the diffusive motion of alkali atoms. An extra benefit of this approach is that it prevents alkali atoms from depolarizing collisions with the walls of the cell which otherwise could lead to a rapid decoherence.  The lifetime of the atomic memory in cells with a buffer gas can reach milliseconds \cite{Novikova:2007}. Note that for some protocols, collisions of atoms in the excited state with the buffer gas lead to an energy redistribution of scattered photons which may lead to large errors in the absence of careful spectral filtering \cite{Manz:2007}. Despite the difficulties, alkali atom cells with a buffer gas has been successfully used for experiments on quantum memory using EIT \cite{Eisaman:2005}.

The effect of atomic motion for the Faraday interaction can be almost completely eliminated. As follows from the propagations equations (\ref{eq:faraday_nodecay_result}) the Faraday interaction couples light to a symmetric atomic mode defined in Eq. (\ref{eq:xp_sym}). In this case atoms with different coordinates $z$ along the direction of light propagation couple to light in the same way. Hence the atomic motion along $z$ does not affect the interaction. The transverse motion of atoms along $x$ and $y$ axes will affect the performance if the spatial profile of the light beam is inhomogeneous which is almost always the case. This effect can be reduced in two extreme cases, either for times short compared to the motion time, or in case when the duration of the light pulse is so long that atoms have a chance to cross the beam many times during the interaction and the effect of motion averages out. The latter was the case of the experiments by \citet{Julsgaard:2001,Julsgaard:2004} and \citet{Sherson:2006}, where pulses of about $1$ msec duration have been used.

\begin{figure}[ht]
  \includegraphics[width=8.6cm]{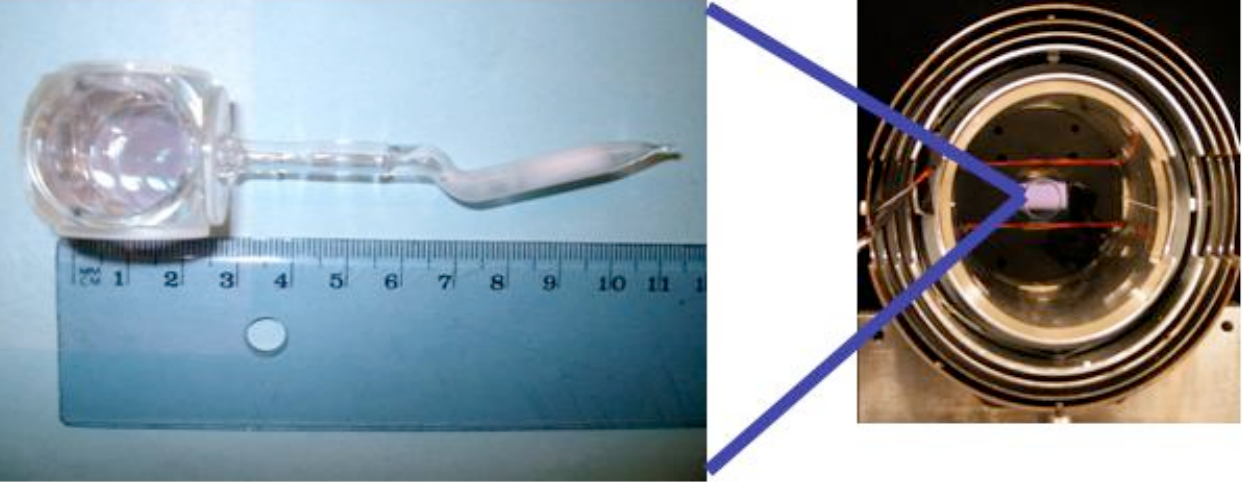}
  \caption{Paraffin coated Caesium cell.}\label{fig:cell}
\end{figure}

 The possibility to eliminate the effect of atomic motion on the efficiency of the interface based on the Faraday interaction has allowed to conduct high-fidelity experiments with room temperature Cesium atoms \cite{Julsgaard:2001,Julsgaard:2004,Sherson:2006}. Atoms were contained in cells with a paraffin coating of the internal walls (Fig. \ref{fig:cell}). Such coating has been used in precision magnetometers for the past three decades \cite{Alexandrov:2003,Weis:2006}, and ground state coherence times of up to a second have been demonstrated.  In paraffin coated cells atoms can withstand tens of thousands collisions with the cell walls before significant spin depolarization occurs. Since it is the number of collisions with walls that matters, the larger is the cell, the longer is the quantum memory lifetime. As discussed in details in Sections IV, V, and VI, quantum memory time of the order of several milliseconds has been achieved in cells with dimensions $25\times25\times25$ mm$^{3}$.

 Room temperature ensembles of Cesium atoms of a few cubic cm size contain about $10^{11}-10^{12}$ atoms.
 For  off-resonant Faraday-type interfaces the Doppler broadening does not affect the ensemble effective resonant optical depth (see sec. \ref{sec:inhom}) which is the same as for atoms at rest reaching the values  of the order 50 or even higher. The experimental challenge lies with the fact that the quantum spin noise of such an ensemble is $\sqrt{N_{A}}^{-1}\approx10^{-6}$. In order to reach the level of the spin quantum noise, all types of technical spin fluctuations, such as driven by stray magnetic fields or fluctuations of the lasers used for optical pumping,  have to be reduced below this level. The solution to this problem used in \textcite{Julsgaard:2001,Julsgaard:2004,Sherson:2006} has been to apply a bias magnetic field along the direction of the collective atomic spin. A field of the order of one Gauss provides the Zeeman splitting of the ground state $\omega_L$ of a few hundred kHz. As discussed in Sec. \ref{sec:magnetic field}  this means that the collective transverse components of the spin which correspond to the atomic canonical variables rotate at the Zeeman frequency. At the same time, as described in Sec. \ref{sec:measurement}, canonical variables of light which couple to the rotating atomic spin can be measured via homodyne detection also at the Zeeman frequency. Thus all relevant variables for light and atoms are now encoded at a frequency of a few hundred kHz. At these frequencies technical noise can be reduced below the  $10^{-6}$ level, so that both spin and light fluctuations are dominated by quantum noise. In practice the photocurrent detected in the homodyning process is measured with lock-in amplifiers which allow  access to the light variables 
(\ref{eq:modulation modes}) encoded at the frequency $\omega_L$. In the experiments by \textcite{Julsgaard:2001,Julsgaard:2004,Sherson:2006} the bandwidth of the memory has been reduced to around $1$ kHz for the reasons discussed in Sec.~\ref{sec:Theory_summary}.

\subsection{Cold and trapped atoms}
\label{sec:cold}

Cold and trapped ensembles of alkali atoms have been among the first atomic objects to be used for quantum interfaces with light. The first experiment mapping quantum properties of light onto atoms was performed with Caesium atoms in a MOT, a magneto-optical trap \cite{Hald:1999}. A MOT provides a relatively simple way to achieve a cold atom sample suitable for the quantum interface, however it also has its limitations. A typical resonant optical depth in a MOT lies in the range between 2 and 10 which is not very high. Another consideration concerns the transverse crossection of the light beam which couples to the atoms. In most of the experiments on interfaces which use a MOT, light is focused down to a few tens of microns which is much less than a typical MOT crossection of a millimeter \cite{Simon:2007,Chaneliere:2005,Chen:2008,Chou:2004,Dantan:2005}. Such geometry limits the atomic memory lifetime to the transient time it takes the atoms to leave the probe volume. For a typical MOT temperature of 100 $\mu$K 
this time is around hundreds of microseconds, as was demonstrated by \textcite{Zhao:2009}. If, on the other hand, the beam crossection is such that the light couples to the entire MOT the transient effects become irrelevant. However, in this case the number of photons in the strong driving field $N_P$ grows proportionally to the cross section $A$, as evident from Eq. (\ref{eq:scaling}). 
When a photon number of the strong field is too high it becomes more difficult to implement protocols based on separating and photon counting of the quantum mode. Protocols based on homodyne detection also place limits on the maximal number of photons in the driving field. In most cases the driving field is also used as the local oscillator for homodyne detection. This implies that in order for the detection to be shot noise limited the classical fluctuations of the strong field should be suppressed to better than $N_P^{-1/2}$. In practice this places the limit $N_P\leq10^{10}$. For higher values of $N_P$ modulation techniques similar to those used with thermal ensembles allow to get down to shot noise limited detection.

Another difficulty of working with a MOT is due to the presence of gradient magnetic fields and the associated difficulty of optical pumping and magnetic states decoherence. This problem can be overcome by switching off the MOT fields which in turn limits the lifetime of the interface, and by using a clock transition as was done in \cite{Zhao:2009} where the memory lifetime up to $1$ msec was achieved. An advantage of using a small sub-ensemble of a large MOT is that a single MOT can then serve as a source of two or more atomic ensembles \cite{Choi:2008,Chen:2007a,Matsukevich:2004}.

Using a far detuned dipole trap allows to overcome a number of problems associated with the MOT. A dipole trap forms an atomic ensemble with a typical transverse size of around 10-50 $\mu$m which is a good size for the interface. The resonant optical depth can reach 20 or more. Dipole trapped atomic ensembles demonstrate coherence times exceeding 10 msec \cite{Windpassinger:2008} and storage times for single excitations exceeding 100~$\mu$s \cite{Chuu:2008} or even a few msec \cite{Zhao:2009a}. Dipole trapping can also be insensitive to the magnetic quantum number and, in some cases, to the hyperfine quantum number. A detailed investigation of dipole trapped atoms as the medium for the spin squeezing and the interface is given in \cite{Oblak:2005,Windpassinger:2008}.

A Bose-Einstein condensate (BEC) is an attractive medium for the quantum interface due to its very high resonant optical depth. Indeed, BEC has been the medium where classical coherent storage of light, the so called stopped light has been first demonstrated \cite{Liu:2001}.  It is important to note that it is not only the on-axis optical depth that defines the strength of the interface coupling. For example, a "one dimensional" sample will not couple efficiently to a focused Gaussian beam because the diffraction of the beam means that it will not be overlapping with the ensemble if the Fresnel number is less than unity \cite{Muller:2005}.

One problem with a BEC-based quantum interface is the low rate at which experiments can be performed. A typical BEC requires tens of seconds to be created. Then the sample can be used for  interface experiments a few times, after which a new sample should be created. BEC on a chip \cite{Hansel:2001,Schneider:2003} offers an attractive alternative where much faster loading times can be combined with very efficient optical coupling.

\subsection{Solid state}
\label{sec:solid}
An optically dense collection of atom-like impurities in a solid state host is an excellent candidate for the quantum interface. Both $\Lambda$ schemes as well as the photon echo-based memory (see Sec. \ref{sec:echo}) have been investigated. The absence of motion in solid state means that complex spatial structures can be generated by light and stored. Recently substantial progress has been achieved with crystals or glasses doped with rare-earth elements, such as erbium (Er), thulium (Tm), praseodymium (Pr), neodymium (Nd), and europium (Eu). The rare-earth ions doped into glass and crystal materials (REIC) display up to a second coherence times of the ground state at liquid Helium temperatures. The ions experience strong inhomogeneous broadening up to $100$ GHz due to local lattice fields. The technique of spectral hole burning \cite{Kroll:2004} followed by spectrally selective anti-hole populating allows to create a sub-ensemble of ions with nearly natural optical transition bandwidth. A substantial optical depth can be created, although its value is usually limited by the ion-ion interaction at high density of doping. Note, however, that one cannot directly compare the measured optical depth in the presence of inhomogeneous broadening with the optical depth introduced in the theoretical derivation in Sec. \ref{sec:theory}, see sec. \ref{sec:inhom}.
EIT- and Raman- based memory \cite{Longdell:2005} has been explored in REIC materials as well as a
photon echo approach based on Controlled Reversible Inhomogeneous Broadening (CRIB)  \cite{Moiseev:2001,Kraus:2006,Hetet:2008,Staudt:2007,Riedmatten:2008}

The optical Raman coupling to the nuclear spin coherence has been investigated for the past decades in  REIC, mostly in praseodymium- or europium-doped crystals. These materials seem particularly suitable for the EIT- and Raman-based quantum memory protocols since they exhibit a hyperfine structure where a $\Lambda$-system can be found, together with long optical coherence lifetimes, and also long hyperfine coherence lifetimes (15ms for Eu:YSO and $550~\mu$s, that can be extended up to 30s by dynamic decoherence control techniques,  for Pr:YSO \cite{Fraval:2005}). The absorption wavelength of these materials is in the domain of dye lasers (606 nm for Pr and 580nm for Eu). In order to take advantage of their long optical coherence lifetimes, the dye laser sources must be stabilized down to less than 1 kHz. The absorption wavelength of Tm lies in the convenient range of diode lasers (793 nm). It also exhibits long optical coherence lifetimes, similar to that of Pr.  It has been recently shown that it is possible to build a $\Lambda$-system in thulium by applying a magnetic field in a very specific orientation \cite{Louchet:2008}.  In rare-earth ion-doped crystals, the transitions are not polarization-selective, so the only way to address them separately is to use a source whose bandwidth is smaller than the ground state sublevel splitting. In Pr and Eu the splittings are fixed (10 and 17 MHz for Pr:YSO, 75 and 102 MHz for Eu:YSO) whereas in Tm they can be adjusted with magnetic field  (36MHz/T in Tm;YAG). The hyperfine coherence lifetime of up to $300 \mu $sec has been measured in Tm:YAG.


Er-doped materials are studied with photon echo techniques \cite{Staudt:2007}. The prime interest to this ion is due to the optical wavelength in the telecom band $1.5~\mu$m. The optical coherence lifetime of this material is very short (a few $\mu$sec at most), but can be dramatically increased by applying a very intense magnetic field. The most promising results up to date have been achieved
in Pr-doped crystals \cite{Longdell:2005,Hetet:2008}.

Materials containing a high concentration of quantum dots may be interesting candidates for the ensemble-based interface. Experiments with spin polarized dots show sufficient ground state coherence times and possibility of optical pumping and quantum nondemolition coupling \cite{Atature:2007}. However, up to now the work with ensembles of dots has not reached quantum limits probably due to an insufficient optical depth.
A different solid state medium which can be used for a quantum interface is Nitrogen vacancies in diamond, where EIT has been observed by \textcite{Hemmer:2001}.

\subsection{Other possible media}
Optical lattices have attracted a lot of attention lately due to the exciting possibilities for generation of entanglement by controlled atom-atom interaction \cite{Mandel:2003}. The lattices can also display high optical depth since structures of up to $100\times100\times100$ atoms spaced by half a micron can be created. In the recent experiment by \textcite{Zhao:2009a} long memory lifetimes exceeding 6~ms were achieved in a one dimensional optical lattice. A quantum interface with such a lattice would offer an exciting possibility to transfer entanglement from atoms to light and to combine the quantum information processing capabilities of lattices with the quantum networking provided by the quantum interface. Theoretical studies of quantum interfacing of light with lattices have recently appeared \cite{Muschik:2008,Eckert:2008} and the first EIT-based storage of classical light with the coherence time of $240$ msec has been demonstrated \cite{Schnorrberger:2009}.

Another system where a collection of atoms can be efficiently interfaced with light is a large ion crystal in an ion trap. Very clean and large ion crystals have been created and first attempts towards achieving quantum coupling to light have been undertaken \cite{Herskind:2008}.

%
%

\section{Entanglement of Atomic Ensembles}
\label{sec:Entanglement}

In this chapter we describe generation of entangled states of two distant macroscopic objects. The first method, which is based on QND interaction, measurement and feedback, generates an EPR (two-mode squeezed) state of atomic spins. The second method, which relies on parametric gain and beam splitter interactions and single photon counting, creates Bell states in two collective spins.

\citet{Enk:2007} gave a useful classification of the various types of entanglement, which are generated in experiments. They distinguish {\it a priori entanglement}, which can be deterministically generated, {\it a posteriori entanglement}, which is generated probabilistically and destroyed when measured, and finally {\it heralded entanglement}, which is as well probabilistically generated, but success can be testified by measuring an auxiliary system, such that the entangled state is still available for use. Using post-selection of successful cases, all types of entanglement are in principle equally useful. When combining a large number of entangled states, e.g. via entanglement swapping in a quantum network, the overall success probability will be dramatically different for a posteriori entanglement as compared to heralded or a priori entanglement. This observation lies at the heart of the original quantum repeater protocol \cite{Duan:2001}, which is based on entanglement of atomic ensembles heralded by detection of single photons, as will be described in Sec.~\ref{sec:probabilistic}. Our main focus in Sec.~\ref{sec:deteministic_ent} will be on the a priori entanglement achieved via a QND-Bell measurement and feedback on two ensembles. To introduce this method, we first explain how a single atomic ensemble can be prepared in a spin squeezed state by means of a QND interaction, homodyne detection of light and feedback on atoms. Note that in Sec.~\ref{sec:raman_eit} we describe the memory experiment \cite{Choi:2008} which involves a {\it heralded entanglement} as an intermediate step.

\subsection{Spin-Squeezing in a Single Ensemble}\label{sec:SpinSqu}

Spin squeezed states of atomic ensembles were introduced by \textcite{Kitagawa:1993} in analogy to squeezed states of the radiation field and suggested by \textcite{Wineland:1992,Wineland:1994} to be of use for enhancing the sensitivity in atomic spectroscopy, Ramsey interferometry, and atomic clocks. Accordingly, \citet{Kitagawa:1993} define the state of a collective spin $J$ to be squeezed if the variance of one spin component $J_\perp$ transverse to its mean polarization is smaller than the transverse variance corresponding to an atomic coherent (Bloch) state \cite{Arecchi:1972}, that is a product state of fully polarized atoms. With this definition, a state is squeezed if $\xi_S=\Delta J_\perp/\sqrt{J/2}<1$ and necessarily consists of correlated atoms. \citet{Wineland:1992,Wineland:1994} on the other hand show that the figure of merit for the suppression of quantum fluctuations, which ultimately limit the sensitivity of atomic Ramsey interferometry, is $\xi_R= (2J)^{1/2}\Delta J_\perp/|\la\vec{J}~\ra|<1$
and provides an alternative, stronger definition of spin squeezing.

We will here follow yet another definition and refer to a spin state as squeezed if $\xi=\Delta J_\perp/(|\la\vec{J}~\ra|/2)^{1/2}<1$,
which is stronger than the definition due to \citeauthor{Kitagawa:1993} but weaker than the one due to \citeauthor{Wineland:1992}, because $\xi_R\geq\xi\geq\xi_S$ as can be easily seen. The conditions are the same for nearly fully polarized states $\la \vec J\ra\approx J$.
If we take the mean polarization along $x$, and assume that the transverse component with minimal variance is along $z$, the definition of spin squeezing adopted here is
\begin{equation}\label{eq:spinsqu1}
\Delta P_A^2<\frac{1}{2},
\end{equation}
where we use the Gaussian approximation \eqref{eq:xp_sym}. We use this definition, because it immediately translates into the entanglement criterion for a bipartite state of two ensembles in Sec. \ref{sec:deteministic_ent}. 

Various ways to create squeezed states in ensembles of two level systems were proposed. They involve direct interaction of spins \cite{Andre:2002,Pu:2000,Sorensen:1999,Sorensen:2001}, mapping of squeezed light onto atoms \cite{Hald:1999,Kuzmich:1997,Appel:2008,Honda:2008,Dantan:2006a}, multiple passes of light through atoms \cite{Takeuchi:2005,Hammerer:2004} or a projective, Faraday interaction based QND measurement \cite{Braginsky:1996}, as used by \textcite{Kuzmich:1998,Kuzmich:2000}. The main idea in the last method is that light correlated with a collective atomic spin via a Faraday interaction can be used as a meter system, reading out one of the spin components. Homodyne measurement of light, as discussed in section \ref{sec:measurement}, then provides information about this spin component, projecting the collective spin into a state with reduced fluctuations in this component.

Mean values and variances of transverse spin components conditioned on a homodyne measurement of light can be easily evaluated by means of the following {\it classical} formulas for the mean value and variance of a Gaussian random variable $\xi$ conditioned on the measurement of another (possibly correlated) Gaussian random variable $\zeta$ with an outcome $z$,
\begin{align*}
  \langle\xi\rangle{\big|_{\zeta=z}}&=\langle\xi\rangle-\frac{\langle\xi\zeta\rangle}{\langle\zeta^2\rangle}\,z,
  &
  \Delta
  \xi^2{\big|_{\zeta=z}}&=\Delta\xi^2-\frac{\langle\xi\zeta\rangle^2}{\langle\zeta^2\rangle}.
\end{align*}
If we assume that both light and atoms are initially prepared in their vacuum states, i.e., atoms are completely polarized along $x$, then the states of atoms and light after the Faraday interaction still have Gaussian statistics, as given by \eqref{eq:faraday_nodecay_result},  and the above formulas apply. Suppose that $X_L$ is measured on the state after the interaction with an outcome $x_L$. According to the formulas above, the conditional variances of atomic spin components are then given by
\begin{align}\label{eq:vars}
  \Delta X_A^2\big|_{X_L=x_L}&=\frac{1+\kappa^2}{2},&
  \Delta P_A^2\big|_{X_L=x_L}&=\frac{1}{1+\kappa^2}\frac{1}{2},
\end{align}
and exhibit spin squeezing.

Conditioned on the measurement outcome $x_L$, $P_A$ is squeezed with the mean value given by
\begin{align*}
  \langle P_A\rangle\big|_{X_L=x_L}&=
  %
  %
  \frac{\kappa}{1+\kappa^2}\,x_L.
\end{align*}
If we ignore the measurement outcome the evolution is given by Eq. \eqref{eq:faraday_nodecay_result} and there is no squeezing since $P_A$ is conserved.
If a feedback operation is applied to the atoms, e.g., by applying a pulse of magnetic field, 
the atomic spin can be tilted such that $P_A$ is displaced by $-\kappa x_L/(1+\kappa^2)$, and the mean value is zero $\la P_A\ra=0$.
 The variances  \eqref{eq:vars} for the (anti)squeezed variances then hold also in the ensemble average.

In Sec. \ref{sec:measurement} we  introduced another method for describing the linear feedback and it is instructive to apply it here. 
The result $x_L$ is fed back on atoms by displacing $P_A$ by an amount $gx_L$, where $g$ is a suitable gain factor. By Eq.  \eqref{eq:feedback} and \eqref{eq:faraday_nodecay_result}, the $P_A$ component after feedback is in the ensemble average  given by,
\begin{align}\label{eq:FeebackSqueezing}
  P_{A,\rm{final}}&=P_{A,\rm{out}}+gX_{L,\rm{out}}=(1+g\kappa)P_{A,\rm{in}}+gX_{L,\rm{in}}.
\end{align}
Minimizing the variance of $P_{A,\rm{final}}$ with respect to the gain $g$, yields an optimal feedback gain $g=-\kappa/(1+\kappa^2)$ and the reduced variance of Eq. \eqref{eq:vars}, in agreement with the discussion above. An experiment along these lines has been reported in \textcite{Kuzmich:2000}. Spin squeezing of 1.8\,dB in the collective spin of a cold ensemble of Ytterbium atoms was reported by \textcite{Takano:2009}. Recently spin squeezing on atomic clock transitions has been demonstrated for Cesium in \textcite{Appel:2009}, with $\xi=-4.5$dB and $\xi_{R}=-3.4$dB, and for Rubidium in \textcite{Schleier-Smith:2009}, with $\xi_{R}=-3.2$dB.

The discussion so far ignores the impairing effects of spontaneous emission and light absorption. In order to take it into account we have to resort to Eq. (\ref{eq:faraday_weakdecay_result}). For the relevant case of small atomic decay, $\eta_A\ll 1$, and dominant light losses due to reflection at glass cells and detector inefficiency parametrized by $\epsilon$ ($1\gg\epsilon\gg\eta_L$) these equations read
\begin{equation*}
\begin{split}
X_{A,{\rm out}}&=\sqrt{1-\eta_A}\big(X_{A,{\rm in}}+\kappa P_{A,{\rm in}}\big)+\sqrt{\eta_A}f_{X_A},\\
P_{A,{\rm out}}&=\sqrt{1-\eta_A}P_{A,{\rm in}}+\sqrt{\eta_A}f_{P_A},\\
X_{L,{\rm out}}&=\sqrt{1-\epsilon}\big(X_{L,{\rm in}}+\kappa P_{A,{\rm in}}\big)+\sqrt{\epsilon}f_{X_L},\\
P_{L,{\rm out}}&=\sqrt{1-\epsilon}P_{L,{\rm in}}+\sqrt{\epsilon}f_{P_L},
\end{split}
\end{equation*}
where we explicitly included Langevin noise operators for atoms $f_{{X_A}({P_A})}$ and light $f_{{X_L}({P_L})}$. For both systems one can to a good approximation assume vacuum properties $\mean{f_\alpha f_\beta}=\delta_{\alpha,\beta}/2$. Using these expressions 
and  minimizing the variance with respect to the gain $g$ yields a minimal variance
\begin{equation}\label{eq:ssslosses}
\Delta P_{A,\rm{final}}^2=\frac{1+\eta_A(1-\epsilon)\kappa^2}{1+(1-\epsilon)\kappa^2}\frac{1}{2}\geq\frac{\eta_A}{2}.
\end{equation}
The bound on the achievable squeezing is not so surprising, given that the state of atoms suffered essentially a 
decay by a fraction $\eta_A$. 
Due to the relation $\kappa^2=d\,\eta_A$, cf. Eq.~\eqref{eq:eta_kappa}, there is always an optimal choice for the decay $\eta_A$ given a certain optical depth $d$ \cite{Hammerer:2004}. For the decoherence model adopted here the limit to spin squeezing by QND measurement and feedback is $\Delta P_{A,\rm{final}}^2\geq (1+\sqrt{1+d})^{-1}\sim d^{-1/2}$, for large optical density. For a detailed discussion on the limits of spin squeezing by means of Faraday interaction and QND measurement we refer to \textcite{Bouchoule:2002}. The limits strongly depend on the particulars of the QND scheme. For example, for the, so-called, two-color probing \cite{Windpassinger:2008} the $1/d$ scaling of the squeezing limit is achievable.

Our description here covers only feedback where the {\it integrated} photocurrent of the homodyne detection is taken as the measurement result and used to correct the atomic state after the probe pulse has passed. It is of course possible to perform a continuous feedback of the photocurrent while the probe pulse is still on. An exhaustive theory for this procedure giving a description in terms of the stochastic Schr\"odinger equation can be found in \textcite{Thomsen:2002,Thomsen:2002a}.

It is interesting to note that, beyond atomic interferometry, spin squeezed states received renewed interest in the theory of many particle entanglement. It was shown in \textcite{Sorensen:2001} that spectroscopic squeezing $\xi_R<1$
is  a sufficient condition for bipartite entanglement within each pair of spins in the atomic ensemble, see also \textcite{Wang:2003}. More general spin squeezing inequalities were fruitfully studied in the context of experimental verification of multipartite entanglement \cite{Sorensen:2001a,Korbicz:2005,Korbicz:2005a,Toth:2007}.

\subsection{Deterministic entanglement}\label{sec:deteministic_ent}

The procedure discussed in the previous section can be applied to deterministically create entanglement between two, spatially separated atomic ensembles, as proposed by \textcite{Duan:2000}. A precursor to this proposal involving entangled light resource has been put forward by \textcite{Kuzmich:2000a}. Each of the ensembles is described by a collective spin $\vec{J}_i\,(i=1,2)$, or, taking polarizations along $x$ for both systems and adopting the Gaussian approximation, by a pair of canonical operators $[x_{A_i},p_{A_j}]=i\delta_{i,j}$. Consider a probe pulse undergoing Faraday interaction with both ensembles, first with ensemble 1 and then, after propagating some distance, with ensemble 2. By linearity, the state of light is desribed by
\begin{align*}
X_{L,{\rm out}}&=X_{L,{\rm in}}+\kappa (P_{A_1,{\rm in}}+P_{A_2,{\rm in}}),\\
P_{L,{\rm out}}&=P_{L,{\rm in}},
\end{align*}
c.f. Eq. \eqref{eq:faraday_nodecay_result}. Note that for both ensembles $P_{A_i,{\rm out}}=P_{A_i,{\rm in}}$ is a conserved quantity in the Faraday interaction. Just as for the single ensemble, a measurement of $X_{L,out}$ will then give a reduced variance of the {\it non-local} observable $P_{A_1}+P_{A_2}$,
\[\Delta(P_{A_1}+P_{A_2})^2=\frac{1}{1+\kappa^2}<1,\]
where the bound corresponds to uncorrelated ensembles in coherent states. Using feedback  this non-local squeezing can be achieved unconditionally. In a second step, the spins of both ensembles are rotated by an angle of $\pi/2$ about the $x$-axis in such a way that
\begin{align*}
  P_{A_1}&\rightarrow X_{A_1}, & X_{A_1}&\rightarrow -P_{A_1},\\
  P_{A_2}&\rightarrow -X_{A_2}, & X_{A_2}&\rightarrow P_{A_2}.
\end{align*}
A second light pulse interacting with both ensembles as before, will then read out the observable $X_{A_1,{\rm in}}-X_{A_2,{\rm in}}$, i.e.
\begin{align*}
X_{L,{\rm out}}&=X_{L,{\rm in}}+\kappa (X_{A_1,{\rm in}}-X_{A_2,{\rm in}}),
\end{align*}
such that a measurement and feedback procedure will produce a squeezed variance of $X_{A_1}-X_{A_2}$, as before. Note that simultaneous squeezing of these two observables is possible only, because we are dealing here with commuting observables, $[P_{A_1}+P_{A_2},X_{A_1}-X_{A_2}]=0.$ The counterwise rotation of the two spins about $x$ is therefore crucial. Overall, this will produce a state which fulfills the inequality,
\begin{equation}\label{eq:EntanglementCondition}
  \Delta(P_{A_1}+P_{A_2})^2+\Delta(X_{A_1}-X_{A_2})^2<2.
\end{equation}
Losses and decoherence will affect this result similarly as the single ensembles spin squeezing in \eqref{eq:ssslosses}. The significance of this inequality is that it constitutes a {\it necessary and sufficient entanglement condition} for symmetric (with respect to $1\leftrightarrow2$) Gaussian states of two systems, \cite{Duan:2000a,Simon:2000}.


In the limit of large squeezing, where the variances of both non-local observables vanish, the corresponding state approaches the (unphysical) ideally correlated state with Wigner function $W\sim\delta(X_{A_1}-X_{A_2})\delta(P_{A_1}+P_{A_2})$, which was considered by Einstein, Podolsky and Rosen (EPR) in their famous Gedankenexperiment \cite{Einstein:1935}, speculating about the incompleteness of quantum mechanics. See \textcite{Keyl:2003} for comments on whether and how this limit can be understood in a more rigorous mathematical sense. Because of this connection, the quantity on the left-hand side of Eq. \eqref{eq:EntanglementCondition} is sometimes termed {\it EPR-Variance} and denoted by $\Delta_{EPR}$. Its importance is supported by the fact that for symmetric Gaussian states the quantity $r=-\frac{1}{2}\ln\frac{\Delta_{EPR}}{2}$ provides an entanglement measure and uniquely determines the entanglement of formation of the state  \cite{Wootters:1998,Giedke:2003,Wolf:2004} via $E_{\rm oF}=\cosh^2(r)\log_2(\cosh^2r)-\sinh^2(r)\log_2(\sinh^2r)$. As follows from the discussion of losses and decoherences, we have to assume a lower limit on the EPR variance $\Delta_{EPR}\gtrsim 2d^{-1/2}$ and thus an upper bound on the bipartite entanglement between the two ensembles of e.g. $E_{\rm oF}\lesssim1.15\,{\rm ebits}~(E_{\rm oF}\lesssim2.77\,{\rm ebits})$ for an optical density of $d=10~(d=100)$.
A useful and comprehensive review of the theory of entanglement in systems of continuous variables was recently given by \citet{Adesso:2007}, for a concise introduction to the basic facts on the same topic see \textcite{Eisert:2003}.

\paragraph*{Protocol with counter-rotating spins} As explained in Sec.~\ref{sec:magnetic field}, for ensembles at room temperature containing a very large number of atoms a constant magnetic field helps to reduce technical noise, because scattered light can be detected at sideband Zeeman frequencies. In the following we will show, that application of an external magnetic field provides in fact an elegant and efficient way to achieve entanglement of two atomic ensembles with a single probe pulse, as was demonstrated in \textcite{Julsgaard:2001}.

In order to show this, we have to resort to the Maxwell-Bloch equations (\ref{eq:faraday_rotframe}). It is straightforward to generalize these equations to the case of two atomic ensembles and to include Larmor precession of the two spins. We assume that the two collective spins precess in opposite directions, which can be achieved by either using oppositely oriented fields or by using parallel fields with atoms prepared in opposite Zeeman substates. Replacing $\omega_L\rightarrow-\omega_L$ for the second ensemble when using Eqs.~(\ref{eq:faraday_rotframe}), the equations of motion for light quadratures are,
\begin{equation}
\begin{split}
\frac{\partial}{\partial z}x_L(z,t)&=
- \frac{g^*(z)\Omega(t)}{\sqrt{2}\Delta}\Big[\cos(\omega_L t)\big(p_{A_1}(z,t)+p_{A_1}(z,t)\big)\\
&\hspace{50pt}+\sin(\omega_L t)\big(x_{A_1}(z,t)-x_{A_2}(z,t)\big)\Big],\\
\frac{\partial}{\partial z}p_L(z,t)&=0.
\end{split}
\end{equation}
Obviously light reads out sums of momenta and differences of positions, which are commuting observables for oppositely rotating spins. It will be instructive to study directly the evolution of these global observables. One finds,
\begin{equation}
\begin{split}
\frac{\partial}{\partial t}\big(x_{A_1}(z,t)+x_{A_2}(z,t)\big)&=
-\frac{g^*(z)\Omega(t)}{\sqrt{2}\Delta}\cos(\omega_L t)p_L(z,t), \\
\frac{\partial}{\partial t}\big(p_{A_1}(z,t)+p_{A_2}(z,t)\big)&=0,\\
\frac{\partial}{\partial t}\big(x_{A_1}(z,t)-x_{A_2}(z,t)\big)&=0,\\
\frac{\partial}{\partial t}\big(p_{A_1}(z,t)-p_{A_2}(z,t)\big)&=
-\frac{g^*(z)\Omega(t)}{\sqrt{2}\Delta}\sin(\omega_L t)p_L(z,t).
\end{split}
\end{equation}

The solutions fall into two groups,
\begin{align}
X_{L_c,{\rm out}}&=\!X_{L_c,{\rm in}}\!+\!\kappa P_{A_+,{\rm in}}, & X_{L_s,{\rm out}}&=\!X_{L_s,{\rm in}}\!-\!\kappa X_{A_-,{\rm in}},\nonumber \\
P_{L_c,{\rm out}}&=\!P_{L_c,{\rm in}}, &  P_{L_s,{\rm out}}&=\!P_{L_s,{\rm in}}, \nonumber \\
X_{A_+,{\rm out}}&=\!X_{A_+,{\rm in}}\!+\!\kappa P_{L_c,{\rm in}}, & X_{A_-,{\rm out}}&=\!X_{A_-,{\rm in}}, \nonumber \\
P_{A_+,{\rm out}}&=\!P_{A_+,{\rm in}}, & P_{A_-,{\rm out}}&=\!P_{A_-,{\rm in}}\!+\!\kappa P_{L_s,{\rm in}}.
\label{eq:two_cell_faraday_out}
\end{align}
which involve non-local atomic variables
\begin{align*}
  X_{A_\pm}&=\frac{1}{\sqrt{2L}}\int dz \big(x_{A_1}(z)\pm x_{A_2}(z)\big),\\
  P_{A_\pm}&=\frac{1}{\sqrt{2L}}\int dz \big(p_{A_1}(z)\pm p_{A_2}(z)\big)
\end{align*}
and cosine and sine modulation modes, $X_{L_c},\,P_{L_c}$ and $X_{L_s},\,P_{L_s}$, which were introduced in \eqref{eq:modulation modes} in section \ref{sec:measurement}. From the discussion of squeezing in a single ensemble it is evident that a measurement of the sine and cosine component of the $X_L$ quadrature will produce a two mode squeezed state with reduced EPR variance \eqref{eq:EntanglementCondition}.

\begin{figure*}[ht]
  \includegraphics[width=0.8\textwidth]{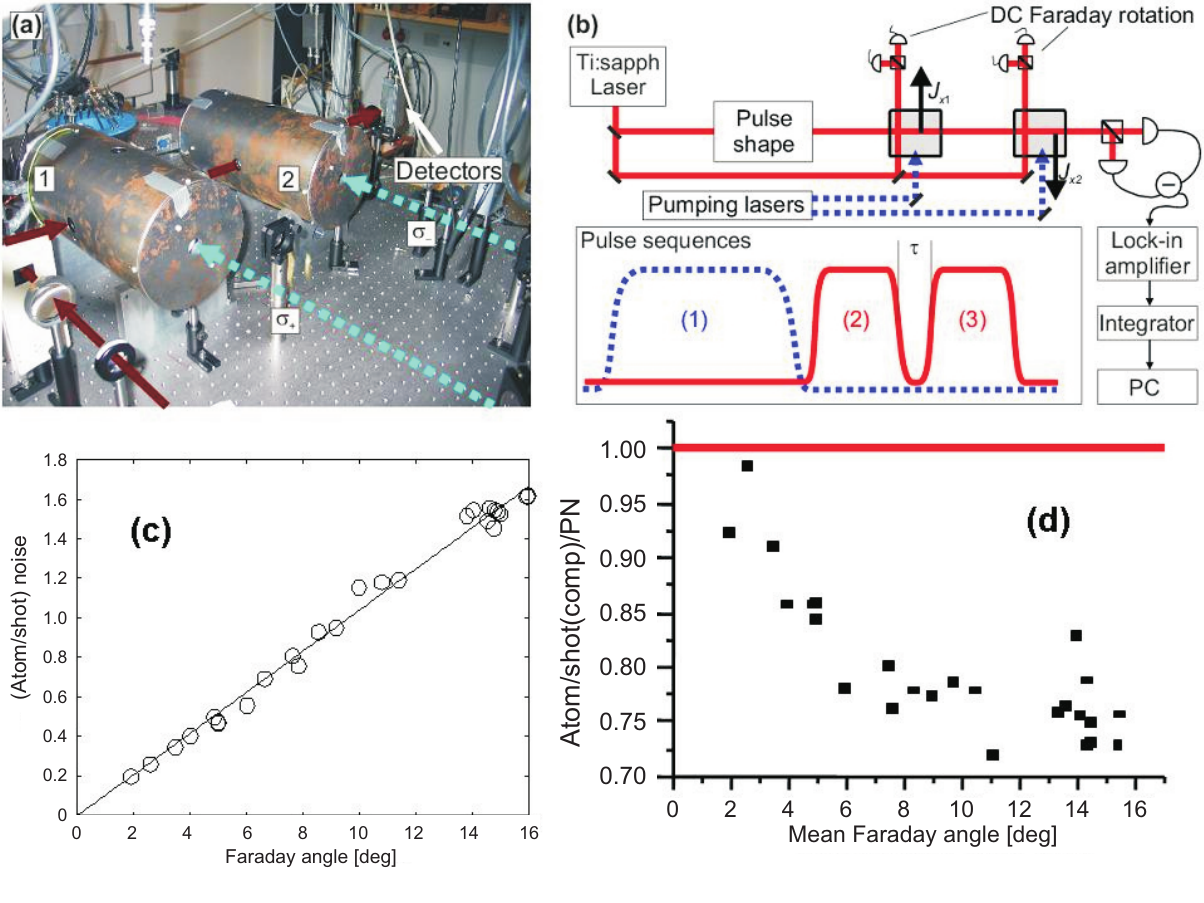}
  \caption{Deterministic entanglement of two atomic ensembles via QND measurement. a) Experimental setup. Dashed lines - optical pumping, solid arrow - entangling light direction, b) Pulse sequence and the layout of the experiment. (1)-optical pumping pulse, (2)-entangling pulse, (3)-verifying pulse, c) Projection noise of atoms d) EPR variance of the entangled state normalized to the projection noise level, i.e., to the variance for a separable coherent state of two ensembles.}\label{fig:entangled_ensembles}
\end{figure*}

\paragraph*{Implementation} Experimental demonstration of deterministic entanglement of two atomic ensembles has been first reported by \textcite{Julsgaard:2001}, with further developments reported by \textcite{Polzik:2003,Sherson:2006c}. The experiments have been performed with Cesium vapor at temperatures in the range 15$^\circ$ - 50$^\circ$ C. Atoms are contained in glass cells coated from inside with a transparent layer of paraffin \cite{Alexandrov:2003}, as discussed in Sec. \ref{sec:room_temp}. The temperature stabilized cells are placed inside cylindrical magnetic shields, see Fig. \ref{fig:entangled_ensembles}. Windows made in the shields allow for optical axis in two directions - one along the axis of the shield used for optical pumping and another in the radial direction used for the probe light. A solenoid produces a homogeneous axial magnetic field inside each shield. Typical cells have the near-cubic shape with the size of $25-30~{\rm mm}$. The magnetic field inhomogeneity is of the order of $10^{-3}$. This rather modest homogeneity is sufficient since the duration of the light-atoms interaction of $1~{\rm msec}$ is sufficiently long so that the atomic motion leads to the effective time averaging of the spatially dependent Zeeman shifts. As a result the magnetic field inhomogeneity has only a quadratic effect on decoherence.

The experimental sequence begins with a few msec pulse of optical pumping along the direction of the axial magnetic field, see Fig.~\ref{fig:entangled_ensembles}. The level scheme and frequencies of light pulses are shown in Figs.~\ref{fig:tensor} and \ref{fig:bfield}. The two cells are pumped with the same lasers with opposite circular polarization of optical pumping in the first and second cell. 99 \% or more of the atoms in $F=4$ state are pumped into the $m_{F}=4$ magnetic substate in one cell and into $m_{F}=-4$ in the other cell, as verified by the magneto-optical resonance method \cite{Julsgaard:2004a,Sherson:2006c}. The total angular momentum $J_{x}$ of $F=4$ state is calibrated by measuring the Faraday rotation angle of a weak linearly polarized light pulse propagating in the direction of the optical pumping. After optical pumping a probe pulse linearly polarized in $x-y$ plane is fired and its polarization rotation, i.e., the value of the operator $S_{y}$, is measured by two detectors via a balanced polarization measurement, see Fig. \ref{Fig:pol_homodyning}. The detected photocurrent pulse is sent into the lock-in amplifier which detects the $\cos(\omega_L t)$ and $\sin(\omega_L t)$ components, $X_{L_c}$ and $X_{L_s}$ introduced in Sec. \ref{sec:measurement}.

The critical condition for the implementation of the deterministic interface based on homodyne mesaurements is quantum noise limited measurement of light and atoms. This is made possible by employing light and atomic detection at high frequency, typically around $\omega_L=320~{\rm kHz}$. Homodyne detectors utilize silicon photodiodes with quantum efficiency more than $98-99$ \% and low noise photo-amplifiers with the response peaked around $\omega_L$. The photodetectors have dark noise equal to shot noise of light for the light power as low as $100~\mu{\rm W}$. This means that with a few mW probe pulse, the detection can be almost perfectly shot noise limited. Light losses have been dominated by reflection off the inner surfaces of the cell windows (the outside surfaces are antireflection coated) amounting to $15\%$ and propagation losses from cells to detectors of about $8\%$.


The duration of the interaction is chosen to fulfill the condition of optimal entanglement which, analogous to the case of spin squeezing, is $\kappa^{2}_{opt}=\sqrt{d}$ obtained with $\eta_{A}=1/\sqrt{d}$ (cf. Eq.~\eqref{eq:eta_kappa}). In the experiment the detuning has been chosen within the range $800-1000~{\rm MHz}$ to be larger than the Doppler width and the hyperfine splitting of the excited state. Together with the optimal value of the optical power set by the detectors around a few mW, the above conditions lead to the minimal pulse duration of the order of a msec and the corresponding bandwidth in the kHz range. This value can be in principle increased by reducing the transverse size of the sample and/or using different detectors.


The first experimental run in the presence of atoms aims at the establishment of the projection noise level of the atomic ensembles. As seen from Eq. \eqref{eq:two_cell_faraday_out} the sum variables $X_{A_{-}}$, $P_{A_{+}}$ can be measured in a QND way using a single probe pulse. In the experiment the sequence of the optical pumping followed by the QND measurement is repeated several thousand times and the variance of the measured photocurrent pulses is calculated. The variance is then plotted as a function of the macroscopic spin of the sample, see Fig.~\ref{fig:entangled_ensembles}(c). The linear dependence along with the almost perfect spin polarization proves that the spin noise is at the projection noise level. The projection noise level has been also independently calculated from the macroscopic collective angular momentum of the sample measured via Faraday rotation of the auxiliary probe pulse \cite{Sherson:2006c}. The calculated value agreed with the measured projection noise to within $10$ \% which is well within the uncertainty of this calculation. The projection noise level defines the right hand side in Eq. \eqref{eq:EntanglementCondition}. When normalized to the shot noise of the probe the projection noise value is equal to the total $\kappa^{2}$ of the two samples according to \eqref{eq:two_cell_faraday_out}.

After the projection noise level is established the experiment proceeds with generation and verification of the entangled state. In the original paper \cite{Julsgaard:2001} no feedback was applied to atoms and hence a conditional entangled state was demonstrated, that is a non-local state with reduced
variance but with a non-deterministic mean value. For possible applications, e.g., teleportation this entanglement is as good as the unconditional one because the knowledge gained with the measurement on the first entangling pulse
 can be applied to achieve teleportation. The creation of a deterministically and unconditional entangled state has been achieved subsequently \cite{Polzik:2003}. The experimental sequence which realizes such entanglement \cite{Sherson:2006c} involves a feedback applied to the atoms in between the two probes (Fig.  \ref{fig:entangled_ensembles}b). The feedback pulse of $320~{\rm kHz}$ magnetic field with the cosine and sine components proportional to  $X_{L_c,{\rm out}}$ and $X_{L_c,{\rm out}}$ respectively is applied to the rf magnetic coils surrounding the cells. An appropriate electronic gain must be chosen so that the feedback pulse rotates the atomic collective spins such as to generate the minimal EPR variance that is the minimal variance of the angle between the spins. The choice of the gain $g=-\kappa/(1+\kappa^2)$ minimizes the EPR variance in the absence of decoherence. In the experiment the optimal gain has been chosen operationally by minimizing the EPR variance (see \textcite{Sherson:2006c} for details). The results for the EPR entangled state of two atomic ensembles are shown in Fig. \ref{fig:entangled_ensembles}. The variance of the entangled state obtained after applying the feedback pulse is measured by the verifying pulse. Fig. \ref{fig:entangled_ensembles} shows this variance normalized to the projection noise variance as a function of the number of atoms. A higher number of atoms leads to a higher value of $\kappa$ and hence to a higher degree of entanglement. The mimimal EPR variance observed in these experiments was $\Delta_{EPR}=1.3$. This variance corresponds to an entanglement of formation of $E_{\rm oF}=0.28~{\rm ebits}$.

\subsection{Probabilistic entanglement}
\label{sec:probabilistic}

One of the main motivations for studies of atom light quantum interfaces is an application for quantum repeaters, which would enable long distance quantum communication \cite{Briegel:1998}. A protocol (also known as the DLCZ-protocol) for such a repeater based on atomic ensembles and linear optics was first  presented by \textcite{Duan:2001}. Several improvements of the protocol have been suggested \cite{Zhao:2007,Simon:2007a,Chen:2007,Jiang:2007,Sangouard:2007a,Sangouard:2008}. We leave the detailed discussion of the DLCZ-protocol and quantum repeaters for a dedicated review. In this section we provide a basic discussion of the probabilistic entanglement generation.


\begin{figure}[htbp]
\begin{center}
\includegraphics[width=8.5cm]{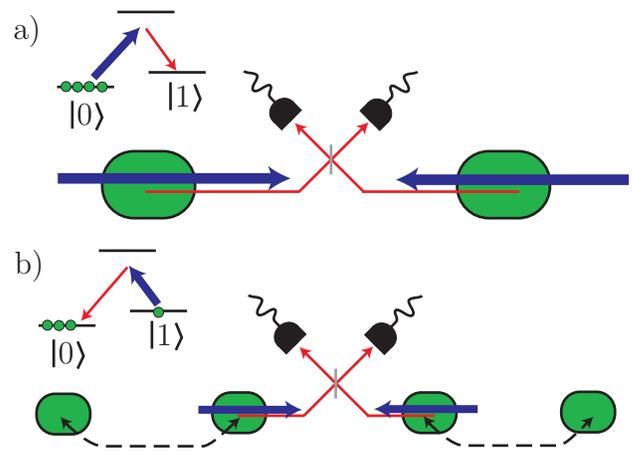}
\caption{DLCZ-protocol. a) Entanglement between light and atoms is generated by a parametric gain interaction in two distant ensembles. Detection of a photon by one of the two photodetectors after the optical beam splitter probabilistically generates entanglement between the two ensembles. b) The atomic excitation in one half of an entangled pair of atomic ensembles (dashed line) is read out onto light with the light-atoms beam splitter interaction and mixed on an optical beam splitter with its counterpart from another entangled pair of ensembles. Photo detection swaps the entanglement so that now the outmost ensembles become entangled.}
\label{fig:DLCZ}
\end{center}
\end{figure}

The entanglement generation in the DLCZ-protocol uses the parametric gain interaction Eq. (\ref{eq:gain_rescale}) between the light and the atoms as shown in Fig. \ref{fig:DLCZ}a. In principle this interaction could be used in the strong coupling regime ($\kappa\sim 1$ in the notation below) to generate continuous variable entanglement along the lines of the entanglement generation protocol used for quantum teleportation in Sec. \ref{sec:teleportation}. Instead the DLCZ-protocol works in the weak coupling limit ($\kappa\ll 1$) and generates probabilistic entanglement. The first term in  Eq. (\ref{eq:gain_rescale}) for $a_A$ is just the AC-stark shift of the ground state,  which can be removed by a simple rescaling, and the phase $\phi$ vanishes for a large detuning.  In the limit of weak coupling the dynamics only involves collective operators analogous to the ones defined in Eq. (\ref{eq:sym_modes}) and is equivalent to the evolution with the ideal two mode parametric gain evolution operator $\exp(i \kappa (a_L^\dagger a_A^\dagger+{\rm H.C.})/2)$. For $\kappa\ll 1 $ the joint state of the collective atomic and light harmonic oscillator degrees of freedom is then
\begin{equation}
|00\ra_{AL}+\frac{\kappa}{2} |11\ra_{AL}+ O(\kappa^2) ,
\label{eq:DLCZ-write}
\end{equation}
where $|mn\ra_{AL}$ describes the state with $m$ ($n$) excitations in the atomic (light) harmonic oscillator.
 This result can be understood rather intuitively from the level scheme in Fig. \ref{fig:DLCZ}a, which shows that the interaction generates simultaneous excitations of the atoms and the light as described by the expression above.

The state in Eq. (\ref{eq:DLCZ-write}) is in itself an entangled state, but in the DLCZ protocol it is used to probabilistically generate an entangled stated of two atomic ensembles.  The outgoing light modes from two different ensembles are combined on a beamsplitter as shown in Fig. \ref{fig:DLCZ}. Conditioned on a click in one of the two photodetectors the ensembles are prepared in the state
$(|01\ra\pm |10\ra)/\sqrt{2} $, because one cannot know from which ensemble the photon was emitted (the sign in the superposition depends on which detector detected the photon).  If two pairs of ensembles are independently  entangled in this fashion, as shown in Fig. \ref{fig:DLCZ}b, the entanglement can be extended to a larger distance. Towards this end the atomic states from the two closest atomic ensembles, each belonging to a different entangled pair, must be read out onto light modes which are mixed on another beam splitter. Detection of a photon after the beamsplitter extends the entanglement to twice the distance by entanglement swapping. This read out process can for instance be done using the beam splitter interaction as discussed in \ref{sec:raman_eit}. Note, however, that the atomic mode functions (\ref{eq:aatom_transverse_lambda}) and (\ref{eq:aatom_transverse_gain}) suitable for the parametric gain and beam splitter interactions are different leading to a mode mismatch if the excitations are read out in the same direction as the entanglement generation. This problem can be avoided by reading out in the backward direction \cite{Andre:2005} (this analysis assumes that the atoms retain their positions throughout the experiment; see also \textcite{Duan:2002} for a discussion of the limit, where the motion of the atoms leads to an averaging over the atomic position).


Following the initial DLCZ-protocol several experiments have been performed, which demonstrate a number of important ingredients of the repeater, of which here we only mention a few.  The first experiments \cite{vanderWal:2003,Kuzmich:2003,Matsukevich:2004,Chou:2004} demonstrated quantum correlations in pairs of photons generated via the creation of pairs of atomic excitations and photons in Raman scattering, followed by a beam-splitter interaction converting the atomic excitations into another photon after a programmable delay. \textcite{Chaneliere:2005} and \textcite{Eisaman:2005} reported the storage and controlled release of single photons, which are deterministically created from a first atomic ensemble, in a second one via electromagnetically induced transparency. \textcite{Chou:2005,Felinto:2006,Chaneliere:2007,Yuan:2007} demonstrated the quantum interference of photons emitted from two independent atomic ensembles in Raman scattering, with \cite{Chou:2005} proving entanglement of the two atomic ensembles. \textcite{Chou:2007} implemented an elementary link for a DLCZ-type quantum repeater as depicted in Fig.~\ref{fig:DLCZ}. Other steps towards the DLCZ quantum repeater were performed by \textcite{Chen:2008}, where the teleportation of photonic to atomic qubits held in an atomic ensemble has been demonstrated, and by \textcite{Choi:2008} where storage and release of photonic entanglement from two atomic ensembles has been reported.

%
%
\section{Quantum Memory for Light}\label{sec:memory}

Atomic quantum memory for light is an important ingredient for a number of quantum information routines. It is implicit in many quantum communication protocols, in particular in those which require local operations on more than just a single optical pulse, so that storage is necessary. It is required for linear optics quantum computing and for scalable cluster state quantum computing with photons. Quantum memory is a necessary ingredient of a quantum repeater. Different applications demand quantum memories fulfilling different requirements. Some applications, such as those which include local operations and classical communication (LOCC), benefit from a high-fidelity deterministic write-in operation into the memory. By deterministic we mean a protocol which works with probability one, so that the fidelity is calculated for every try. Others can tolerate lower probability of write-in and read-out, provided the success is heralded. Nonetheless, the latter protocols, such as a quantum repeater, would also benefit from deterministic write-in which would lead to a higher efficiency, higher rate, and eventually to longer distances.

In this chapter we review several main approaches to the quantum memory for light. We first discuss a figure of merit and a classical benchmark for determining the quality of a quantum memory. We present the protocol based on a QND interaction and feedback, which demonstrates a quantum memory channel with the fidelity higher than the classical benchmark. We then discuss the memory based on the $\Lambda$ scheme, concentrating on the recent achievement of the EIT-based memory experiments. We conclude with a discussion of memories based on various types of the photon echo.

\subsection{Figure of Merit}
\label{sec:figure_of_merit}

From a fundamental perspective a quantum memory can be analyzed as a quantum channel, acting in time. A perfect quantum memory is nothing but the identity map taking arbitrary states as input and returning them unchanged, some time later. A realistic memory will be imperfect and the question arises what figure of merit to use in order to evaluate its performance. One sensible measure characterizing the performance of a memory is the  fidelity, i.e. the average state overlap \cite{Nielsen:2000}, which can be achieved between an input state drawn from a predefined set of states according to a predefined probability distribution and the state which is finally read out of the memory. The fidelity is of fundamental relevance if it exceeds the best classical fidelity, in which case the channel is thus outperforming the best classical channel. The classical fidelity relies on the simple strategy of measuring the given quantum state, storing the resulting classical data and on demand reconstructing the quantum state as good as possible.

For example, in a special case where both input and output states are Gaussian states with amplitudes $\mean{X_{\rm in(out)}},~\mean{P_{\rm in(out)}}$ and variances $\Delta X^2_{\rm in(out)},~\Delta P^2_{\rm in(out)}$ the fidelity is given by
\begin{multline}
F=\left[\left(\Delta X^2_{\rm in}+\Delta X^2_{\rm out}\right)\left(\Delta P^2_{\rm in}+\Delta P^2_{\rm out}\right)\right]^{-1/2}\\
\times\exp\left[-\frac{(\mean{X_{\rm in}}-\mean{X_{\rm out}})^2}{2(\Delta X^2_{\rm in}+\Delta X^2_{\rm out})}-
\frac{(\mean{P_{\rm in}}-\mean{P_{\rm out}})^2}{2(\Delta P^2_{\rm in}+\Delta P^2_{\rm out})}\right].
\end{multline}
In experiments the average of this fidelity can be taken with respect to a Gaussian distribution of coherent states centered at vacuum with a mean occupation number, $\bar{n}$. If the average fidelity for a flat (infinitely broad) distribution of coherent input states equals unity, the memory is ideal and would store also any Non-Gaussian state perfectly. This follows from the fact that coherent states provide an (overcomplete) basis for the Hilbert space.

It is worth emphasizing that the performance of a quantum memory can be, in principle, tested with coherent states only. Knowing the performance of the memory (a quantum channel) for all coherent states, that is performing a quantum tomography of the memory process with coherent states, one can predict the fidelity of the memory channel for an arbitrary class belonging to a single mode. In this sense the often used division between continuous variable memory and discrete variable memory is not quite justifiable. It is more appropriate to speak about the protocols which are based on continuous variable \emph{measurements} (homodyning) and discrete variable \emph{measurements} (photon counting).

The question then remains what exactly a measured average fidelity smaller than unity guarantees. The benchmark maximum fidelity of a classical channel $F_{\rm class}$ is known for a limited number of quantum states including qubit states and coherent states with a Gaussian distribution in phase space of width $\bar{n}$. For the latter case \textcite{Braunstein:2000} conjectured and \textcite{Hammerer:2005a} proved that
\begin{equation}
  \label{eq:ClassFidel}
  F_{\rm class}=\frac{1+\bar{n}}{1+2\bar{n}}\to \frac{1}{2}~,\qquad \bar{n}\to \infty
\end{equation}
In the former case, for a class of arbitrary qubit states the maximum classical fidelity is $F_{\rm class}=2/3$ \cite{Massar:1995}. Very recently a classical benchmark fidelity for a third class of states, namely the displaced squeezed states, has been found \cite{Owari:2008,Adesso:2008}. For a class of pure squeezed states with the variance $s$ in vacuum units, arbitrary orientation and arbitrary displacements the classical benchmark approaches zero for large $s$ as $F_{\rm class}=\frac{\sqrt{s}}{1+s}$. Quantum memory which exceeds these classical benchmark fidelities thus shows  performance which is classically impossible. For similar fidelity benchmarks for finite dimensional systems see the work by \textcite{Keyl:1999}.

The fidelity is not necessarily the one and only figure of merit for a quantum memory protocol. A relatively high fidelity may still be compatible with errors which are hard to correct. On the other hand, a lower fidelity protocol with particular kinds of errors may be more appropriate for a specific application. For instance the analysis in \textcite{Brask:2008} shows that different types of errors can have a very different effect on the  repeater protocol of \textcite{Duan:2001}, and \textcite{Surmacz:2006} argue that in certain applications it might be more important to preserve entanglement, when one partner of an entangled pair is stored, than to conserve the quantum state itself.

For an important class of protocols discussed in Secs. \ref{sec:raman_eit} and \ref{sec:echo}, the performance is theoretically described by a simple beam splitter relation between the input and output operators $\hat{a}_{{\rm out}}=\sqrt{\eta}\hat{a}_{{\rm in}}+\sqrt{1-\eta}\hat v$, where $\eta$ is the efficiency, e.g, for mapping the input light intensity to the output light intensity, and $\hat{v}$ is a vacuum operator. The memory performance  is then completely characterized by the single parameter $\eta$, and quantities such as the fidelity for a given distribution of states may later be derived from it. The performance is therefore often discussed in terms  of the single parameter $\eta$ and we will use this characterization in Secs. \ref{sec:raman_eit} and \ref{sec:echo}. In an assessment of a given experiment one should, however, verify that the simple beam splitter relation is indeed  applicable for this experiment.

It is possible to define other figures of merit  and also to consider different benchmarks than the one given by a classical measure and prepare strategy, in order to quantify the quality of storage -- in quantum memories -- but also of transmission of quantum states -- as in quantum teleportation, cf. Sec.~\ref{sec:teleportation}. For experiments with single photons, it is common to consider the conditional fidelity, which characterizes the fidelity conditioned on the detection of a photon after the interface. Since this conditioning suppresses  the effect of losses, the conditional fidelity is often much higher than the unconditional fidelity discussed above. When a conditional fidelity is used, another parameter, often called efficiency is introduced which describes the probability of success of the protocol. In the context of teleportation of coherent states, the question of the ''right'' figure of merit was subject to considerable debate in the literature, see \citet{Ralph:1998,Braunstein:2001,Grosshans:2001,Bowen:2003} and references therein. In particular, \citet{Grosshans:2001} emphasize the importance of a benchmark $F_{1\rightarrow2}$ given by the maximal fidelity achievable in a 1 to 2 cloning machine. For $F>F_{1\rightarrow2}$ the memory output is guaranteed to be the best possible copy of the input state. For coherent input  $F_{1\rightarrow2}\simeq0.68$, as shown by \cite{Cerf:2005}, and is thus more demanding than the classical fidelity benchmark $F_{\rm class}$, which can also be interpreted as the maximal fidelity of a 1 to $\infty$ cloning machine \cite{Hammerer:2005a}. Figures of merit different from fidelity were used in \citet{Ralph:1998,Bowen:2003a,Bowen:2003,Hetet:2008a} to characterize both quantum storage and teleportation. There is, however, a consensus that 
the most sensible figure of merit ultimately depends on the specific application of the quantum memory or  teleportation link within a quantum network for e.g. quantum cryptography or optical quantum computation.

\subsection{QND \& Feedback Protocol}
The first demonstration of a quantum memory \cite{Julsgaard:2004} beating a classical benchmark \cite{Hammerer:2005a} was based on the QND-Faraday interaction of a pulse of light, carrying the quantum state to be stored, with two collective spins counter-rotating in an external magnetic field. The basic input-output equations describing the interaction are given by \eqref{eq:two_cell_faraday_out}. Each of these constitutes a realization of the simpler input-output relations \eqref{eq:faraday_nodecay_result} for a pulse interacting with a single atomic ensemble without the magnetic field. For simplicity, we will base the theoretical discussion of the main idea on this single ensemble setup, and will return to the actual implementation based on the setup involving two atomic ensembles in the experimental part.

 The input light is described by canonical operators $X_{L,{\rm in}}$ and $P_{L,{\rm in}}$ while the collective spin is prepared in the fully polarized state $\langle X_{A,{\rm in}}\rangle=\langle P_{A,{\rm in}}\rangle=0$ and $\Delta X_{A,{\rm in}}^2=\Delta P_{A,{\rm in}}^2=1/2$. With a choice of $\kappa=1$ in \eqref{eq:faraday_nodecay_result} the entangled state of atoms and light after the interaction is described by
\begin{align*}
X_{L,{\rm out}}&=X_{L,{\rm in}}+P_{A,{\rm in}}, &
P_{L,{\rm out}}&=P_{L,{\rm in}},\\
X_{A,{\rm out}}&=X_{A,{\rm in}}+P_{L,{\rm in}}, &
P_{A,{\rm out}}&=P_{A,{\rm in}}.
\end{align*}
Following the interaction the light quadrature $X_{L,{\rm out}}$ is measured and the corresponding measurement result $\xi$ is fed back displacing the atomic state such that $P_{A,{\rm out}}\rightarrow P_{A,{\rm out}}-\xi$. As shown in section \ref{sec:measurement}, the final transformation of the collective atomic spin in the ensemble average is given by
\begin{equation}\label{eq:efeed_memory}
\begin{split}
X_{A,{\rm out}}&=X_{A,{\rm in}}+P_{L,{\rm in}}, \\
P_{A,{\rm out}}&=P_{A,{\rm in}}-X_{L,{\rm out}}=-X_{L,{\rm in}}.
\end{split}
\end{equation}
This concludes the mapping of the quantum state of light onto atoms: the mean values are transmitted faithfully (apart from an unimportant phase change) as $\langle X_{A,{\rm out}}\rangle=\langle P_{L,{\rm in}}\rangle$ and $\langle P_{A,{\rm out}}\rangle=-\langle X_{L,{\rm in}}\rangle$. The operator $X_{L,{\rm in}}$ is mapped perfectly onto the atomic collective spin, while the operator $P_{L,{\rm in}}$ is mapped with the addition of one unit of vacuum operator which comes from the initial coherent state of the atomic ensemble. This latter imperfection can be remedied if the initial atomic spin state is squeezed before the memory operation, such that $\Delta X_{A,{\rm in}}^2\rightarrow 0$. If such squeezing operation is performed by, for example, an additional QND measurement, the fidelity of the quantum memory operation can, in principle, approach $100\%$.

In the experiment the quantum memory performance has been tested with a set of coherent states of light taken from a Gaussian distribution of coherent states centered at vacuum with a mean photon number, $\bar{n}$. Given the measured gains and the measured variances, $\Delta X_{A,{\rm out}}$ and $\Delta P_{A,{\rm out}}$ of the state of the memory, the fidelity can be calculated as:
\begin{multline*}
  F = (\bar{n}(1-\kappa)^2 + 1/2 + \Delta X_{A,{\rm out}}^2)^{-1/2}\\
  \times(\bar{n}(1-g)^2+1/2+\Delta P_{A,{\rm out}}^2)^{-1/2}
\end{multline*}
As discussed above, if the protocol starts with the atomic ensemble in a coherent state $\Delta X_{A,{\rm out}}^2=1$, $\Delta P_{A,{\rm out}}^2=1/2$, and hence $F=\sqrt{2/3}\approx82\%$ with the choice of $\kappa=g=1$ which is optimal for the class of arbitrary coherent states. For an unknown qubit state $(\alpha |0\rangle+\beta |1\rangle/\sqrt{2}$
the same protocol yields the fidelity of $80\%$ for optimal values of $g$ and $\kappa$.

\begin{figure*}[ht]
  \includegraphics[width=0.6\textwidth]{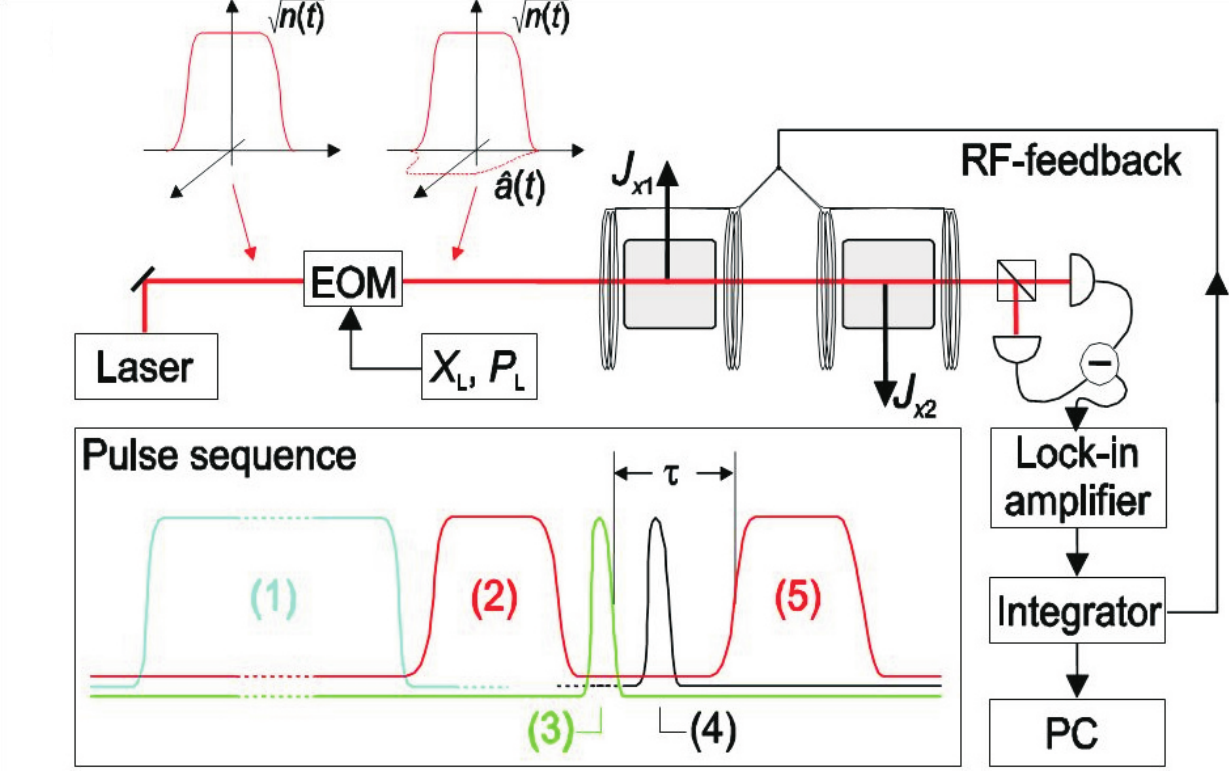}
  \caption{QND+feedback memory experimental setup and pulse sequence \cite{Julsgaard:2004}. The state of light is encoded by the EOM in the sidebands of the strong pulse $\sqrt{n(t)}$. Two cells serve as the quantum memory unit. The feedback pulse proportional to the photodetector signal is applied to the RF magnetic coils. Inset. (1) - optical pumping, (2) - input pulse, (3) - feedback RF pulse, (4) - RF pulse rotating atomic P into X used for half of verification pulses, (5) - verifying pulse}
  \label{fig:memory_sequence}
\end{figure*}

 The experimental setup and the sequence of operation for writing into the quantum memory are similar to the sequence described in the chapter on deterministic entanglement. The quantum light mode which corresponds to the $\omega_L$ sidebands in the polarization orthogonal to the strong field now carries the quantum state of light to be mapped. The state is generated by an electro-optical modulator (EOM) as shown in Fig. \ref{fig:memory_sequence}. The quantum sidebands together with the strong pulse propagate through the atomic cells and are analyzed with the polarization homodyning technique (Fig. \ref{Fig:pol_homodyning}). The strong pulse thus serves a dual purpose, first as a strong driving pulse for the interaction with atoms, and second as a local oscillator for the homodyne measurement.

 In the experiment two cells in a magnetic field play the role of one quantum memory unit. As discussed above this approach allows to achieve quantum limited noise for ensembles of trillions of atoms because quantum information is encoded and processed at $\omega_L=320$ kHz sidebands where classical noise can be strongly suppressed. For two cells with magnetic field according to Eq. \eqref{eq:two_cell_faraday_out} the light variable $X_{L_c}=\sqrt{\frac{2}{T}}\int dt\cos(\omega_L t)x_L(t)$ should be measured and the result fed back to the atomic variable $P_{A_+,{\rm out}}$. The method for measuring $X_{L_c}$ by the homodyne measurement of the Stokes parameter $S_{z}$ of the light and the subsequent processing of the photocurrent by the lock-in amplifier indicated in Fig. \ref{fig:memory_sequence} is discussed in section \ref{sec:measurement}, Eq. \eqref{eq:modulation modes}. Note that at the same time the variable $X_{L_s}=\sqrt{\frac{2}{T}}\int dt\sin(\omega_L t)x_L(t)$ can be measured and fed back into the $X_{A_-,{\rm out}}$ variable of atoms. The memory could hence be used as a two mode memory, although this direction has not been pursued.

\begin{figure}[ht]
  \includegraphics[width=8.3cm]{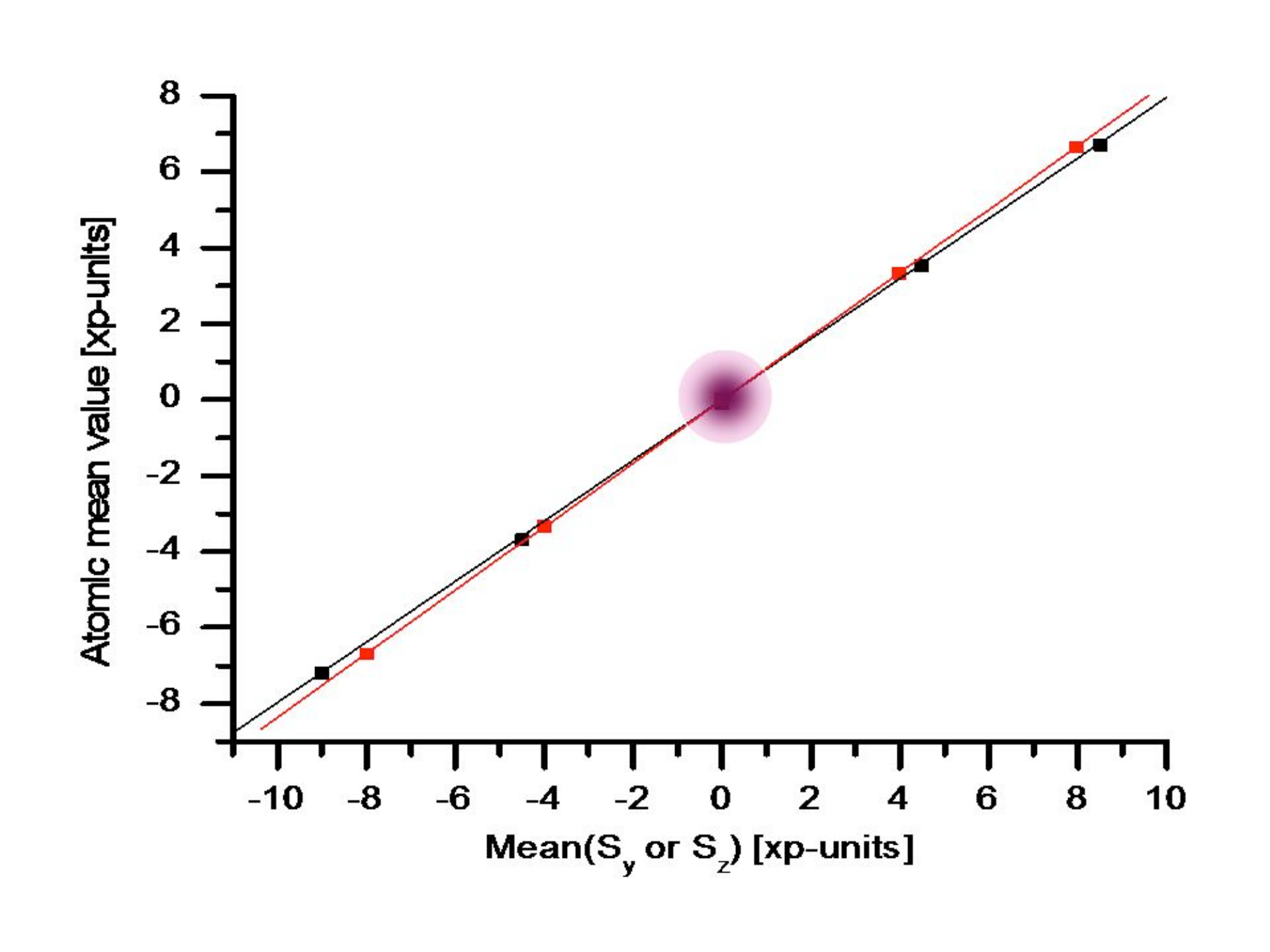}
  \caption{Atomic coherent memory results  \cite{Julsgaard:2004}. The mean values for both quadratures of the input light, atomic memory and the output light are identical to within a chosen factor (here $0.8$).}\label{fig:memory_mean}
\end{figure}

 After the projection noise level of atoms is established, as described in the section on entanglement, the optimal feedback gain must be determined. The gain is chosen to optimize the fidelity for the class of states to be stored in the memory. An example corresponding to the class of coherent states distributed around vacuum with $\bar{n}=8$ is shown in  Fig.~\ref{fig:memory_mean}. The optimal values for this class of states are $\kappa=0.8,g=0.8$. After the memory sequence, cf. Fig.~\ref{fig:memory_sequence}, is over the atomic variable  $P_{A_+,{\rm out}}$ is measured with a strong QND verifying pulse. The quantum sideband modes of this pulse are initially in the vacuum state.  After propagating through the memory the quantum sideband modes contain the atomic memory variable $P_{A_+,{\rm out}}$ according to \eqref{eq:two_cell_faraday_out}. Fig.~\ref{fig:memory_mean} shows that the mean values of this variable for various light input states are very satisfactory proportional to the mean values of the input light canonical variables. As also shown in the figure the same is true for the other canonical variable of light $P_{L_c}$ stored in the memory variable $X_{A_+,{\rm out}}$. $X_{A_+,{\rm out}}$ has been measured in another experimental sequence, where an RF pulse rotating the collective atomic spins by $\pi/2$ and thus converting $P_{A_+,{\rm out}}$ into $X_{A_+,{\rm out}}$ has been applied just before the verifying pulse (Fig.~\ref{fig:memory_sequence}). The memory is thus shown to work very well as a classical memory for light since the mean amplitude and phase of the input light pulse and the retrieved light pulse are equal to within a chosen factor.
 Note that as a classical coherent memory this Faraday+feedback memory can have unity retrieval efficiency because the gain is adjustable, whereas for the Raman, EIT, and photon echo-based memories discussed below the efficiency is less than unity in the presence of losses.

\begin{figure}[ht]
  \includegraphics[width=8.3cm]{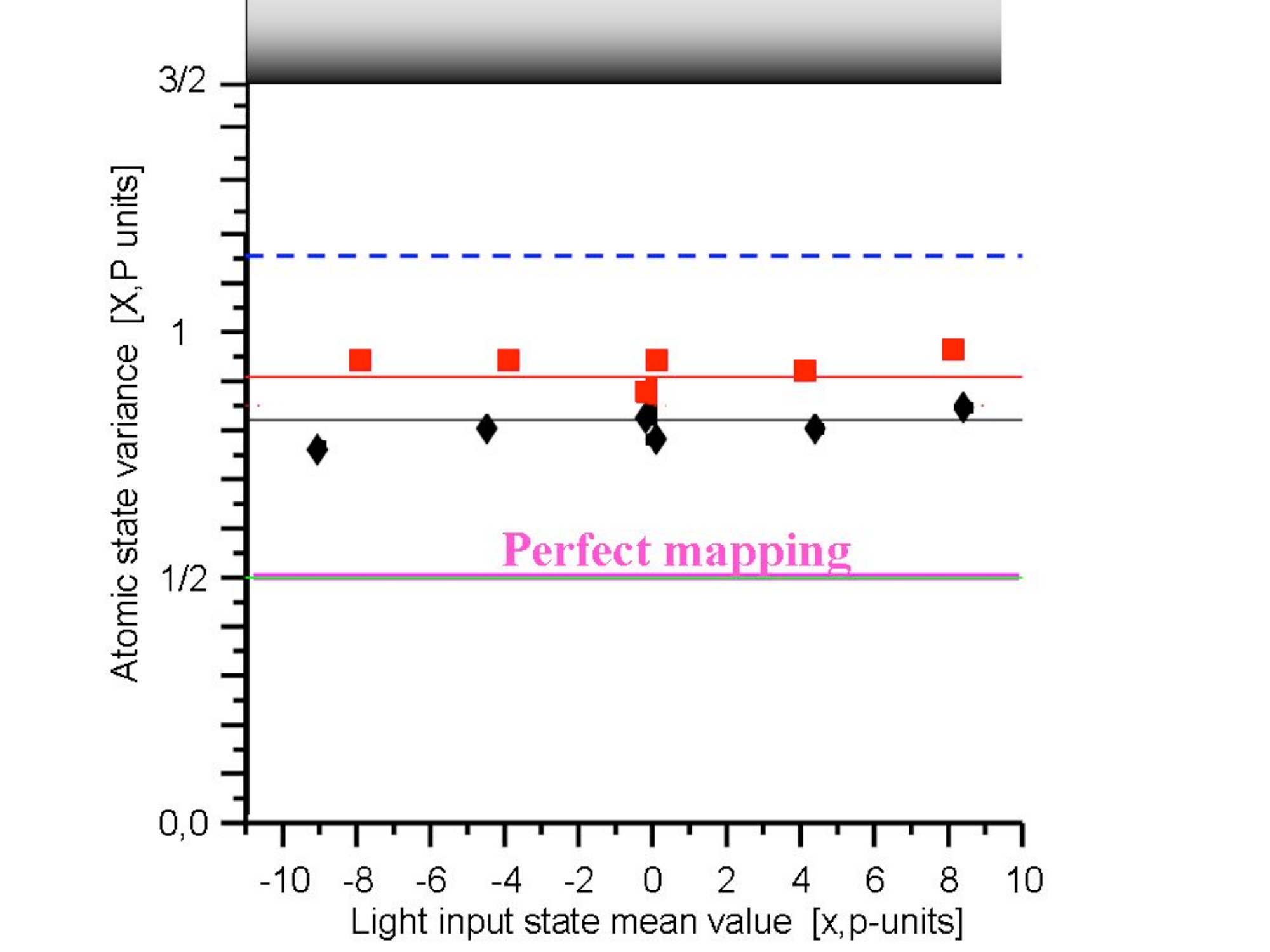}
  \caption{Variance of the atomic memory state for input coherent states with $n\leq8$. The variance of $1/2$ corresponds to the memory variance equal to the input light variance. Above the $3/2$ level is the classical memory performance for arbitrary coherent states. The dashed line is the best classical performance for the states within the $n\leq8$ class. Diamonds and squares are experimental results from \textcite{Julsgaard:2004}.}\label{fig:memory_variances}
\end{figure}

 The decisive demonstration of the quantum character of the memory follows from the analysis of the variances of the stored quantum state. Fig.~\ref{fig:memory_variances} shows the experimental variances of the state of the atomic memory for input light states with the mean photon number between zero and eight. From these values the experimental fidelity of $64\%$ has been calculated, which is higher than the benchmark classical fidelity $52\% $ for this class of states.

\subsection{Multipass Approaches}

The Faraday+feedback quantum memory and the protocols for entanglement and spin squeezing in Sec.~\ref{sec:Entanglement} rely on a pulse of light (or two pulses in the case of \textcite{Duan:2000}) interacting with the medium {\it once}. It is of course possible to have a pulse of light interacting several times with one (or more) atomic ensembles, as suggested in several theoretical studies. This section is intended to provide an overview over proposals relying on multiple passes of light through atoms.

Common to all these proposals is that they take advantage of the possibility to perform phase shifts on light and to rotate atomic spins in between the passes. A relative phase shift $\phi$ between the classical field and the quantum field, in $x$- and $y$-polarization respectively, will give rise to a rotation of field quadratures,
\begin{equation}\label{eq:local_rots}
  X_L\rightarrow\!\cos\phi X_L\!+\sin\phi P_L,~ P_L\rightarrow\!\cos\phi P_L\!-\sin\phi X_L.
\end{equation}
Rotation of atomic spins about the axis of polarization, via e.g. fast RF  pulses, allows for an analogous rotation of $X_A$ and $P_A$. Alternatively light can be sent through atoms from a different direction. As light is sensitive to the projection of the collective atomic spin along the axis of propagation, this is equivalent to a rotation of the atoms. Especially for room-temperature atoms in a cell, optical access from two orthogonal directions can be afforded trivially, while light impinging from different sides still talks to the same symmetric mode of atoms due to thermal averaging.

\begin{figure}[ht]
  \includegraphics[width=8.3cm]{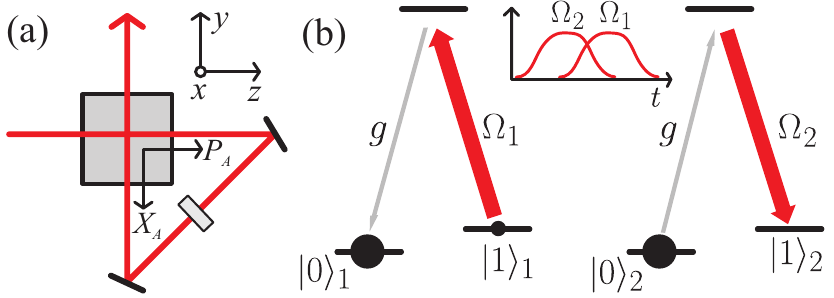}
  \caption{(a) Setup for two pass protocol for read-out of atomic states as suggested by \textcite{Kuzmich:2003a} and, with an additional magnetic field applied along $x$, for full state exchange or creation of  entanglement between light and atoms as suggested by \textcite{Muschik:2006}. (b) State swap between two ensembles coupled to a common cavity mode based on adiabatic passage as demonstrated by \citet{Simon:2007}. The classical pulses $\Omega_i$ are applied in a counterintuitive sequence.}\label{fig:multipass}
\end{figure}

The first proposal along this line is due to \textcite{Kuzmich:2003a} and presents a protocol for atomic state read-out, i.e. mapping of the atomic spin state onto the polarization of  light. As shown in Fig.~\ref{fig:multipass}, in a first pass a pulse of light propagating along $x$ interacts with atoms in a QND fashion, generating a state described by \eqref{eq:faraday_nodecay_result}. A phase shift of $\phi=\pi/2$ then changes $X_{L,{\rm in}}'= P_{L,{\rm out}}$ and $P_{L,{\rm in}}'=-X_{L,{\rm out}}$, where primed variables refer to the second pass of light. The pulse is redirected to the ensemble along the negative $y$ direction, such that the input output relations become, taking $\kappa=1$,
\begin{equation}
\begin{split}
X_{L,{\rm out}}'&\!=\! X_{L,{\rm in}}'\!-\!X_{A,{\rm in}}' = P_{L,{\rm in}}\!-\!(X_{A,{\rm in}}\!+\!P_{L,{\rm in}}) = -X_{A,{\rm in}},\\
P_{L,{\rm out}}'&\!=\!P_{L,{\rm in}}'=-X_{L,{\rm in}}-P_{A,{\rm in}}.
\end{split}
\end{equation}
Aside from an unimportant phase change, the state of atoms is mapped on light. The spurious effect of light noise $X_{L,{\rm in}}$ in the second line can be removed by using squeezed light. The overall input-output relations are similar to the ones for quantum memory for light in Eq. \eqref{eq:efeed_memory}, but require neither measurement nor feedback.

The previous protocol naturally raises the question whether in principle a perfect state transfer, or state swapping, could be achieved in several passes and specific rotations on atoms and light without using squeezed light. \textcite{Kraus:2003} addressed this question in full generality and gave necessary and sufficient conditions for what types of quadratic Hamiltonians can be achieved in two modes, given a specific interaction -- such as e.g. the Faraday interaction $\sim P_LP_A$ -- when combined with 'local' operations of the type \eqref{eq:local_rots}. It was found there, that the Faraday, or QND, interaction is most capable for simulating, in this sense, other interactions, while the beam splitter and parametric gain interaction have no potential to emulate any other interaction. Furthermore, optimal strategies for generation of squeezing and entanglement were devised in the same paper. \textcite{Fiurasek:2003} extended these results asking what types of unitary transformations (instead of Hamiltonian interactions) can be achieved and showed in particular that a perfect state swap can be performed with three passages of light, and not for less, see also the work by \textcite{Takano:2008}. \textcite{Hammerer:2004} applied these general results to the specific situation of a light-matter quantum interface, showing that atomic decay and light losses can be tolerated. See also \textcite{Kurucz:2008} for a discussion of memory conditions and a comparison of these multipass approaches to Raman and EIT based memories.

These protocols all assume a QND interaction in light and matter, which is practical for large cells with room-temperature atoms only when two cells and counter-rotating spins are used, as discussed in Sec.~\ref{sec:deteministic_ent}. This makes multiple passes with different directions of propagation a difficult issue. Another complications is due to the fact that room-temperature ensembles require msec pulses, implying an unreasonable long delay line in the loop of Fig.~\ref{fig:multipass} in order to prevent the pulse meeting itself in the atomic medium. These problems were overcome by  \textcite{Sherson:2006}, \textcite{Fiurasek:2006} and \textcite{Muschik:2006}, who showed that protocols for quantum memory and entanglement generation involving multiple passes of one or more pulses can be matched to Larmor precessing ensembles and that light traversing atoms simultaneously from different directions can in fact be advantageous. In particular, \textcite{Muschik:2006} consider a single cell Larmor precessing in a magnetic field oriented along $x$ in a setup as shown in Fig.~\ref{fig:multipass}, assuming a loop length much smaller than the pulse length, such that the pulse ''meets'' itself in the medium. For atoms rotating in the sense of the light propagating along the loop, solution of the corresponding Maxwell-Bloch equations, taking care of propagation effects, yields input-output relations for atomic variables,
\begin{align*}
  X_{A,{\rm out}}&=e^{-\kappa^2/2}X_{A,{\rm in}}+\sqrt{1-e^{-\kappa^2}}X_{L,{\rm in}}^+,\\
  P_{A,{\rm out}}&=e^{-\kappa^2/2}P_{A,{\rm in}}+\sqrt{1-e^{-\kappa^2}}P_{L,{\rm in}}^+,
\end{align*}
where $X_{L,{\rm in}}^+,~P_{L,{\rm in}}^+$ refer to a light mode centered at the upper side band frequency $\omega_0+\omega_L$ ($\omega_0$ is the carrier frequency of the classical driving pulse and $\omega_L$ the Larmor frequency) with a weakly exponentially decaying mode function. $\kappa$ is again given by \eqref{eq:kappa}. An analogous input-output relation holds for light, such that this scheme realizes an exponentially efficient state exchange of atoms and light.

The Faraday interaction was achieved as a sum of the beam splitter and the parametric gain interaction, c.f., Fig. \ref{fig:lambda}.    What we effectively achieve by having multiple passes is that we add two Faraday interactions with different relative phases between the beam splitter and parametric gain interactions in Fig. \ref{fig:lambda} (c). A suitable choice of geometry and phase shifts leads to cancelation of the parametric gain interactions after the two passes, and we are left with the beam splitter interaction making an ideal memory transformation. With other configurations it is the beam splitter interaction which cancels and the parametric gain interaction takes effect, generating an entangled state of light and atoms, whose EPR variance, cf. Sec.~\ref{sec:deteministic_ent}, scales asymptotically as $\Delta_{EPR}\sim\exp(-\kappa^2)$ \cite{Muschik:2006}.


\subsection{Raman  and EIT approach}
\label{sec:raman_eit}
The beam splitter interaction is a quite natural choice for a quantum memory since the interaction maps excitations from light to atoms and back. The beam splitter interaction was proposed for a quantum memory by \textcite{Kozhekin:2000} who considered the far-detuned Raman limit $\Delta\gg d \gamma$. The same Raman limit was also considered by \textcite{Nunn:2007}.
Most of the attention to the beam splitter interaction, however,  came with the realization that on resonance with the atomic transition, Electromagnetically Induced Transparency (EIT) can be used for quantum memory for light. The principles of EIT and its applications for coherent memory for light have been reviewed earlier \cite{Lukin:2003,Fleischhauer:2005}, hence we will concentrate here mostly on the recent advances on quantum memory. The EIT is achieved when a strong control optical pulse renders the $\Lambda$ system transparent for the signal pulse, see Fig. \ref{elemlevels}(b). This transparency is accompanied by a strong reduction in the group velocity of the signal pulse. The result is that the signal pulse entering the atomic ensemble is spatially compressed. If the compression is sufficient to make the signal pulse fit inside the sample, the control field can be turned off at this point and the signal pulse is "frozen" into the atomic ground state coherence -- the dark state polariton wave. The process can be inverted by turning on the control field after some delay, which leads to the generation of a signal pulse which "remembers" the classical and quantum properties of the input signal pulse.

To see what happens in this situation let us now consider the $\Delta
\rightarrow 0$ limit of the solution in Eqs. (\ref{eq:light_out_bs}),
(\ref{eq:lambda_kernel}), and  (\ref{eq:atom_bs_out}).  If we assume
a sufficiently large incident classical driving field so
that $h(0,t)z/L\gg 1$ we may use the asymptotic form of the Bessel
function and  write the integral kernel (\ref{eq:lambda_kernel}) as
\begin{equation}
m(\Omega;t,z)\approx \sqrt{v_g} \sqrt{\frac{d}{L}} \frac{1}
{2\sqrt{2\pi}\sqrt[4]{v_gtz}} {\rm e}^{-d(\sqrt{v_g t }-\sqrt{z})^2/L},
\end{equation}
where we have for simplicity assumed that the classical driving field
is time independent and have introduced the group velocity
\begin{equation}
v_g=\frac{\Omega^2L}{\gamma d}.
\label{eq:vgroup}
\end{equation}
For a large optical depth this kernel is centered around $z=v_g t$.
If the spin wave mode is slowly varying on a length
scale $L/\sqrt{d}$, or if the incoming field is slowly varying on a
time scale $L/\sqrt{d}v_g$, i.e., the input field is inside the EIT-transparency window
\cite{Fleischhauer:2005}, the
expression above can be approximated by a delta function $m=\sqrt{v_g} \delta(z-v_g t)$.
The solutions of  Eqs. (\ref{eq:light_out_bs}) and
(\ref{eq:atom_bs_out}) then become
\begin{equation}
\begin{split}
a_{A,{\rm out}}(z)&=\frac{1}{\sqrt{v_g}} a_{L,{\rm in}}(T-z/v_g)\\
a_{L,{\rm out}}(t)&=\sqrt{v_g} a_{A,{\rm in}}(L-v_g t).
\end{split}
\end{equation}
The first line describes the writing of the input field into the atomic memory, whereas the second one describes the readout of the atomic memory back into light. To accomplish the writing (storage) process the control field must be turned off when the entire input pulse is inside the medium, which requires $T_{{\rm signal}}<L/v_g$. This prompts the reduction of $v_g$ to zero and hence "stopping" of light. The process of the readout or the retrieval is accomplished according to the second line by turning the control field on again which leads to the mapping of the atomic memory operator back onto the light operator.

Recently several experimental implementations of the EIT atomic memory \cite{Chaneliere:2005,Eisaman:2005,Kuzmich:2003,Choi:2008} have demonstrated that such quantum features of light as violation of a Cauchy-Schwarz inequality and entanglement can be preserved by the memory. Very recently the EIT based memory for quantum fluctuations has been also demonstrated using squeezed vacuum light in \cite{Honda:2008,Appel:2008}. As mentioned in \cite{Honda:2008} a fidelity calculation along the lines of the discussion in Sec. \ref{sec:figure_of_merit} would be misleading for the case studied in these two papers since only one particular state has been used and, in addition, the overlap of this weakly squeezed state with vacuum is not far from unity. An overall efficiency of storage and retrieval of around $10-15 \%$ has been achieved, and weakly squeezed light (around $-0.2$dB) has been retrieved from the memory.

The experiments \cite{Chaneliere:2005,Eisaman:2005,Kuzmich:2003,Choi:2008} have had a relatively low overall efficiency of the storage-retrieval process, however it did not preclude observation of the storage of non-classical light. The reason is that the violation of the Cauchy-Schwarz inequality is based on the measurement of the normally ordered second order correlation function \mbox{$g^{2}(1,2)\equiv\langle :\hat{n}_{1}\hat{n}_{2}:\rangle/\langle \hat{n}_{1}\rangle\langle \hat{n}_{2}\rangle$} where $\hat{n}_{1,2}$ denote photon number operators measured at space-time points 1,2. $g^{2}(1,2)$ which describes the normalized probability of detecting a photon at point 2 conditioned on the detection at point 1 is insensitive to losses because they affect the numerator and the denominator in the same proportion. For an ideal single photon state $g^{2}(1,2)=0$, whereas any $g^{2}(1,2)<1$ is a signature of the non-classical character of the field which for the case of a stationary photon flux is referred to as photon anti-bunching.

\begin{figure*}[ht]
  \includegraphics[width=0.7\textwidth]{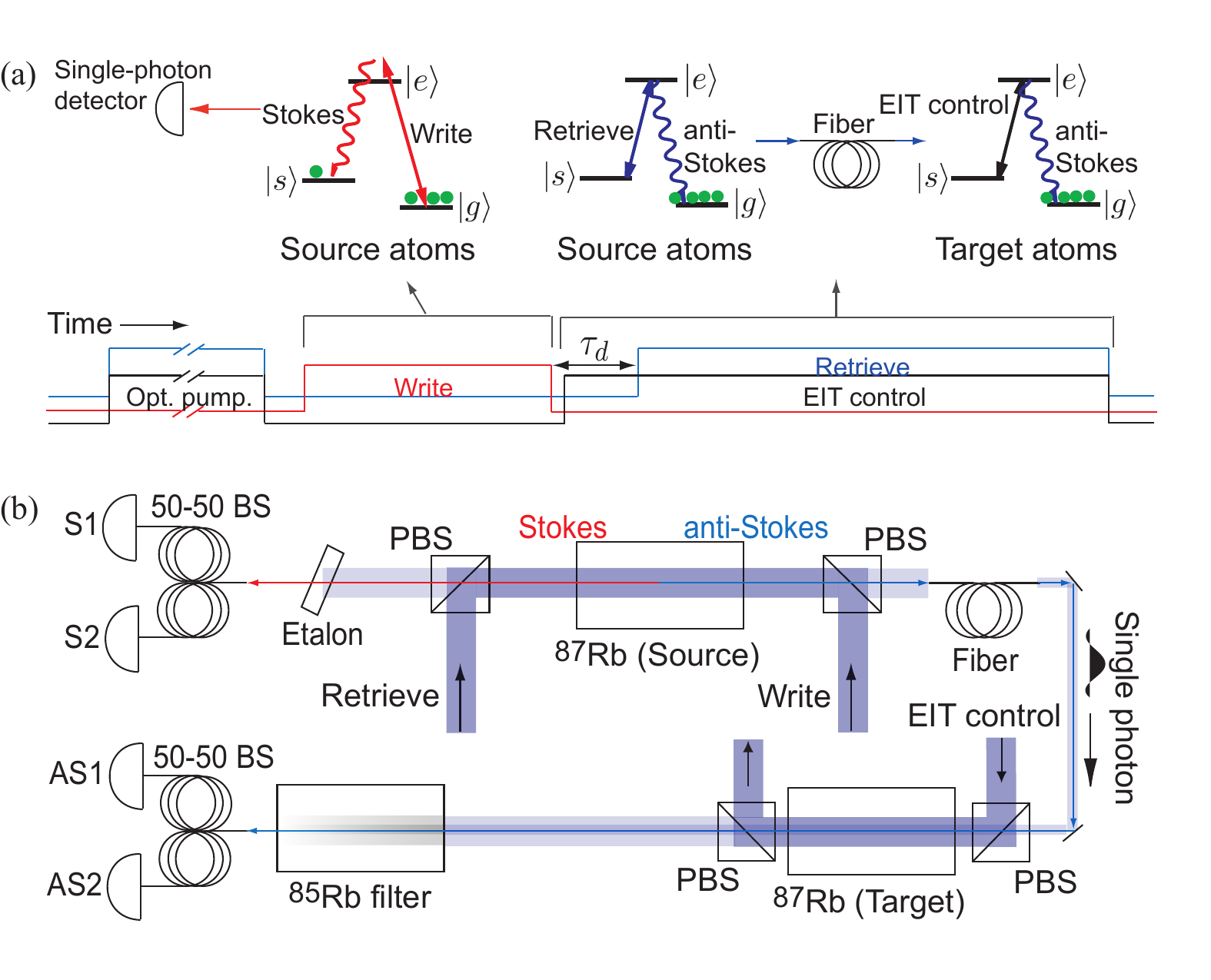}
  \caption{EIT-based memory setup after \textcite{Eisaman:2005}. a) Level scheme and sequence of control pulses. Left -- parametric interaction for a probabilistic generation of an excitation in the source ensemble, center and right -- EIT-based beam splitter interactions for storage and retrieval. b) Setup showing the source ensemble of the non-classical photon flux and the memory ensemble (target atoms) . A count in detector S1 or S2 is a condition for starting the process. Detectors AS1 and AS2 analyze the statistics of the input and output memory photons.}\label{fig:Lukin_setup}
\end{figure*}

The source of the non-classical field used in \cite{Chaneliere:2005,Eisaman:2005,Matsukevich:2006,Yuan:2007}, and \cite{Choi:2008} has been an atomic ensemble prepared in an approximate atomic single excitation state by a weak parametric gain interaction (see Sec. \ref{sec:probabilistic}) which was then retrieved onto light by a beam splitter process. Non-classical states prepared in this way have been reported by \textcite{vanderWal:2003,Kuzmich:2003,Chou:2004,Chaneliere:2005,Eisaman:2005,Matsukevich:2004,Balic:2005,Du:2008}. The sequence is similar to that described in Sec. \ref{sec:probabilistic} except for only one atomic ensemble is involved at this stage. The layout of the experimental setup of \textcite{Eisaman:2005} is shown in Fig.~\ref{fig:Lukin_setup}. The weak parametric gain interaction  driven by the "write" classical field so that $\kappa\ll 1$ creates a single atomic excitation in the "source" ensemble conditioned on the detection of a photon in a particular spatial "Stokes" mode.
The requirement of low gain $\kappa\ll 1$ is necessary so that multi-photon pulses are  suppressed, which is a typical situation for the parametric-type interaction. The "retrieve" strong pulse converts the atomic excitation into an "anti-Stokes" light pulse. The non-classical character of this field is manifested by the correlation function conditional on the detection of one Stokes photon $g^{2}(AS\parallel n_{S}=1)<1$. This condition means that if a Stokes photon has been detected and if there is a photon in the anti-Stokes pulse, then the probability of having a second photon in the anti-Stokes pulse is less than that for a random process. In the ideal case this probability is zero and the anti-Stokes pulse contains either no photons or just a single one.

\begin{figure}[ht]
  \includegraphics[width=8.3cm]{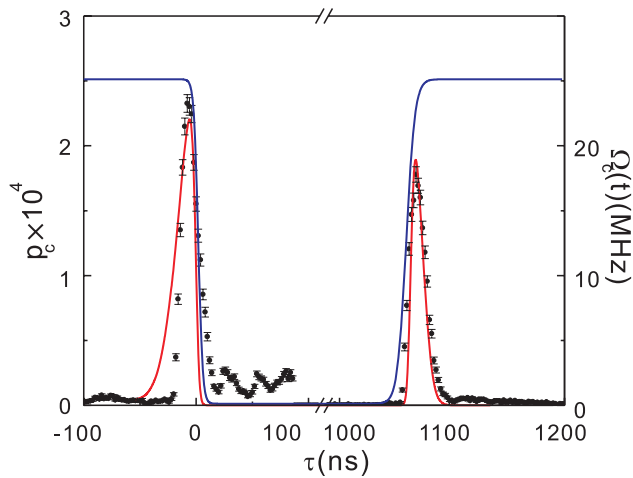}
  \caption{EIT-based memory (\textcite{Choi:2008}). Probability of detecting a photon downstream the memory ensemble (left axis). The points around $\tau=0$, the moment when the input pulse is launched are due to the imperfect memory leading to the "leakage" of photons through it. The strong field (right axis) is turned off around $\tau=0$ and turned on again at $\tau=1\mu$sec leading to the retrieved pulses with the overall storage retrieval probability of $17\%$}.\label{fig:Kimble_pulses}
\end{figure}

The non-classical light pulses produced by the "source" ensemble are then directed towards the memory ensemble. There they are stored and retrieved using the EIT pulse sequence as shown in Fig.~\ref{fig:Kimble_pulses} from \textcite{Choi:2008}. The figure shows the probability of detecting a photon after the storage medium. The strong "storage" driving EIT field has a constant amplitude until the quantum input pulse appears at $\tau=0$. Then the driving field is turned off leading to the storage of the light in the medium. For ideal storage there should be no counts corresponding to the input pulse. The counts around $\tau=0$ hence correspond to the "leakage" of light through the memory ensemble. After a delay time of $1~\mu$sec the strong "read" control field is applied and the atomic excitation is read out generating a light pulse as prescribed by Eq. (\ref{eq:atom_bs_out}). An overall storage-retrieval probability of $17\%$ has been shown in \textcite{Choi:2008}. This experiment and the one by \textcite{Chaneliere:2005} used, respectively, Cs and Rb atoms cooled and trapped in a magneto-optical trap (MOT), whereas \textcite{Eisaman:2005} used Rb atoms in a neon buffer gas cell at room temperature. In the two former experiments a small sub-ensemble of atoms contained within the volume of the focused optical beam inside the sample served as the memory. The storage time in these experiment was restricted by the motion of atoms and/or by magnetic field inhomogeneity to about  $1~\mu$sec. The two lower levels of the $\Lambda$ system were the two ground state $S_{1/2}$ hyperfine levels and the upper level was the $P_{3/2}$ level.

\begin{figure}[ht]
  \includegraphics[width=8.8cm]{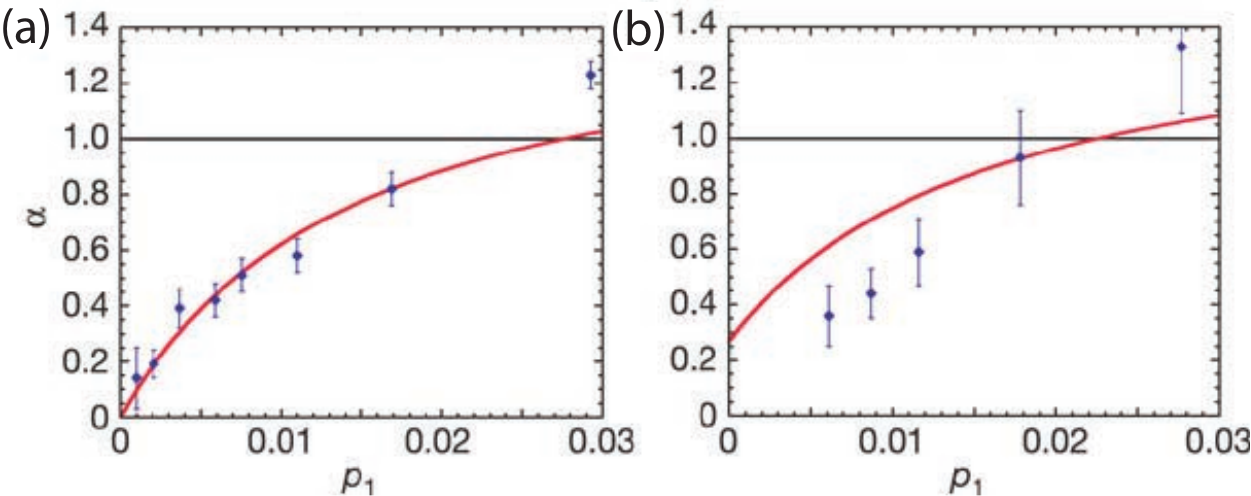}
  \caption{The correlation function for the input (b) and output (d) light for the EIT-based memory from \textcite{Chaneliere:2005}. $g^{2}(0)<1$ is a signature of non-classical character corresponding to the single photon in the limit of $g^{2}(0)=0$.}\label{fig:Kuzmich_g2}
\end{figure}

The results for the conditional correlation function $g^{2}(out\parallel n_{S}=1)<1$ for the memory output field obtained by \textcite{Chaneliere:2005} are shown in Fig.~\ref{fig:Kuzmich_g2}. The results clearly show the nonclassical character of the retrieved light. The overall efficiency of the storage/retrieval process of a photon was $6.4$ \%. Similar results have been obtained by \textcite{Eisaman:2005}.

\begin{figure*}[ht]
  \includegraphics[width=0.8\textwidth]{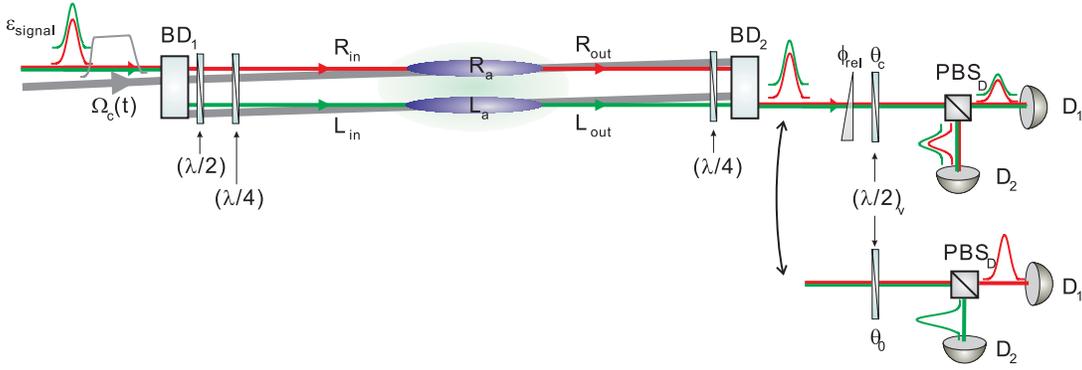}
  \caption{EIT-based memory for an entangled state \cite{Choi:2008}. Two memory ensembles store two components of an entangled state which are later retrieved and analyzed by two pairs of coincidence detectors $D_{1,2}$.}\label{fig:Kimble_setup}
\end{figure*}

The experiment by \textcite{Choi:2008} have taken the EIT approach one step further by demonstrating it for an entangled state, more precisely for a superposition state of a photon being in one of two possible pathways (Fig. \ref{fig:Kimble_setup}). In this experiment a conditional non-classical state with $g^{2}(AS\parallel n_{S}=1)<1$, an approximate single photon state, was split on a beam splitter in two parts. Conditioned on the registration of the Stokes photon in the source ensemble (not shown in Fig.~\ref{fig:Kimble_setup}) the input consisted of a single photon with $15\%$ probability. The rest was mostly vacuum with a small addition of a two-photon component with the probability of $9$\% of that for a Poisson source with the same average photon number. The input state with these properties has the concurrence of $0.10$ corresponding to an entanglement of formation of $E_{\rm oF}=0.025~{\rm ebits}$. (A concurrence of unity corresponds to $E_{\rm oF}=1~{\rm ebits}$, i.e. to a maximally entangled Bell state.) The input state was linearly polarized at $45^{\circ}$ with respect to the polarizing beam splitter (Fig.~\ref{fig:Kimble_setup}). The photonic state of the two outputs of the beam splitter conditioned on the successful preparation of the single photon state is $\Psi_{in}=(|0_{L}\rangle|1_{R}\rangle+e^{i\varphi}|1_{L}\rangle|0_{R}\rangle)/\sqrt{2}$. The two components of this state have been directed into two atomic ensembles. The two ensembles were two groups of atoms within a Magneto-optical trap (MOT). Rb atoms were optically pumped into a particular magnetic sublevel of the ground state $F=4, m_{F}=0$ which contributed to a better performance of the memory.

\begin{figure}[ht]
  \includegraphics[width=8.3cm]{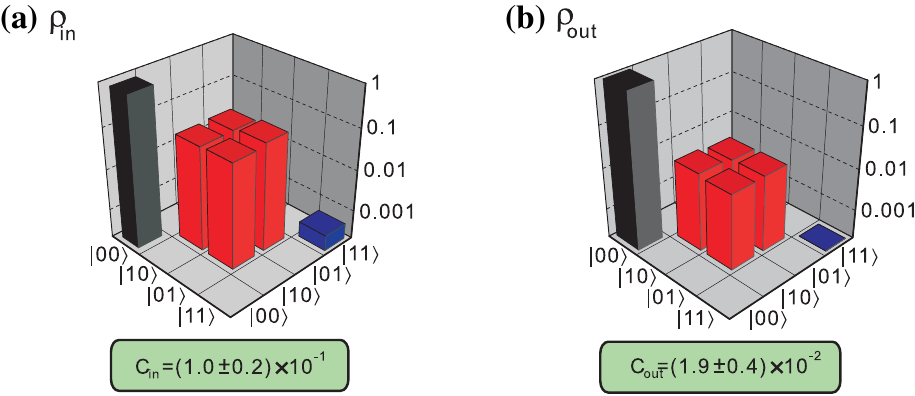}
  \caption{Characterization of the retrieved entangled state \cite{Choi:2008}. The reconstructed density matrices of the input (a) and the output (b) states of the light.}\label{fig:Kimble_entanglement}
\end{figure}

After the EIT storage  and retrieval steps similar to those described earlier in this section the two retrieved components were combined on a polarizing beam splitter. With a suitable choice of the phase the state recreated from an ideal memory would make the single linearly polarized photon, just like the input one. Changing the phase between the two components of the state stored in the two ensembles by adjusting the $\lambda/2$ wave plate \textcite{Choi:2008} performed tomography of the entangled state and obtained the density matrix of it. The results of this procedure are shown in  Fig.~\ref{fig:Kimble_entanglement}. From these results the concurrence of the retrieved state of $0.017$ ($E_{\rm oF}=0.001~{\rm ebits}$) has been inferred demonstrating that the retrieved state has retained entanglement after the storage process.

The overall efficiency of the storage/retrieval of entanglement measured by the ratio of the concurrence of the output to the concurrence of the input is $20$ \% after the storage time of $1.1~\mu$sec. It is limited by the finite optical depth of the sample and can be improved by optimization of the pulse shapes as discussed below.

These experiments represent exciting experimental progress, however a higher efficiency is still desirable for future applications. Higher efficiency can be achieved with an increased optical depth, but also by optimizing the shape of the classical driving field $\Omega(t)$. This optimization problem is the subject of a series of papers by \textcite{Gorshkov:2007a,Gorshkov:2007b,Gorshkov:2007c,Gorshkov:2007d,Gorshkov:2008} (see also related work by \textcite{Dantan:2004,Dantan:2005,Dantan:2006} and by \textcite{Nunn:2007}). These studies show that for any slowly varying pulse of duration $T$ such that $Td\gamma\gg 1$ (bandwidth $BW\ll d\gamma$) the optimal storage and retrieval efficiency is the same and is independent of the detuning from the excited atomic state. Furthermore, the optimized inefficiency  only depends on the optical depth $d$ and scales as $1/d$. These results can be understood from a new "universal" physical picture of the storage and retrieval process: First of all for a given stored spin wave, the retrieval process is essentially a constructive interference effect similar to super radiance, where the radiated fields from all atoms interfere constructively in a certain direction. As a result there is a fixed branching ratio between the decay into the desired quantum field mode and the decay into all other modes, which is independent of the control field shape as long as sufficient optical power is used. Secondly, the optimal  storage procedure is the time reversal of the retrieval procedure, and, by time reversal symmetry, the optimal storage efficiency is identical to the retrieval efficiency.

The retrieval efficiency does, however, depend on the spin wave mode, and for optimal storage the classical drive field $\Omega(t)$ has to be chosen so that it maps  an incoming field mode into the optimal spin wave. This optimal field shape $\Omega(t)$ can be found by a direct calculation but an alternative experimental procedure for finding optimal shapes is demonstrated by \textcite{Novikova:2007}. In this experiment which worked in the classical regime with many photons in the signal beam, the shape of the pulse to be stored was optimized for a given classical drive field.
\textcite{Novikova:2007} first stored a given pulse and recorded the shape of the retrieved pulse. Then the time reverse of the recorded pulse was used as an input for the next round of storage, retrieval and measurement. This procedure rapidly converges and yields the optimal efficiency \cite{Gorshkov:2007c}, which in the experiment was in the range 42--45\%. In \textcite{Novikova:2008} on the other hand the full theoretical optimization of the classical field shape $\Omega(t)$ was used to  store arbitrary field shapes and retrieve them into a possibly differently shaped mode with a similar efficiency.

Because the optimal strategy for storage and retrieval is based on time reversal, higher efficiency can actually be achieved if the excitation is read out in the  backward direction compared to the direction of storage \cite{Gorshkov:2007c}. This change of direction, however, requires a redefinition of the atomic operators (\ref{eq:aatom_transverse_lambda}). The mode functions $u_m(z;\rvec_\perp)$ in Eq. (\ref{eq:aatom_transverse_lambda}) are  solutions to the 3-dimensional Maxwell equation in the forward direction but it may not be so in the backward direction. This will complicate the dynamics unless the mode function
 can be chosen real, which requires  a Fresnel number much bigger than unity (for a discussion of a related problem see \textcite{Andre:2005}).  Furthermore a finite energy difference $\omega_{01}$ between the two ground states introduces a momentum difference $\Delta k=\omega_{01}/c$ which reduces the achievable memory efficiency unless $\Delta k L\lesssim 1$ \cite{Gorshkov:2007c}.   A way to cope with this problem is presented by \textcite{Surmacz:2008}.

For applications in a quantum network, stored excitations will have to be processed in the quantum memory. A first step in this direction was performed in recent experiments by \citet{Simon:2007b} and \citet{Simon:2007}. The latter experiment involved
two atomic ensembles in a medium Finesse ($\mathcal{F}=240$) cavity. It demonstrated adiabatic transfer of a single excitation stored in one ensemble to the other ensemble with the cavity mode serving as a quantum bus, as well entanglement of the two ensembles by a partial transfer. First, a single excitation is generated in one ensemble by driving a weak parametric gain interaction with subsequent detection of a Stokes photon, as described above. Conditioned on the successful generation of a single excitation in ensemble 1, beam splitter interactions are switched on and couple both ensembles to a common cavity mode, as indicated in Fig.~\ref{fig:multipass}(b). This is done adiabatically, first for the ''empty'' ensemble 2 and then for ensemble 1, in a counterintuitive sequence generating an adiabatic dark--state passage $\ket{1}_1\ket{0}_2\rightarrow\ket{0}_1\ket{1}_2$. After the transfer, the single excitation was read out from ensemble 2, demonstrating a transfer efficiency between 10\% and 25\%, depending on the optical depth. \citet{Simon:2007} also demonstrated  a partial swap of the excitation, generating in the ideal case an entangled state of the ensembles, $\ket{1}_1\ket{0}_2\rightarrow\cos\theta\ket{1}_1\ket{0}_2+\exp(i\phi)\sin\theta\ket{0}_1\ket{1}_2$, where $\theta$ is controlled via the intensities and $\phi$ by the relative phase of the laser fields $\Omega_i$ in Fig.~\ref{fig:multipass}(b). Reading out the collective excitations and measuring photon correlation functions, similar to what was done by \citet{Chou:2005}, a lower bound on the entanglement of the ensembles was determined, giving a concurrence larger than 0.0046 corresponding to an entanglement of formation of $E_{\rm of}\geq 0.0001~{\rm ebits}$ \cite{Wootters:1998}.

\subsection{Photon Echo}\label{sec:echo}

It has long been known that the photon echo technique \cite{Kurnit:1964} could be used to store classical light pulses. Recently it has been realized that one can extend these
 techniques into the quantum interface domain \cite{Moiseev:2001,Kraus:2006}. If a light pulse is absorbed by an ensemble of two-level atoms, the quantum properties of light will be stored in atoms as shown already in the early experiment by \textcite{Hald:1999}. The problem is that for a meaningful memory the coherence time of the optical transition should be longer than the duration of the pulse. However, this means that the bandwidth of the interaction which is set by the inverse coherence time is narrower than the bandwidth of the light set by its inverse duration, so that the entire pulse cannot be stored. A way out of this problem is to use a medium with inhomogeneous broadening. Then different frequency components of the light pulse will be effectively stored in different sub-groups of atoms. To avoid the dephasing of the stored state caused by the broadening a photon echo technique is used which also allows to control the release of the stored excitation.

The essence of the photon echo approach is to have an inhomogeneously broadened line which is then reversed. In the original approach an incoming light field is absorbed by a two level system with a broadened optical
transition. Due to the inhomogeneous broadening the optical coherence from each atom ($i$) precesses at different frequencies $ \exp(-i\omega_i t)$. These different precession frequencies dephase the optical polarization such that it does not radiate because the radiation from different atoms interfere destructively. In the simplest version of the photon echo, a strong $\pi$-pulse, which interchanges the ground and excited states, is applied after a time
 $T/2$. This strong $\pi$-pulse effectively reverses the phase acquired by each atom such that the subsequent time  evolution causes a rephasing of the optical coherence. At time $T$ all the atomic polarizations are again in phase causing an echo signal to be emitted.

 A modification of the photon echo technique  called Controlled Reversible Inhomogeneous Broadening (CRIB) which  in principle allows an ideal memory efficiency  was introduced by \textcite{Nilsson:2005} and \textcite{Moiseev:2001} and further developed by \textcite{Kraus:2006}. \textcite{Moiseev:2001} considered a  $\Lambda$-system
 similar to Fig. \ref{elemlevels}(b). The photon is absorbed while an applied electric field broadens the $|0\ra$--$|e\ra$ absorption line.  A $\pi$ pulse traveling in the same  (forward) direction is applied, which takes the population from the excited state to the initially empty   state $|1\ra$ for long term storage. Later on another $\pi$-pulse traveling in the opposite (backward) direction is applied, which releases the excitation. During the retrieval process the broadening with the applied electric field is reversed, which reverses the time evolution resulting in an ideal retrieved signal in the backward direction, provided that the excited state decoherence is negligible.  Experimentally this scheme was realized by \textcite{Alexander:2006}. \textcite{Staudt:2007} demonstrates that similar photon echo techniques can preserve the coherence of time bin qubits  conditioned on having an outgoing photon. There are by now many versions of this CRIB approach, differing, e.g., by whether the broadening arises from differences in the response of the atoms to a homogeneous field (transverse broadening) or from spatially dependent broadening (longitudinal broadening). A full account of all different types is beyond the scope of this review (see e.g. \textcite{Longdell:2008} for a theoretical discussion of several different schemes).

\begin{figure}[ht]
  \includegraphics[width=8.3cm]{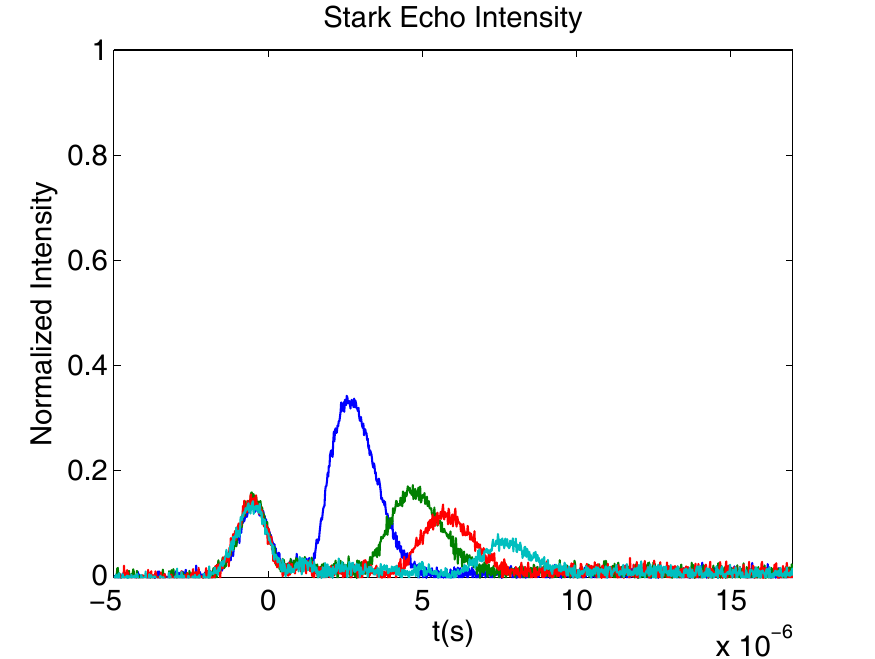}
  \caption{Transmitted and retrieved pulses for different storage times for an echo based memory (private communication from \textcite{Hetet:2008}). The small peak at $t\approx 0$ represents a small leakage (non stored component) of the incoming field, whereas the later peaks are the retrieved field. The emission time of these light pulses is controlled by reversing the sign of  an applied electric field halfway during the storage period. See comments in the text. }
  \label{fig:Sellars_echo}
\end{figure}

In a recent experiment \cite{Hetet:2008} CRIB has been implemented in a particularly elegant and simple way. A similar technique was also considered theoretically by \textcite{Sangouard:2007}. In contrast to most other techniques considered in this paper, no strong classical control pulse is required in this scheme. \textcite{Hetet:2008} applied a gradient electric field to the crystal used for the memory. Light interacted with Pr$^{3+}$ ions which were doped into a Y$_{2}$SiO$_{5}$ host. Due to the gradient field the resonance frequencies of the ions at different parts of the crystal were distributed by the Stark effect within the $2$ MHz bandwidth. The homogeneous line width of the ions is $100$ kHz. Hence the ions should have been in principle capable of storing a pulse of light for this time which is $20$ times longer than the pulse duration. The stored light has been released by reversing the sign of the electric field gradient.  In the experiment a classical pulse of $2~\mu$sec duration was stored for $2~\mu$sec. An overall efficiency of the storage and retrieval of $15$ \% has been achieved. The results for the echo memory for classical pulses are shown in Fig.~\ref{fig:Sellars_echo}. The figure shows the  transmitted part of the input pulses (centered around zero) and stored and retrieved pulses for different storage times. All pulses are normalized to the amplitude of the input pulse.

An important question is to which extent the addition and reversal of broadening improves the memory performance compared to the other approaches considered here. For the approaches of \textcite{Hetet:2008} and \textcite{Sangouard:2007} a major practical  advantage is that it does not rely on any classical laser fields. This,  however, limits the attainable storage time to the coherence time of the optical transition, which is typically shorter than for the ground states. The  advantage of reversible broadening for the storage of a single mode in the ground state coherence was investigated by \textcite{Gorshkov:2007d,Gorshkov:2008} where it was found that little was gained. The situation is, however, different for the storage of multiple modes. In particular, it was found by \textcite{Simon:2007} and \textcite{Nunn:2008} that CRIB changes the scaling of the number of modes which can be stored from $\sqrt{d}$ to $d$, thus significantly increasing the multi mode capacity. Furthermore a novel type photon echo approach using atomic frequency combs was recently proposed \cite{Afzelius:2008}. This proposal takes advantage of a large inhomogeneous broadening in a solid state system. By exploiting atoms with different resonance frequencies one can increase the effective number of atoms participating in the memory and thereby achieve a very efficient quantum memory with high multimode capacity. Using this approach \textcite{Riedmatten:2008} demonstrated the coherent mapping of light at the single photon level onto a neodymium ion doped crystal, and collective release of the stored light at a pre-determined time. Moreover, this experiment demonstrated the storage of pulses in different temporal modes, proving the multimode capacity of this approach.

%
%

\section{Quantum teleportation between light and atoms}
\label{sec:teleportation}

\subsection{Quantum Teleportation}

Quantum teleportation is a means for sending quantum states from A to B in a disembodied fashion using two separate channels, a quantum channel connecting A and B and a classical one. It makes use of entanglement shared via the quantum channel and classical communication. Apart from being one of the most surprising and mind-boggling discoveries in quantum information theory, quantum teleportation has become an essential primitive in quantum computation and quantum communication.

Shortly after the first theoretical layout for quantum teleportation of states of a qubit \cite{Bennett:1993} the protocol was extended to states of continuous variables \cite{Vaidman:1994,Braunstein:1998}. Both sorts of protocols were first demonstrated with light, utilizing either probabilistically generated Bell states of photon pairs \cite{Bouwmeester:1997} or deterministically generated EPR beams \cite{Furusawa:1998}, both obtained from parametric down conversion. The first teleportation involving massive particles was performed recently in the ion trap experiments at Innsbruck \cite{Riebe:2004} and NIST \cite{Barrett:2004}. For a recent review on teleportation of states of continuous variables see \cite{Furusawa:2007}.

Our focus here is on teleportation protocols involving both matter and light. In such a scenario, quantum states carried by traveling pulses of light are teleported onto a stationary quantum memory, for example, an ensemble of neutral atoms, using the quantum interface. The first realization of teleportation involving matter and light utilized the entangled state created via the quantum Faraday - QND interaction of a pulse of light with the collective spin of an atomic ensemble \cite{Sherson:2006a}. In this experiment deterministic teleportation with the fidelity higher than any classical state transfer can achieve has been demonstrated. Very recently probabilistic teleportation between light and matter has been also demonstrated using parametric gain- and $\Lambda$-type interactions \cite{Chen:2008} and has been extended to entanglement swapping, that is the teleportation of entanglement \cite{Yuan:2008}.

\begin{figure*}[ht]
  \includegraphics[width=0.7\textwidth]{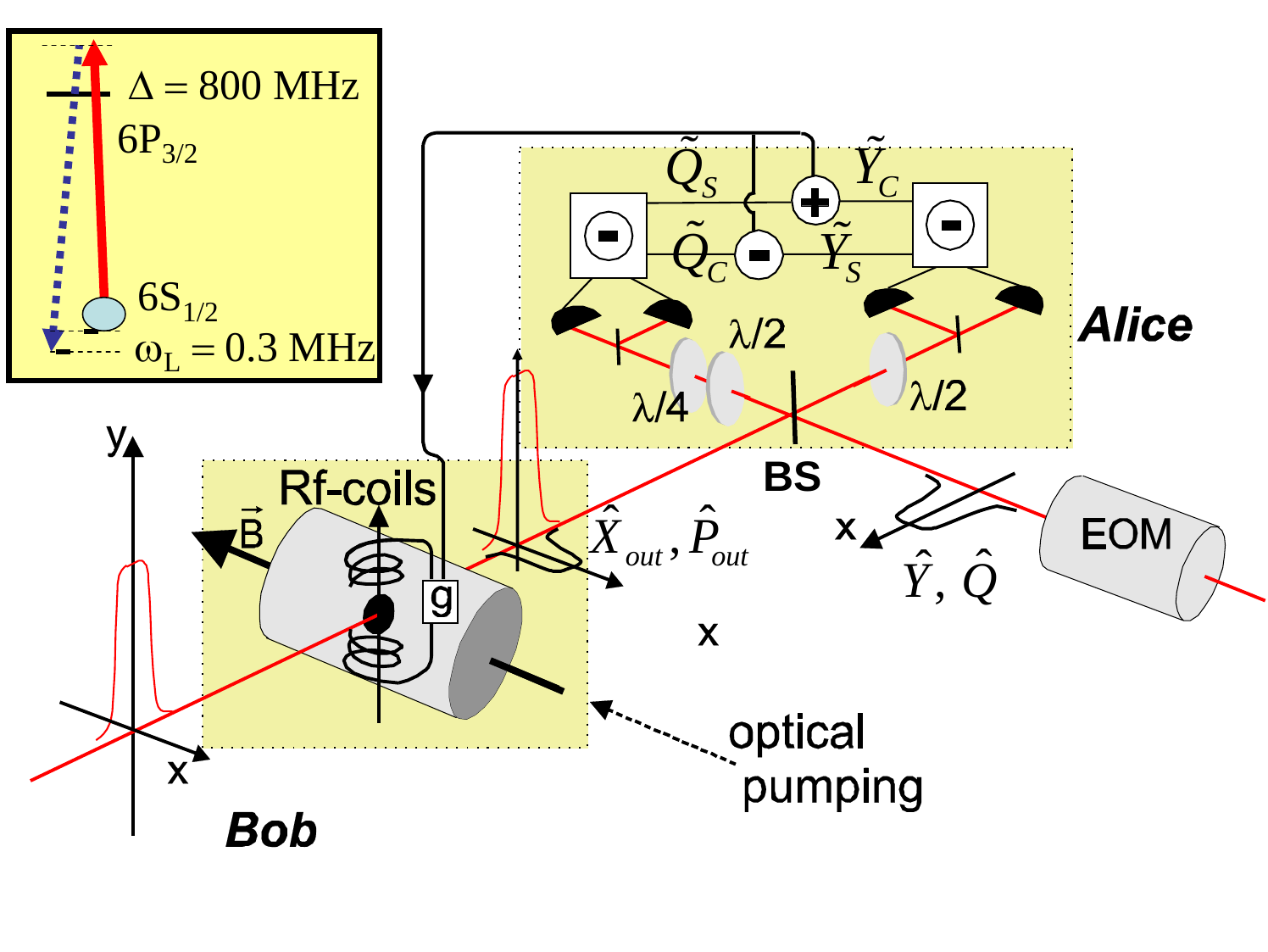}
  \caption{Experimental setup for the teleportation of light to atoms \cite{Sherson:2006a}. Strong y-polarized pulse interacts with atoms and generates an entangled x-polarized mode at the upper sideband frequency (inset). This entangled light is overlapped with the pulse to be teleported on a 50/50 beam splitter BS after which the Bell measurements are performed. The results of the Bell measurements are fed back onto atoms via RF magnetic coils. Detailed comments in the text.}\label{fig:telesetup}
\end{figure*}

In a nutshell, a teleportation protocol involves three steps: First, an entangled state is created and shared as a resource between two stations (usually termed ''Alice'' and ''Bob''). In the cases considered below, the entanglement between atoms (Bob) and light is generated by sending a strong driving pulse through atoms. As a result of this interaction the forward scattered photons of the orthogonal polarization sent to Alice become entangled with atoms kept by Bob as shown in Fig.~\ref{fig:telesetup}. The ideal continuous variable EPR entanglement corresponds to $ \Delta(X_A+X_L)^2=\Delta(P_A-P_L)^2\rightarrow 0$. Alice also receives another, unknown quantum state described by canonical variables $Y,Q$ sent by a hypothetical sender ''Victor'', which is to be teleported to Bob. For this, Alice performs a joint measurement of $X_L+Y$ and $P_L-Q$, called a Bell measurement, on the photonic part of the entangled state that she has received and the unknown quantum state of light to be teleported. This measurement is performed by mixing two light pulses on a beamsplitter as in Fig.~\ref{fig:telesetup}. The results of this measurement are communicated to Bob. Bob uses this classical information to perform a correcting operation on his quantum system -- atoms -- by shifting $X_{A,{\rm fin}}=X_A+(X_L+Y)\rightarrow Y$ and $P_{A,{\rm fin}}=P_A-(P_L-Q)\rightarrow Q$, thereby recovering the original unknown state.

We see that in the hypothetical case of vanishing EPR variances, first and second moments and thus any Gaussian state (i.e. a state with Gaussian wave function or Wigner function) is transmitted perfectly. This in turn implies that any (non-Gaussian) state, including a qubit of the form $\alpha\ket0+\beta\ket1\in L^2(\mathds{R})$, would be teleported faithfully, as the set of coherent states is a subset of all Gaussian states and provides a basis for the full Hilbert space $L^2(\mathds{R})$ (for the proof of this in the Schr\"odinger picture and for the Wigner function, see \textcite{Braunstein:2005}). In this sense the distinction between teleportation protocols for qubits and continuous variables is superficial. This said, for realistic cases of imperfect teleportation where fidelity is not perfect, performance of each teleportation protocol should be evaluated in detail having in mind a particular application.

If EPR variances do not vanish, as is necessarily the case due to energy restrictions, teleportation will not be perfect, and it is necessary to evaluate the performance of the teleportation. The teleportation is essentially a protocol mapping the state of one system to another. We can therefore use the fidelity as the figure of merit as discussed in Sec. \ref{sec:figure_of_merit}, i.e., find out when the performance of the protocol becomes better than that of the best classical protocol for a given class of input states.

\subsection{Teleportation based on Faraday interaction in magnetic field}

In \textcite{Sherson:2006a} the entangled state of light and atoms for teleportation is obtained via the Faraday interaction of light with a single collective atomic spin, precessing in an external magnetic field. The relevant level scheme is shown in Fig.~\ref{fig:bfield} and in the inset to Fig. \ref{fig:telesetup}. The situation is described by the Hamiltonian and Maxwell-Bloch equations given in section \ref{sec:magnetic field}. In this case, the Hamiltonian does not fulfil the QND criteria \cite{Holland:1990,Poizat:1994} because the Faraday interaction $H_F$, c.f. Eq. \eqref{eq:H_Faraday_1}, does not commute with the free Hamiltonian \eqref{eq:magnetic_field} describing Zeeman splitting of ground states. In fact, as seen from the inset in Fig. \ref{fig:telesetup}, the interaction resembles the Raman process. The Maxwell-Bloch equations \eqref{eq:faraday_rotframe} were integrated by \textcite{Hammerer:2005,Hammerer:2006a} and the solutions are again conveniently expressed in terms of experimentally measurable cosine and sine modulation modes \eqref{eq:modulation modes},
\begin{align*}
\xa{out}&=\xa{in}+\frac{\kappa}{\sqrt{2}}\p{c}{in},\\
\pa{out}&=\pa{in}+\frac{\kappa}{\sqrt{2}}\p{s}{in},\\
\p{c}{out}&=\p{c}{in},\\
\x{c}{out}&=\x{c}{in}+\frac{\kappa}{\sqrt{2}}\pa{in}\\
&\hspace{30pt}+\left(\frac{\kappa}{2}\right)^2\p{s}{in}
+\frac{1}{\sqrt{3}}\left(\frac{\kappa}{2}\right)^2\p{s,back}{in},\\
\p{s}{out}&=\p{s}{in},\\
\x{s}{out}&=\x{s}{in}-\frac{\kappa}{\sqrt{2}}\xa{in}\\
&\hspace{30pt}-\left(\frac{\kappa}{2}\right)^2\p{c}{in}
-\frac{1}{\sqrt{3}}\left(\frac{\kappa}{2}\right)^2\p{c,back}{in}.
\end{align*}
The terms proportional to $\kappa^2$ are a new feature, specific for this setup, and represent atom-mediated back-action of light onto itself. This effect involves the previously defined cosine and sine modes as well as yet another pair of canonical independent ''back-action modes'' $\p{c(s),back}{in}$ which can be treated as vacuum noise operators.
To see that this interaction can be used for teleportation let us inspect the EPR type correlations between the atomic mode $\xa{out},\,\pa{out}$ and the light mode of the upper sideband $X_{+},P_{+}$ with the frequency $\omega+\omega_{L}$
\begin{align*}
\x{+}{out}&=\frac{1}{\sqrt{2}}(\x{s}{out}-\p{c}{out}),\\
\p{+}{out}&=\frac{1}{\sqrt{2}}(\x{c}{out}+\p{s}{out}).
\end{align*}
One easily finds an EPR variance of
\begin{multline}\label{eq:Etele}
  \Delta_{EPR}=\Delta(\xa{out}+\x{+}{out})^2+\Delta(\pa{out}-\p{+}{out})^2\\
  =\frac{1}{2}\left[1+\left(1-\frac{\kappa}{2}\right)^2 \right]^2+\frac{1}{3}\left(\frac{\kappa}{2}\right)^4\gtrsim 0.66
\end{multline}
where the lower bound is achieved for $\kappa\simeq 1.48$. Hence indeed the modes of light and atoms are in an entangled state which can be used for teleportation.

The state to be teleported is encoded in the lower sideband $\omega-\omega_{L}$ with respect to the carrier frequency (Fig.~\ref{fig:telesetup}), expressed in terms of measurable cosine and sine modulation modes as
\begin{equation}\label{eq:inputstate}
Y=\frac{1}{\sqrt{2}}\left(\y{s}+\q{c}\right),\quad
Q=-\frac{1}{\sqrt{2}}\left(\y{c}-\q{s}\right),
\end{equation}
where $[Y,Q]=i$. The Bell measurement of the commuting observables after combining the entangled and the to-be-teleported states yields
\begin{align}
\tilde{X}_{\mathrm{c}}&=\frac{1}{\sqrt{2}}\left(\x{c}{out}+\y{c}\right),
&\tilde{X}_{\mathrm{s}}&=\frac{1}{\sqrt{2}}\left(\x{s}{out}+\y{s}\right),\label{measurement}\raisetag{-0pt}\\
\tilde{Q}_{\mathrm{c}}&=\frac{1}{\sqrt{2}}\left(\p{c}{out}-\q{c}\right),
&\tilde{Q}_{\mathrm{s}}&=\frac{1}{\sqrt{2}}\left(\p{s}{out}-\q{s}\right).\nonumber
\end{align}
Conditioned on these results the atomic state is then
displaced in order to get in the ensemble average the final state
\begin{align*}\label{eq:finalstateatoms1}
\xa{fin}&=\xa{out}+\tilde{X}_{\mathrm{s}}-\tilde{Q}_{\mathrm{c}}=(\xa{out}+\x{+}{out})+Y\\
\pa{fin}&=\pa{out}-\tilde{X}_{\mathrm{c}}-\tilde{Q}_{\mathrm{s}}=(\pa{out}-\p{+}{out})+Q.
\end{align*}
In the hypothetical case of vanishing EPR variances of $(\xa{out}+\x{-}{out})$ and $(\pa{out}-\p{-}{out})$ atoms would correctly display the statistics of the $Y,Q$ mode, reproducing any input state (coherent, Fock, etc) as desired. For the given minimal EPR variances of 0.66, that is for a variance of 0.33 in each EPR variable $\xa{out}+\x{+}{out}$ and $\pa{out}-\p{+}{out}$, teleportation will not be perfect, but still below the classical limit corresponding to the total EPR variance of 2.
 This discussion ignores atomic decay and light absorption which can be included as described in section \ref{sec:decoherence}.

The experimental implementation of this teleportation protocol by \textcite{Sherson:2006a} was performed with a new generation of paraffin coated cells filled with Caesium similar to those shown in Fig.~\ref{fig:cell}. Atoms are initially prepared in a coherent spin state by a $4$ msec circularly polarized optical pumping pulse propagating along the direction of the magnetic field, into the sublevel $F=4,m_{F}=4$ of the ground state (insert in Fig.~\ref{fig:telesetup}). Then an entangled light-atoms state is generated by sending a strong pulse polarized along the $y$ axis (Fig.~\ref{fig:telesetup}).  The initially vacuum state of the $x$ polarization of this pulse is populated after the interaction with the field $X_{L,{\rm out}},P_{L,{\rm out}}$ which is entangled with the atomic variables $X_{A,{\rm out}},P_{A,{\rm out}}$ according to \eqref{eq:Etele}. (Compared to the theoretical derivation of the Faraday interaction in Sec. \ref{sec:theory} the polarizations of the classical and quantum fields are interchanged. This is of minor importance since it merely swaps the vertical and diagonal transitions in Fig. \ref{fig:lambda}.)

The nearly optimal value of the coupling constant $\kappa\approx 1$ was achieved with $4\times 10^{13}$ photons in the strong pulse with the duration of $1$ msec and a crossection of $4.4$ cm$^{2}$ detuned by $825$ MHz and a number of atoms on the order of $10^{12}$ corresponding to the Cs temperature of $25^{\circ}$C. At Alice's location the mode of light entangled with atoms is combined with the input pulse to be teleported on a $50/50$ beamsplitter BS (Fig.~\ref{fig:telesetup}). The strong $y$ polarized pulse which travels along with the entangled quantum field conveniently serves as the local oscillator for the polarization homodyne measurements of the Stokes operators $S_{y}$ and $S_{z}$ performed at two outputs of BS. The output $\cos(\omega_{L}t)$ and $\sin(\omega_{L}t)$ components of the photocurrent are processed by the lock-in amplifiers to produce the feedback signals $\widetilde{Q}_{c,s},\widetilde{Y}_{c,s}$.
The two feedback signals at $\omega_{L}=322$ kHz phase shifted by $\pi/2$ with respect to each other are fed into the RF magnetic coils surrounding the atoms with a variable electronic gain. The gain is chosen so that the atomic variables are shifted by one vacuum unit if the input light mode contains one vacuum unit of excitation. This condition corresponds to the "unity gain" teleportation.

\begin{figure}[ht]
  \includegraphics[width=8.3cm]{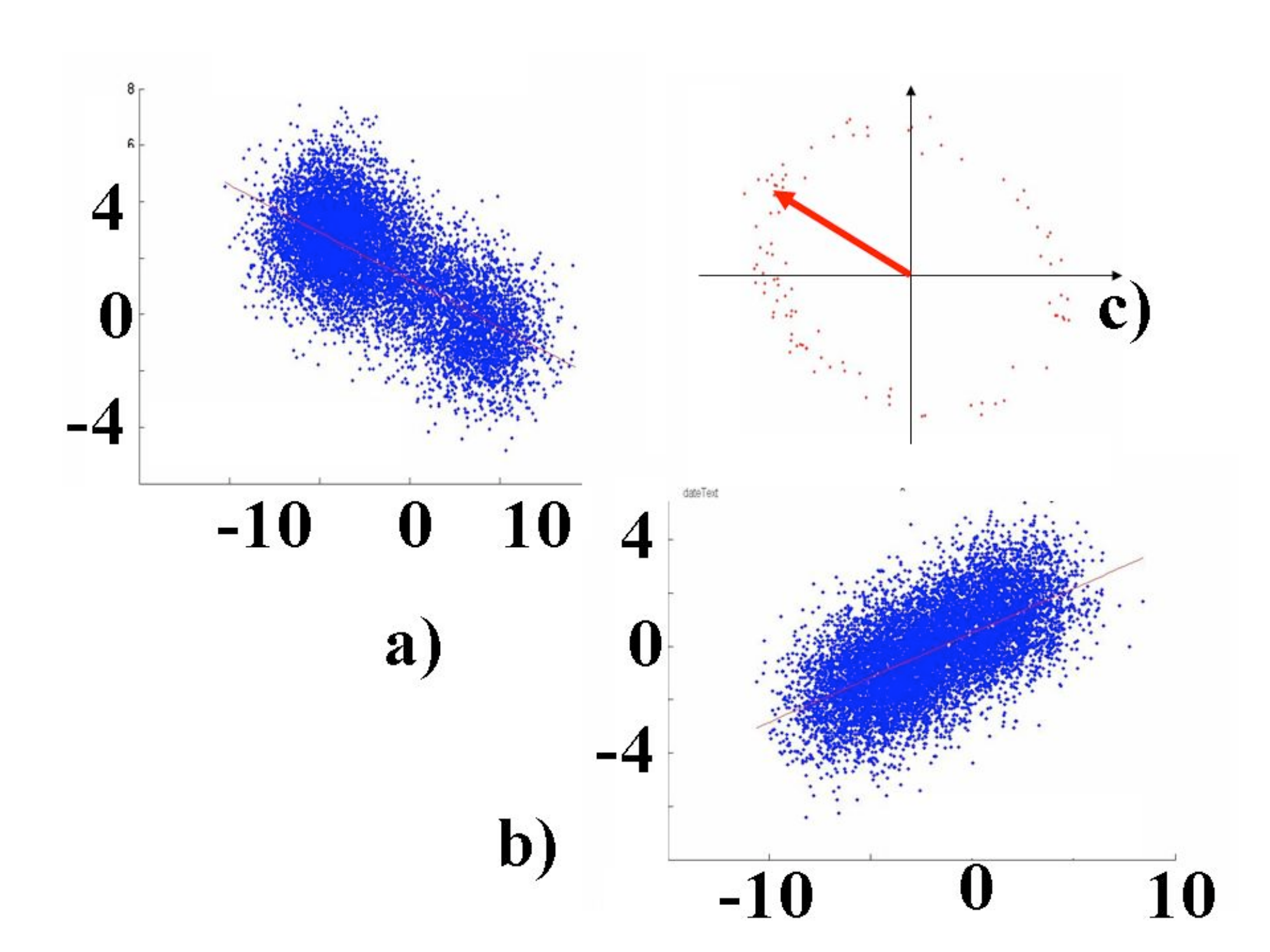}
  \caption{Teleportation results \cite{Sherson:2006a}. a) and b) Two canonical operators of the verifying pulse plotted as a function of the corresponding canonical operators of the input pulse for $2\cdot10^{3}$ realizations (vacuum units). The slope which is close to $1/2$ should be multiplied by $2$ to account for the attenuation of the verifying pulse on the beam splitter. From the variances of the distributions along the vertical axis the atomic state variances and the fidelity are obtained. c) A set of input states with $\langle n\rangle=5$ and random phases.}\label{fig:telecoh5}
\end{figure}

To prove the success of the teleportation protocol a strong verifying pulse reads out the atomic operators (collective spin projections). The same homodyne polarization measurement setup is used for that. An example of the atomic state readout is shown in Fig.~\ref{fig:telecoh5}. The gain of $0.95$ is found from the slope of the linear fit to the data, whereas the variances of the final atomic state after teleportation $\Delta X_{A,{\rm out}}=\Delta P_{A,{\rm out}}=1.2$ are found from the variance of the distribution of the data. These variances are below the classical teleportation limit corresponding to three units of vacuum noise, that is to $3/2$. Accordingly the fidelity of the teleportation calculated as
\begin{multline}
  F = (\bar{n}(1-g)^2 + 1/2 +\Delta X_{A,{\rm out}}^2)^{-1/2}\\
  (\bar{n}(1-g)^2+1/2+\Delta P_{A,{\rm out}}^2)^{-1/2}
\end{multline}
for a distribution of coherent states with a width of $\bar{n}$ is greater than the best classical fidelity. Indeed the fidelities of, e.g., $F=0.60\pm0.02$ and $F=0.58\pm0.02$ have been obtained for sets of input states with $\bar{n}=5$ and $\bar{n}=20$ respectively whereas best classical fidelities for these cases are $0.54$ and $0.51$.

As shown in the Supplementary Notes to \textcite{Sherson:2006a} and in \citet{Hammerer:2006a} the knowledge of $\Delta X_{A,{\rm out}},\Delta P_{A,{\rm out}}$ for the teleportation of the coherent states corresponds to the complete knowledge of the teleportation map and allows to calculate the fidelity of the qubit teleportation as
\begin{multline}\label{eq:fidelityqubit}
  F_{q}=(6+16s^{2}+24s^{4}+4(g-1)(1-2s^{2})+\\
  (g-1)^{2}(1-6s^{2}))/6(1+2s^{2})^{3},
\end{multline}
where $s^{2}=2\Delta X_{A,{\rm out}}^{2}-1$. The quality of mapping for the two canonical operators is assumed to be equal $\Delta X_{A,{\rm out}}=\Delta P_{A,{\rm out}}$. Direct demonstration of the qubit teleportation under the conditions of \textcite{Sherson:2006a} has not been possible due to the absence of a light qubit source with the pulse duration of $1$ msec. As shown theoretically in \citet{Sherson:2006a}, a qubit fidelity of $F_{q}=0.74$ is achievable for $\kappa=1$.

The protocol can in principle be improved by properly taking back-action modes, treated here simply as noise terms, into account. This is to a large extent a question of detector bandwidth and improved post processing of photocurrents. In \textcite{Hammerer:2005} it is shown that in this way the fidelity can be increased up to 80\% corresponding to half a unit of added vacuum noise. This noise stems from the initial vacuum fluctuations of light before light-atom interaction which can in turn be reduced by using a squeezed light for entanglement with atoms. One ends up with a similar situation as in the standard protocol \cite{Braunstein:1998,Vaidman:1994}: the quality of teleportation is in the end limited by the amount of available squeezing, which emerges here once more to be an irreducible resource \cite{Braunstein:2005a}.

A number of alternative proposals for deterministic teleportation involving light and atomic ensembles have been suggested. \citet{Horoshko:2000,Mista:2005} study the application of the state of light and atoms created in a QND interaction, cf. \eqref{eq:faraday_nodecay_result}, as a resource for teleportation and show that the use of squeezing of light and atoms as well as unbalanced beam splitters in the Bell measurement can improve the fidelity. Teleporation of states of light to atoms, based on entanglement between motional degrees of freedom of a Bose-Einstein condensate and light 
has been suggested by \citet{Cola:2004,Paris:2003}.


%
%

\section{Errors and fidelity for different interfaces
}
\label{sec:errors}
\label{sec:inhom}

Despite impressive successes the unconditional fidelity and/or efficiency of the quantum interfaces demonstrated so far does not exceed $70$ \% and in many cases is at the level of $20$ \%. Several factors, some more fundamental and some more technical contribute to this.

\paragraph*{Scaling with optical depth.} In the theory section we have shown that spontaneous emission can be avoided for large optical depth $d\gg1$. Here we discuss the errors due to a finite $d$ for different approaches. In the entanglement section we have shown that the single ensemble squeezing  $\Delta P_A^2$ or the two ensemble correlation $\Delta_{EPR} $ is $\sim 1/\sqrt{d}$. This limit, however, depends on the exact decay mechanism, and a better scaling of single ensemble squeezing $1/d$ can be achieved if the QND interaction through phase shift measurements using two closed transitions (Sec. \ref{sec:other_strat}), is used, since in this case spontaneous emission does not add noise to $P_A$.

For a quantum memory the inefficiency of storage with the beam splitter interaction (Raman or EIT) scales as $1/d$ if suitably shaped spatial mode functions are used. If storing or reading out the spatially symmetric mode is preferable, as in the repeater case where the probabilistic entanglement is generated in this mode (Sec. \ref{sec:probabilistic}), the scaling is again weak, $1/\sqrt{d}$ \cite{Gorshkov:2007a,Gorshkov:2007c}. However, a much better performance can be achieved with ensembles in optical cavities \cite{Gorshkov:2007b}, where the scaling is $1/\mathcal{F}d$ ($\mathcal{F}$ is the cavity finesse), see in particular the experiments by \citet{Simon:2007}, \citet{Simon:2007b}. The protocols based on the Faraday interaction naturally couple to the symmetric modes and may be better suited for this mode than the beam splitter interaction. These memory protocols have a coupling constant $\kappa^2 \sim 1$, which yields $\sim 1/d$ or $\sim \log(d)/d$ \cite{Sherson:2006} error due to  spontaneous emission .

Different protocols are also characterized by different \emph{kinds} of errors. The protocols based on the single pass QND - Faraday interaction have variable gain, faithfully reproduce mean values of the input states, and thus give good fidelity over a large phase space.  The error in the beam splitter protocols corresponds to a loss on a beam splitter and thus gives bad fidelity for states with a large amplitude.
On the other hand the low fidelity beam splitter (Raman and EIT) protocols usually add only vacuum noise, whereas single pass QND-based protocols add multi-photon errors.
Which kind of error is less harmful depends on a particular application. For instance \textcite{Brask:2008} show that memories based on the single pass QND - Faraday interaction are not very well suited for the DLCZ-repeater protocol \cite{Duan:2001}, because the repeater protocol is specifically designed to correct only for the photon loss errors. A full evaluation of the performance thus depends on the particular application one has in mind.

\paragraph*{Optical losses}
Losses of photons obviously affect the performance of memory, and different memories are affected to a different extent. Clearly, in the protocols insensitive to the vacuum component optical losses only lead to lower efficiency, that is to a lower probability of success. On the contrary, if the goal of a protocol is an     unconditionally high fidelity, as in the case of protocols based on homodyning, optical losses directly affect the fidelity. Beside usual losses due to reflection on windows of cells and chambers which can be reduced by coating, there are losses due to absorption of light by atoms of the memory. These losses were included already in our theoretical discussion in  Sec.~\ref{sec:decoherence}, where we found that the  decoherence of the quantum fields and the atoms vanished in the limit of large $d$. In addition one should also account for the damping of the classical fields. In Sec.~\ref{sec:decoherence} we also discussed that the probability of photon absorption $\eta_L$ is linked to the probability of spontaneous emission of an atom $\eta_A$ by $\eta_L=\eta_A N_{A}/N_{L}$ where $N_{L}, N_{A}$ are the number of photons of the driving field and the number of atoms respectively. If $N_{L}\gg N_{A}$ can be satisfied then the photon absorption can be made very small.

\paragraph*{Inhomogeneous broadening}
 Another source of possible errors is inhomogeneous broadening of the optical transitions. Solid state systems have a strong inhomogeneous broadening because each of the emitters sits in a slightly different environment, while inhomogeneous broadening in room temperature atomic ensembles is due to the Doppler broadening of the atomic lines.
In the theoretical derivation we only assumed a homogeneous broadening of the optical transitions (the photon echo approach discussed in Sec. \ref{sec:echo} is a notable exception). One therefore cannot just replace the optical depth $d$ appearing in the formulas derived in the theory section by the measured optical depth in the presence of inhomogeneous broadening.

A detailed study of the effect of inhomogeneous broadening for a quantum memory using the beam splitter interaction is presented by \textcite{Gorshkov:2007d} who have shown that even far off-resonance the inhomogeneous broadening of the line still plays a role. The reason is that the strong field leads to a considerable AC-Stark shift of the ground state $|1\ra$ even far off resonance. The inhomogeneous broadening introduces  energy shifts which are different for each atom and thus causes decoherence of the collective states. As a result the effect of inhomogeneous broadening is as severe off resonance as it is on resonance. Hence the total inefficiency of the memory protocol has two contributions. The first is the spontaneous emission which 
scales as $1/d_{{\rm hom}}$, where $d_{{\rm hom}}$ is the optical depth of the ensemble without the broadening. The second contribution is from the inhomogeneous broadening and scales as $1/d_{{\rm inhom}}^2$, where $d_{{\rm inhom}}$ is the actual optical depth in the presence of broadening (this scaling assumes that the broadened lines fall off sufficiently fast; for other profiles, e.g., Lorentzian, the scaling is $1/d_{{\rm inhom}}$).  If one increases the length of the sample the system will eventually be dominated by the $1/d_{{\rm hom}}$ contribution and thus behaves as if it were homogeneously broadened.

The imperfections induced by inhomogeneous broadening discussed above, can to some degree be reduced by taking a more {\it active} approach where one tries to engineer the broadening. In essence the photon echo approaches discussed in Sec. \ref{sec:echo} provide examples of such engineering of inhomogeneous broadening, where the externally imposed inhomogeneous broadening becomes a useful resource.  An interesting example of  engineering of inhomogeneous broadening is presented by
\textcite{Afzelius:2008}, who considers engineering a frequency comb in the atomic line shape, e.g. using the hole-burning techniques discussed in Sec. \ref{sec:solid}. Essentially this allows for exploiting atoms in a solid state medium which have their resonance frequency shifted far away by inhomogeneous broadening (thus reducing the  $1/d_{{\rm hom}}$ error discussed above), while avoiding the detrimental effects of inhomogeneous broadening (thus reducing the $1/d_{{\rm inhom}}^2$ error)

For the Faraday interaction the situation is different than for the beam splitter interaction. The Faraday interaction is only employed far off resonance, and unlike the case of the beam splitter interaction the AC-stark shift does not cause decoherence. The strong classical light couples to both ground states and shifts them by the same amount. As a result the spin dynamics is not affected by a difference in the level shifts, and the Faraday interaction becomes insensitive to the inhomogeneous broadening for detunings much larger than the hyperfine structure of the excited state and the Doppler width.

In addition to broadening of the optical transitions, inhomogeneous broadening also affects the ground states, e.g., due to magnetic field gradients. This broadening is particularly bad because it leads to decoherence during the  period when information is stored in the ground states of the ensemble, and is a major limitation for the coherence time in many experiments.

\paragraph*{Atomic motion}
As discussed in Sec. \ref{sec:atomic_media} the atomic motion in and out of the optical beam can severely affect the performance of the interface. For the Faraday interaction a strong suppression of this effect is achieved  in the experiments \textcite{Julsgaard:2001,Julsgaard:2004,Sherson:2006a} 
because the light beams cover most  of the atomic volume and the interaction time is much longer than the atomic transient time of flight. This suppression is, however,  not perfect. As analyzed in detail in \textcite{Sherson:2006c} the fact that atoms move accross the interaction volume leads to extra spin noise which should be accounted for when the projection noise level is being established.

For experiments with cold atoms the atomic motion can still be of a problem, in particular when spin waves with very short wavelength are created in the memory.

\paragraph*{Atomic collisions}
For atomic gasses it is also important to consider the effect of atomic collisions. For experiments with paraffin coated cells atom-atom collisions contribute up to $20-40$ Hz to the ground state decoherence  \cite{Sherson:2006c}.
In experiments where a buffer gas is used to suppress atomic motion, the alkali atoms - buffer gas collisions often have little effect on the ground state coherence of the atoms and the memory time. The buffer gas does, however, change the dynamics during the interaction with the light. The homogeneous broadening due to collisions with the buffer gas can be included in the theory by modifying the homogenous line width $\gamma$ \cite{erhard:2001}. At the same time collisions change the velocity of the atoms and thereby their Doppler shift. The collisions thus change the phase spreading which causes decoherence of the collective states from being ballistic to diffusive, and this reduce the inefficiency from the inhomogeneous broadening.

For the write stage of the probabilistic entanglement protocol a different effect related to the collisional broadening has been observed by \textcite{Manz:2007} and has been discussed theoretically in a different context \cite{Childress:2005}. The collisions with buffer gas atoms cause the alkali atoms to emit photons at the resonance frequency of the atoms rather than at the anti-Stokes frequency.  To observe the entanglement it is thus necessary to filter out these incoherent photons with a frequency filter. This effect, however, has little consequences for other protocols.



\paragraph*{Geometry of the ensemble.}
 In most cases the two ground states of the ensemble are nondegenerate, so there is an additional phase factor $\exp[i(k_0'-k_0)z]$ associated with the difference of the $k$ vectors for the classical and quantum fields. This phase can be absorbed into the definition of the mode functions $u_m(\rvec,t)$ in (\ref{eq:aatom_transverse_lambda}) and (\ref{eq:aatom_transverse_gain}), and does not play a role for the beam splitter and parametric gain interaction applied separately. However, if one reads out the memory in the backward direction, which is sometimes advantageous, or combines the two protocols, the atomic operators should be redefined and
 this phase can have a detrimental effect \cite{Andre:2005,Gorshkov:2007c,Duan:2002,Surmacz:2008}, see also Sec. \ref{sec:raman_eit}. For the Faraday interaction on the other hand, the mode functions must have a constant phase and this means that the atomic ensemble must be much smaller than the wavelength corresponding to the ground state splitting. In practice this wavelength varies from $\sim100$ m in case of Zeeman splitting with $B\simeq1$ Gauss to $\sim3$cm in case of hyperfine splitting. In addition the Fresnel number corresponding to the shape of the atomic ensemble must be large in order to stay within a single transverse spatial mode approximation for the Faraday interaction.


\paragraph*{Deviation from a two-level ground state model.}

When the ground state level used for the interface is magnetically degenerate with more than two states and the detuning of the strong field is not sufficiently larger than the hyperfine splitting of the excited state, the interaction of the Faraday type ($a_{1}$ vector term in Sec.~\ref{sec:real}) is modified with the Raman interaction ($a_{2}$ tensor term). The Hamiltonian becomes $\hat{H}=\chi_{BS}\hat{a}_{L}\hat{a}_{A}^{\dag}+ \chi_{P}\hat{a}_{L}\hat{a}_{A}+ {\rm H.C.}$.  Various aspects of this effect have been considered in \textcite{Julsgaard:2003}, \textcite{Kupriyanov:2005}, and \textcite{Sherson:2006c}, as well as in \textcite{Geremia:2006}. Whereas in early work this effect was considered as a source of imperfections, lately it became a subject of intensive studies as a new resource for interfaces \cite{Wasilewski:2009}.

The beam splitter interaction (EIT) experiments mostly use two ground hyperfine levels. Magnetic degeneracy leads to imperfections which can be reduced by careful optical pumping and polarization filtering \cite{Choi:2008}.

%
%
\section{Outlook}
\label{sec:outlook}

The light--matter quantum interface, a term coined in the end of 1990s, is one of the pillars of the field of Quantum Information Processing and Communication (QIPC) \cite{Zoller:2005}. In less than a decade since the first demonstrations of a quantum interface between light and an atomic ensemble the ensemble approach has become one of the most active areas of research in the field. The interactions which seem most promising for interfaces at the moment are discussed in this review: the QND - Faraday and Raman interactions, EIT, and photon echo.

Both fundamental and application driven aspects are obvious within this approach. One of the interesting fundamental issues concerns the multi-particle entanglement necessarily present in the ensemble-based approach. At the same time the interface is a kind of a quantum channel, hence its relation to the theory of quantum channel capacity  should be explored in the future. This issue is connected to the fidelity of the interface since it is known that a quantum channel with $F>2/3$ for coherent states has a non-zero quantum capacity \cite{Wolf:2007,Grosshans:2001}. Quantum interface also allows for storing optimal quantum clones of a state of light as proposed by \textcite{Fiurasek:2004}.
Another feature which is intrinsic for the ensembles of atoms is their multi-mode capacity. This multimode capacity comes both in form a different temporal light shapes which are mapped into different longitudinal or spectral atomic modes \cite{Fleischhauer:2002,Simon:2007a,Nunn:2008,Afzelius:2008} as well as different transverse modes or "quantum holograms" \cite{Vasilyev:2007,Surmacz:2008}, see also \textcite{Tordrup:2009}. Along the lines of the latter the first experiments demonstrating storage of classical images via the EIT approach have recently appeared \cite{Vudyasetu:2008,Shuker:2008}.

Long distance quantum communication is one of the most actively pursued applications of the interface at the moment. It is based on the combination of probabilistic entanglement generation and deterministic entanglement swapping - a quantum repeater with atomic ensembles \textcite{Duan:2001}- which  may serve as the basis for a "quantum internet"
\cite{Kimble:2008}.

New applications for quantum memories and interfaces, such as, e.g., quantum voting and surveying \cite{Vaccaro:2007,Hillery:2006} should be explored.

Advanced architectures for quantum computing may be enabled by highly efficient photon-based connections between small scale atomic processing nodes \cite{Jiang:2007}.
 An interesting  direction in this respect is the combination of photon counting and QND-Faraday continuous variable measurement techniques. It allows to combine the best of the two worlds: the high efficiency of the homodyne measurement and the non-Gaussian states, such as Schr\"{o}dinger cat states which can be generated by photon counting \cite{Massar:2003,Genes:2006}. Progress along these lines depends critically on the development of highly efficient photon counters and photon number resolving detectors \cite{Waks:2006,Achilles:2004}.

For new  applications of quantum interfaces a major challenge for experimentalists  will be to improve the fidelity and efficiency of the interface, and for theorists to find protocols where atomic memories with the fidelity and efficiency at the level of $90-95 \%$ - the likely levels to be achieved within the next few years - can help to achieve goals impossible with classical interfaces.


Besides the fiducial write- and read-processes and long storage times, a quantum interface between light and matter will likely have to show yet another key element: the possibility to process stored quantum information and to allow for quantum logical gate operations for active entanglement purification and error correction \cite{Dur:2007}. Theoretical studies of the requirements on gate operations in quantum repeater architectures have been performed in great detail by \citet{Briegel:1998,Dur:1999,Hartmann:2007}, and \citet{Dorner:2008,Klein:2006} recently investigated the usage of decoherence free subspaces in quantum communication. Proposals for ensemble-based implementations unifying an efficient light matter interface, stable quantum memory and reliable small scale quantum processors are rare. A number of theoretical studies suggest gate operations on stored collective excitations via a Rydberg blockade mechanism \cite{Pedersen:2009,Lukin:2001,Brion:2007,Petrosyan:2008}, via an EIT enhanced optical nonlinearity \cite{Lukin:2001a,Andre:2005a,Ottaviani:2003,Wang:2006} or in hybrid systems, such as in \cite{Rabl:2006}, where a Cooper pair box serves as a saturable, nonlinear element. Initial experiments along those lines are in progress. 


 Efficient ensemble-based quantum memories and matter-light interfaces, small scale quantum processors for error correction, repeaters, possibly satellite based quantum communication \cite{Aspelmeyer:2003,Pfennigbauer:2005}, and even hybrid systems \cite{Hammerer:2009} -- are the goals of today. The research performed towards this end is of both fundamental interest for our understanding of quantum physics and of technological importance. Its highly interdisciplinary character encompasses a broad spectrum of fields in physics as well as in computer science and information theory.

\section*{Acknowledgments}
%
%
We are grateful to our colleagues with whom we have had collaboration and many useful discussions on the subject of quantum interfaces over the past years. In particular we would like to thank

N. Cerf,
J. I. Cirac,
J. Fiur\v a\'sek,
M. Fleischhauer,
A. V. Gorshkov,
H. J. Kimble,
D. Kupriyanov,
A. Kuzmich,
M. Lukin,
S. Massar,
J. H. M\"{u}ller,
K. M{\o}lmer,
I.V. Sokolov,
R. Walsworth, and
P. Zoller.
E.P. would like to especially thank the enthusiastic and talented members of his experimental team for whom the quantum interface has been a buzz word for the last decade.
We acknowledge the financial support by the Danish National Research Foundation, by the Sixth and the Seventh Framework Programmes for Research of the European Commission under Future and Emerging  Technologies (FET) grant agreements COVAQUIAL, QAP, COMPAS, CONQUEST, EuroSQUIP, and HIDEAS (FP7-ICT-221906), by the Austrian FWF through SFB FOQUS and by the Institute for Quantum Optics and Quantum Information of the Austrian Academy of Sciences. 
%

\section{Appendices}

\subsection{Adiabatic Elimination}
\label{sec:adiabat}
The effective ground state Hamiltonian can be obtained by adiabatic elimination.
 We start with the dipole Hamiltonian $H_{\rm int}=-\vec{E}\cdot \vec{D}
$.
 For magnetic sub-levels $\{ |g_m\ra \}$ all matrix elements of the ground state dipole operator vanish  $\la g_m|\vec{D}|g_{m'}\ra=0$, and similarly for the excited state $\la e_m|\vec{D}|e_{m'}\ra=0$. Introducing positively and negatively oscillating components $\vec{D}=\Dplus+\Dminus$ and $\vec{E}=\eplus+\eminus$ we obtain in the rotating wave approximation
 \begin{equation}
H_{\rm int}=-(\eminus\cdot \Dplus+ \Dminus\cdot \eplus),
\label{eq:Hrot}
\end{equation}
where  the first (second) term describe down (up) transitions. Expanding the Hamiltonian we obtain
\begin{equation}
H_{\rm int}=-\sum_{m,m'} \eminus\cdot \Dplus_{mm'}  |g_m\ra\la e_{m'}| + {\rm H.C.}
\label{eq:ham_rot_wave}
\end{equation}
Using the Hamiltonian $H_{\rm A}+H_{\rm int}$ we obtain
\begin{equation}
\begin{split}
\frac{d}{dt}& |g_m\ra\la e_{m'}|=
 -i\Delta_{m'}
  |g_m\ra\la e_{m'}| \\
&+  i \eplus \sum_{m''} {\left(\Dminus_{m',m''} |g_m\ra\la g_{m''}| -\Dminus_{m'',m} |e_{m''}\ra \la e_{m'}| \right)}
\label{eq:deriv_polarization}
\end{split}
\end{equation}
For weak excitation (far below saturation)  we
can neglect the excited state operator $|e_{m''}\ra \la e_{m'}|$. Next, assuming the dynamics is slow
compared to the detuning $\Delta$, we ignore the left hand side compared to the first term on the
right hand side. These approximations are valid provided that
$\eplus\cdot \Dminus_{m,m'} \ll 
\Delta_{m'}
$.  In this limit we
obtain
 \begin{equation}
\begin{split}
|g_m\ra\la e_{m'}|\approx +\sum_{m''}\frac{\eplus\cdot\Dminus_{m',m''}}{
 \Delta_{m'}
 } |g_m\ra\la g_{m''}|.
\end{split}
\label{eq:AdiabaticPolarization}
\end{equation}

To obtain an effective ground state Hamiltonian we substitute the expression
(\ref{eq:AdiabaticPolarization}) back into the Hamiltonian $H_{\rm A}+H_{\rm int}$. Note that we have to choose normal ordering of the operators in Eq. (\ref{eq:ham_rot_wave}) when we insert Eq. (\ref{eq:AdiabaticPolarization}). A detailed discussion of this issue may be found in \textcite{Barnett:1997}.
Secondly, since the terms in $H_{{\rm A}}$ involve the excited state
population one could be tempted to ignore it since the population is
proportional to $1/
\Delta_m
^2$. On the other hand there is
also a factor $\Delta_m
$ in front of this term, so that in
total this is on the order of $1/
\Delta_m
$  and we have to include it.  To treat this term we introduce any
intermediate state $|g_0\ra$
\begin{equation}
|e_{m'} \ra\la e_{m'}|=|e_{m'} \ra \la g_{0}| \cdot |g_{0}\ra \la
e_{m'}|
\end{equation}
and substitute the result (\ref{eq:AdiabaticPolarization}) for
$|g_{0}\ra \la e_{m'}| $. By doing this we arrive at the effective
Hamiltonian  in Eq. (\ref{eq:Hadiabat}).

\subsection{Three-dimensional Hamiltonians}
\label{sec:3d_ham}
In the main text we present the general expression  (\ref{eq:general_3d_ham}) for the full  three-dimensional Hamiltonian. In this appendix we discuss the 3D Hamiltonians for the three model system, beams splitter, parametric gain, and Faraday interaction.

First we consider the atomic beam splitter interaction in \ref{fig:lambda}(a). We assume that the state $|1\ra$ is coupled to an excited state by a classical field, so that $\vec{E}(\rvec)=\la \vec{E}(\rvec)\ra$. Inserting the expression for the quantized electric field in Eq. (\ref{eq:E_multi}) we obtain
\begin{equation}
\begin{split}
H_{BS}=\int  &d^3\rvec\ {\Bigg [}  \frac{-|\Omega(\rvec,t)|^2}{4\Delta
}a_A^\dagger(\rvec)a_A (\rvec)
\\ -& \frac{|g(\rvec)|^2}{\Delta
}\sum_m |u_m(\rperp;z)|^2 a^\dagger_{L,m}(z)a_{L,m}(z)\\
-&
 \sum_m{\Bigg (}\frac{g^*(\rvec)\Omega(\rvec,t)}{2\Delta}   u_m^*(\rperp;z) \\ &  \hspace{1.5cm}
 \times a^\dagger_{L,m}(z)a_A(\rvec)
 + {\rm H.C.}{\Bigg )}{\Bigg ]}
,
\end{split}
\label{eq:H_lambda_3}
\end{equation}
where the coupling constant $g(\rvec)$ and resonant Rabi frequency are defined as in Eq. (\ref{eq:coupling}) except  with $z$ replaced by $\rvec$.
The first term in this Hamiltonian is the AC-Stark of the ground state caused by the classical field, the second term is the change in the index of refraction that the quantum field experiences, and the last term represents the exchange of excitation between the light and the ground state coherence.
 We have here ignored the small AC-Stark shift caused by the weak quantum field.

In case of the parametric gain interaction shown in Fig. \ref{fig:lambda}(b), the fields interact to the transitions which are flipped compared to the case of the beam splitter interaction. The emission of a photon is therefore coupled to an atomic  transition in the opposite direction and we obtain the parametric Hamiltonian by making the replacement  $a_A (\rvec)\leftrightarrow a^{\dag}_A (\rvec\,)$ in the coupling between atoms and light. In addition, the effective  AC-Stark shift has the opposite sign. The Hamiltonian thus reads
\begin{equation}
\begin{split}
H_G=\int  &d^3\rvec\  {\Bigg [} \frac{|\Omega(\rvec,t)|^2}{4\Delta 
}a_A^\dagger(\rvec\,)a_A (\rvec\,)
\\
-& \sum_m{\Bigg (}\frac{g^*(\rvec\,)\Omega(\rvec,t)}{2\Delta}   u_m^*(\rperp;z) \\ & \hspace{2.5cm}a^\dagger_{L,m}(z)a_A^\dagger(\rvec\,)+ {\rm H.C.}{\Bigg )}{\Bigg ]}.
\end{split}
\label{eq:H_gain_3}
\end{equation}
 There is no index of refraction here since now the quantum field couples to the almost empty transition. On the other hand the classical field now sees the atomic population, and we should include the index of refraction in the propagation of the classical light, which enters through the phase of $\Omega$.

 For the Faraday interaction we assume a strong $x$-polarized classical field, and consider the quantum field in the $y$-polarization as shown in  Fig. \ref{fig:lambda}(c). There are now two paths which lead to creation of a  photon: photon creation (or absorption) can appear both through the creation and annihilation of an atomic excitation. As discussed in the main text the Faraday interaction is a combination of the two Hamiltonians in Eqs. (\ref{eq:H_lambda_3}) and (\ref{eq:H_gain_3}) with equal weights. For the $1/2$--$1/2$ transition depicted in Fig. \ref{fig:lambda},
however, the two paths leading to the creation of a photon have opposite signs due to the Clebsch-Gordon coefficients. The coupling constant $g$ is defined for circularly polarized light, so we should include  a factor of $\sqrt{2}$ coming from the expansion of the $y$--polarized quantum field in the circular polarization basis.
The Hamiltonian is then given by $H_F=(H_{BS}-H_G)/\sqrt{2}$.

For the spin $1/2$ system classical and quantum fields experience the same index of refraction, so we can remove the index of refraction from $H_{BS}$ if we also ignore it for the classical field.
An important result for this Hamiltonian is that the AC-Stark shift of the ground state disappears, because the classical field shifts the two atomic states by the same amount.

\subsection{Propagation equations for light}
\label{sec:propagation}

 The light Hamiltonian (for a single transverse mode) is given by
\begin{equation}
 H_L=\sum_k |k|c\, a_k^\dagger a_k,
\end{equation}
 where the sum is over all the longitudinal wave vectors. We now derive the time evolution of the operator $a_L(z)$ defined in Eq. (\ref{eq:a_light_z}). If we assume that the modes contributing to the slowly varying operator $a_L(z)$ are centered around a large positive  value $k_0$, we can ignore the absolute value of the wave vector.
Combining the time derivative caused by the explicit time dependence introduced in  (\ref{eq:a_light_z}) with the  time evolution caused by $H_L$ we find the Heisenberg equation of motion
 \begin{equation}
\frac{\partial a_L}{\partial t}=i\omega_0 a_L-i[a_L,(H_{{\rm L}}+H_{{\rm int}})]=-c \frac{\partial}{\partial z} a_L-i[a_L,H_{{\rm int}}].
\end{equation}

To simplify the equations we introduce a rescaled time variable by defining new operators $\tilde{a}_L(z,\tau)=a_L(z,t=\tau+z/c)$ and $\tilde{a}_A(z,\tau)=a_A(z,t=\tau+z/c)$. The propagation equation for the light can then be simplified by using
\begin{equation}
{\left(\frac{\partial}{\partial t}+c\frac{\partial}{\partial z}\right)} a_L(z,t=\tau+z/c)=c \frac{\partial}{\partial z} \tilde{a}_L(z,\tau)
.
\end{equation}
We also take the classical field to be moving in the positive $z$-direction, in which case the Rabi-frequency is $\Omega(z,t)=\Omega(t-z/c)=\Omega(\tau)$. The equations of motion in the main text  always use this rescaled time and we omit the tilde on the operators and simply denote the rescaled time $\tau$ by $t$.

For the parametric gain the $z$ dependence of the Rabi-frequency does not disappear completely from the equations of motion since we should include the $\exp(i\int dz' \ |g(z')|^2/\Delta
)$ dependence associated with the index of refraction for the classical field. For the Faraday interaction the same factor appears in the propagation of the quantum field and both factors can be omitted.

\subsection{Inclusion of spontaneous emission}
\label{sec:decay_derivation}
In Sec. \ref{sec:model_atoms} we derived equations of motion ignoring spontaneous emission. To include spontaneous emission we should modify Eq. (\ref{eq:deriv_polarization}) such that it contains a decay as well as the Langevin noise operators associated with the decay. The general treatment of the Langevin noise operators is quite complicated because they depend on details of the decay mechanism. In particular, the equations for the parametric gain and Faraday interaction depend on whether the atoms end up in the states $|0\ra$ and $|1\ra$ or some auxiliary states $|a_m\ra$ (Fig.~\ref{fig:lambda}). For the beam splitter interaction, on the other hand, the precise state that the atoms  decay  to is less important \cite{Gorshkov:2007b,Gorshkov:2007c}. Here we will for simplicity only consider the decay to some auxiliary states $|a_m\ra$.

To describe spontaneous emission we assume that each of the excited states $|e_m\ra_j$ of the $j$th atom couples to some state $|a_m\ra_j$ via a continuum of modes described by the annihilation operators $b_j(\omega)$. We assume that each atom couples to its own continuum. By doing so we ignore collective scattering effects such as superradiance and Bragg scattering, and we also ignore dipole-dipole interactions  mediated by other than forward modes. Whether this is a suitable approximation should be evaluated for each particular realization, but to our knowledge little work has been done on this subject.  The decay of the $j$th atom may then be described as
\begin{equation}
\begin{split}
H_{{\rm decay}}^j=&\int d\omega {\bigg [}b_j^\dagger(\omega) b_j(\omega) \\
&+ \rho(\omega)\sum_m {\left( g_m(\omega) |e_m\ra_j\la a_m| b_j(\omega)+ {\rm H.C.} \right)}{\bigg ]},
\label{eq:Hdecay}
\end{split}
\end{equation}
where $g_m(\omega)$ is the coupling constant of the $m$th excited state, and $\rho(\omega)$ is the density of states of the continuum. To arrive at the equations of motion for the atomic operators we first derive the equations of motion for $b_j(\omega)$ using the Hamiltonian (\ref{eq:Hdecay}). We then formally solve this equation by integrating over time and substitute the result into the equation for $|g_m\ra_j\la e_m|$. In the Markov approximation \cite{Barnett:1997}  Eq. (\ref{eq:deriv_polarization}) acquires the additional terms
 \begin{equation}
 \frac{d}{dt}|g_m\ra_j\la e_{m'}|= .... -\frac{\gamma_{m'}}{2}|g_m\ra_j\la e_{m'}|+\sqrt{\gamma_m} F^j_{gm,em'}(t),
\label{eq:decay_pol}
\end{equation}
where we have ignored the Lamb shift. The decay and noise operators are given by
\begin{equation}
\gamma_m= 2\pi |g_m(\omega_m)|^2\rho(\omega_m)
\end{equation}
and
\begin{equation}
\begin{split}
F^j_{gm,em'}(t)=&\frac{-i|g_m\ra_j\la a_{m'}|}{\sqrt{\gamma_{m'}}}\\
&\times \int d\omega g_{m'}(\omega) b_j(\omega,t=0) {\rm e}^{-i(\omega-\omega_{m'})t}
\end{split}
\end{equation}
with $\omega_m$ being the transition frequency for the $|e_m\ra$ to $|a_m\ra$ transition.
Note that if there is a difference between the population decay rate and twice the decay rate of the polarization, e.g., due to collisional broadening, the decay rate $\gamma_{m}/2$ appearing here should be the polarization decay rate.
The correlation functions within the Markov approximation are
\begin{equation}
\begin{split}
\la {F^{j\dagger}_{gm,em'}} F^{j'}_{gm'',em'''}\ra&\approx 0\\
\la F^j_{gm,em'} {F^{j'\dagger}_{gm'',em'''}}\ra&\approx \delta(t-t')\delta_{m',m'''}\delta_{j,j'} \la |g_m\ra\la g_{m''}|\ra.
\end{split}
\end{equation}
Since most of the atoms are always in the state $|0\ra$ the only important noise operators are $F_{g0,em}$.

The equations of motion that we derive in Sec. \ref{sec:model_atoms} are due to the time evolution of  $a_L$ and  the time evolution of $j_+(\rvec\,)$ used in the definition of $a_A$ in Eq. (\ref{eq:cont_aatom}).  Because the decay simply adds a term in Eq. (\ref{eq:decay_pol}) which is similar to the detuning term in Eq. (\ref{eq:deriv_polarization}) we can  obtain the equations of motion for these operators by using the replacement  in Eq. (\ref{eq:substitute_delta}). Note that for operators $|g_{m}\ra\la e_{m'}|$  the detuning and decay appears in the combination $\Delta-i\gamma/2$ so that, e.g., for the ground state operators $|g_m\ra\la g_{m'}|$ coupling to  $|g_{m}\ra\la e_{m''}|$ one should use the minus sign in the substitution. On the other hand if the coupling is to operators $|e_{m''}\ra\la g_{m'}|$ one should use the plus sign. In order to conserve the commutation relation for the operators $a_A$ and $a_L$ one should also include  the Langevin noise operators and the time derivative of $\langle j_x(\rvec\,)\ra$ used in the definition of $a_A$ (\ref{eq:cont_aatom}).

 For the beam splitter interaction the time derivative of $\langle j_x(\rvec\,)\ra$ can be ignored because the strong classical field couples to an almost empty transition. The equation of motion can therefore be obtained simply by making the substitution (\ref{eq:substitute_delta}). The equation of motion for the light (\ref{eq:lambda_nodecay}) arises from the commutator of $a_L$ with the Hamiltonian  (\ref{eq:Hrot}). In the resulting equation of motion $a_L$ couples to $|0\ra\la e|$ and we should therefore use the minus sign in Eq. (\ref{eq:substitute_delta}). The atomic annihilation operator is proportional to $|0\ra\la 1|$, which couples to $|0\ra\la e|$ and $|e\ra\la 1|$. The contribution from $|e\ra\la 1|$, is however very small and has been neglected in Eq. (\ref{eq:lambda_nodecay}), and we should once again use the minus sign in the substitution (\ref{eq:substitute_delta}).  Ignoring the noise operators we arrive at the equations of motion (\ref{eq:lambda}). Including the noise operators gives the additional terms
 \begin{equation}
 \begin{split}
\frac{\partial}{\partial z} a_L(z,t)=& ...+ \frac{\sqrt{\gamma}g^*}{\Delta-i\frac{\gamma}{2}} F(z,t)\\
\frac{\partial}{\partial t} a_A(z,t)=& ... - \frac{\sqrt{\gamma}\Omega^*}{2\Delta-i\gamma} F(z,t).
\end{split}
\label{eq:lambda_noise}
\end{equation}
%
%
 The noise $F(z,t)$ appearing here is defined by first defining
 \begin{equation}
F(\rvec,t)=\frac{1}{\sqrt{n(\rvec,t )}} \sum_j F^j_{g0,e}(t) \delta(\rvec-\rvec_j).
\end{equation}
Integrating the equations of motion over the transverse coordinates we define $F(z,t)$ in analogy with Eq. (\ref{eq:aatom_transverse_lambda}) (ignoring the mode index $m$ for a single mode). The resulting operator has the standard expectation value of vacuum noise $\la F^\dagger(z,t)F(z',t') \ra=0$, $\la F(z,t)F^\dagger(z',t') \ra=\delta(z-z')\delta(t-t')$. The addition of these noise operators ensures that the atom and light operators retains the correct commutation relations.

For the parametric gain the situation is a little different. The light operator $a_L$ couples to the coherence $|1\ra\la e|$ and we should still use the substitution (\ref{eq:substitute_delta}) with the minus sign. The atomic coherence $|0\ra\la1|$ again couples to  $|0\ra\la e|$ and $|e\ra\la 1|$, but now we can no longer neglect  $|e\ra\la 1|$, which gives rise to the AC-Stark shift [the first term in the spin equation in Eq. (\ref{eq:gain_nodecay})]. We should therefore use the plus sign in the substitution (\ref{eq:substitute_delta}) for this term.

To include the change of the mean spin $\la j_x(\rvec,t)\ra$ we include the decay of the densities $n_0$ and $ n_1$ of atoms in states $|0\ra$ and $|1\ra$ given by
 \begin{equation}
\begin{split}
\frac{d}{dt} n_0(\vec{r},t)&\approx -\frac{\gamma|\Omega(\rvec,t)|^2}{4 \Delta^2+\gamma^2}  n_0(\rvec,t).\\
\frac{d}{dt}  n_1(\vec{r},t)&\approx 0
\label{eq:decay_density}
\end{split}
\end{equation}
Since we also assume that the decay takes the atoms to some auxiliary state $|a_m\ra$, an initially fully polarized state will remain fully polarized so that $\la j_x(\rvec,t)\ra \approx  n_0(\rvec,t)/2$ as long as the interaction with the quantum field is weak (note, however, that we are describing a situation leading to superradiant scattering so that this approximation may break down very quickly).
The contribution from the time derivative of the mean spin then exactly cancels the decay of $a_A(z,t)$ arising from the substitution (\ref{eq:substitute_delta}). Since there is no longer any decay, one also finds that the resulting equations of motion (\ref{eq:gain_full}) do not contain any noise.

Finally for the Faraday interaction the easiest way to proceed is again to combine the results for the beam splitter interaction and the parametric gain. Similarly to the discussion in Appendix \ref{sec:3d_ham} the equations of motion for the Faraday interaction can therefore be obtained by subtracting the right hand side of Eq. (\ref{eq:gain_full}) from the right hand side of Eq. (\ref{eq:lambda}) and dividing by $\sqrt{2}$. Furthermore, we  again ignore the change of the propagation caused by the index of refraction  because this is accompanied by a similar change in the propagation of the classical light. The resulting equations in terms of the $x$ and $p$ operators
are given in Eq. (\ref{eq:faraday_full}) without the noise operators.  The noise operators give the additional terms
\begin{equation}
\begin{split}
\frac{\partial}{\partial z}x_L(z,t)=& ... + \frac{g\sqrt{2\gamma}}{\sqrt{4\Delta^2+\gamma^2}} F_x(z,t)  \\
\frac{\partial}{\partial z}p_L(z,t)=&...  + \frac{g\sqrt{2\gamma}}{\sqrt{4\Delta^2+\gamma^2}} F_p(z,t) \\
\frac{\partial}{\partial t}x_A(z,t)=& ... - \frac{\Omega\sqrt{\gamma}}{\sqrt{4\Delta^2+\gamma^2}} F_x(z,t) \\
\frac{\partial}{\partial t}p_A(z,t)=& ... - \frac{\Omega\sqrt{\gamma}}{\sqrt{4\Delta^2+\gamma^2}} F_p(z,t).
\label{eq:faraday_noise}
\end{split}
\end{equation}
Here we have for simplicity assumed $\Omega$ and $g$ to be real and have defined new noise operators
\begin{equation}
\begin{split}
F_x(z,t)=&\frac{1}{\sqrt{2}}{\left({\rm e}^{i\phi} F(z,t)+{\rm e}^{-i\phi}F^\dagger(z,t)\right)}\\
F_p(z,t)=&\frac{1}{\sqrt{2}i}{\left({\rm e}^{i\phi} F(z,t)-{\rm e}^{-i\phi} F^\dagger(z,t)\right)},
\end{split}
\end{equation}
which have the standard commutation relation
$[F_x(z,t),F_p(z',t')]= i\delta(z-z')\delta(t-t').$

\subsection{Dimensionless equations of motion}
\label{sec:dimensionless}
Casting the equations of motion for all three basic interactions in a dimensionless form allows to see that the constant $\kappa$ (\ref{eq:faraday_weakdecay_result}), naturally plays the role of the coupling constant for all protocols. We introduce the dimensionless position and time
coordinates $s=z/L$ and  $v=h(0,t)/h(0,T)$ running from 0 to 1. The rescaled time variable simplifies the equations because it is proportional to the total integrated intensity of the field. In the weak saturation limit the dynamics is completely controlled by the incident number of photons in the classical field. Changing the intensity will thus influence the temporal dynamics of the system, but the final state primarily depends on the total number of incident photons. When using such rescaled coordinates it is desirable to also change the field operators such that, e.g., an incident light field operator which is normalized in time $[a_L(t),a_L^\dagger(t')]=\delta(t-t')$  is now normalized relative to the new time variable $[a_L(v),a_L^\dagger(v')]=\delta(v-v')$. This normalization is achieved with the rescaling
\begin{equation}
\begin{split}
 \tilde a_A(s=z/L)&=\sqrt{L} a_A(z)\\
\tilde a_L(v(t))&=-\frac{\kappa \sqrt{4\Delta^2+\gamma^2}}{\sqrt{d\gamma}\Omega(t)} a_L(t),
\label{eq:rescale}
\end{split}
\end{equation}
where we have used the  dimensionless coupling constant $\kappa=\sqrt{h(0,T)}$, which appeared in the solution for  the Faraday interaction, c.f., Eq. (\ref{eq:kappa_full}) in the far off-resonant limit $\Delta\gg \gamma$.
Below we omit the tilde on the operators on the left hand side. In these new rescaled variables the equations of motion for the beam splitter interaction (\ref{eq:lambda}) become
\begin{equation}
\begin{split}
\frac{\partial }{\partial s} a_L(s,v)&=i \frac{\gamma d}{2(2\Delta-i\gamma)} a_L(s,v) -i \kappa \frac{e^{i\phi}}{2} a_A(s,v)\\
\frac{\partial }{\partial v} a_A(s,v)&=i \kappa^2 \frac{\Delta+i\frac{\gamma}{2}}{\gamma d} a_A(s,v) -i \kappa \frac{e^{i\phi}}{2} a_L(s,v).
\end{split}
\end{equation}

With the same rescaling for the parametric gain interaction (\ref{eq:gain_full}) we find
\begin{equation}
\begin{split}
\frac{\partial }{\partial s} a_L(s,v)&= -i \kappa \frac{e^{i\phi}}{2} a_A^\dagger (s,v)\\
\frac{\partial }{\partial v} a_A(s,v)&=-i \frac{\kappa^2\Delta
}{\gamma d} a_A(s,v) -i \kappa \frac{e^{i\phi}}{2} a_L^\dagger(s,v),
\end{split}
\label{eq:gain_rescale}
\end{equation}
where we have again assumed that $g$ is real.

For the Faraday interaction the equations of motion (\ref{eq:faraday_full}) become
\begin{equation}
\begin{split}
\frac{\partial}{\partial s} x_L(s,v)&= \kappa p_A(s,v)-\frac{\gamma^2}{4\Delta^2+\gamma^2} \frac{d}{2} x_L(s,v)\\
\frac{\partial}{\partial s} p_L(s,v)&= -\frac{\gamma^2}{4\Delta^2+\gamma^2} \frac{d}{2} p_L(s,v)\\
\frac{\partial}{\partial v} x_A(s,v)&= \kappa p_L(s,v)- \frac{\kappa^2}{2d} x_A(s,v)\\
\frac{\partial}{\partial v} p_A(s,v)&= - \frac{\kappa^2}{2d} p_A(s,v),
\end{split}
\label{eq:faraday_rescale}
\end{equation}
where we have ignored small corrections which vanish in the far detuned limit.

\subsection{Tensor Decomposition}\label{Appendix:TensorDecomposition}

The coefficients determining the strength of the irreducible tensor components are
\begin{multline*}
a_k(\Delta)=(-)^{1+F}c_k(2k+1)\sqrt{\frac{2F+1}{3}}\\
\times\left[\sum_{F'}\frac{(-)^{F'}}{1-\delta_{F'}/\Delta}
(2F'+1)\SixJ{J'}{F'}{I}{F}{J}{1}^2\SixJ{F}{k}{F}{1}{F'}{1}\right],
\end{multline*}
where the expressions in curly brackets are $6j$-symbols, $\Delta=\Delta_{F+1}$, $\delta_{F'}=\Delta_{F+1}-\Delta_{F'}$ and
\begin{align*}
c_0&=1,\quad\quad c_1=\sqrt{\frac{2}{F(F+1)}}, \\
c_2&=-\frac{3}{\sqrt{10F(F+1)(2F-1)(2F+3)}}.
\end{align*}
The last line is valid for $F>1/2$, that is nuclear spin $I\neq0$, and has to be replaced by $c_2=0$ for $I=0$.

In the asymptotic limit of large (blue) detuning,
$-\Delta\gg\delta_{F'}$, the sum in square brackets can be
simplified by means of
\begin{multline*}
\sum_{F'=F-1}^{F+1}(-)^{F'}(2F'+1)\SixJ{J'}{F'}{I}{F}{J}{1}^2\SixJ{F}{k}{F}{1}{F'}{1}=\\
(-)^{(2J+2F+J'+I+k)}
\SixJ{J}{I}{F}{F}{k}{J}\SixJ{J}{J}{k}{1}{1}{J'}
\end{multline*}
to get
\begin{multline*}
a_k=\lim_{\Delta\rightarrow-\infty}a_k(\Delta)=(-)^{2J+F+J'+I+k+1}c_k(2k+1)\\
\times\sqrt{\frac{2F+1}{3}}\SixJ{J}{I}{F}{F}{k}{J}\SixJ{J}{J}{k}{1}{1}{J'}.
\end{multline*}
From this expression it is evident that $a_2$ has to vanish because
the triple $\{J,J,k\}=\{1/2,1/2,2\}$ does not satisfy the triangle
inequality. For the particular case of the Cesium ($I=7/2$)
$D_2$-line at $F=4\rightarrow F'=3,4,5$ the asymptotic values of the
non-vanishing coefficients are $a_0=1/6$ and $a_1=1/24$.

%
%

%

\end{document}